\documentclass[prd,aps,twocolumn,floats,floatfix,nofootinbib,superscriptaddress] {revtex4-1}

\usepackage{graphicx}
\usepackage{dcolumn}
\usepackage{bm}
\usepackage{graphics}
\usepackage{slashed}
\usepackage{amssymb}
\usepackage{natbib}
\usepackage{amsmath}
\usepackage{url}
\usepackage[usenames,dvipsnames,svgnames,table]{xcolor}
\usepackage{color}
\usepackage{xcolor}
\usepackage{tabularx}
\usepackage{hyperref}
\usepackage{enumitem}
\usepackage{soul}
\usepackage{grffile}
\usepackage{algorithm}
\usepackage[noend]{algpseudocode}
\usepackage{mathtools}
\usepackage[normalem]{ulem}

\def\beq{\begin{equation}}
\def\eeq{\end{equation}}
\newcommand{\bea}{\begin{eqnarray}}
\newcommand{\eea}{\end{eqnarray}}
\newcommand{\mhduet}{{\texttt{MHDuet}~}} 
\newcommand{\mhduetX}{{\texttt{MHDuet}}} 

\DeclareMathOperator{\sgn}{sgn}
\DeclareMathOperator{\Minmod}{Minmod}
\DeclareMathOperator{\Median}{Median}

\hypersetup{
	colorlinks=true,
	linkcolor=red,
	linkbordercolor=black,
	citecolor=blue,
	filecolor=magenta,      
	urlcolor=magenta,
}

\begin{document}

\title{ \mhduetX: a high-order General Relativistic Radiation MHD code for CPU and GPU architectures}

\author{Carlos Palenzuela}
\address{Departament  de  F\'{\i}sica,  Universitat  de  les  Illes  Balears,  Palma  de  Mallorca,  Baleares  E-07122,  Spain}
\address{Institute of Applied Computing \& Community Code (IAC3),  Universitat  de  les  Illes  Balears,  Palma  de  Mallorca,  Baleares  E-07122,  Spain}

\author{Miguel Bezares}
\address{Nottingham Centre of Gravity,
	Nottingham NG7 2RD, United Kingdom}
\address{School of Mathematical Sciences, University of Nottingham,
	University Park, Nottingham NG7 2RD, United Kingdom}

\author{Steven Liebling}
\address{Long Island University, Brookville, New York 11548, USA}	

\author{Federico Schianchi}
\address{Departament  de  F\'{\i}sica,  Universitat  de  les  Illes  Balears,  Palma  de  Mallorca,  Baleares  E-07122,  Spain}
\address{Institute of Applied Computing \& Community Code (IAC3),  Universitat  de  les  Illes  Balears,  Palma  de  Mallorca,  Baleares  E-07122,  Spain}

\author{Julio Fernando Abalos}
\address{Departament  de  F\'{\i}sica,  Universitat  de  les  Illes  Balears,  Palma  de  Mallorca,  Baleares  E-07122,  Spain}
\address{Institute of Applied Computing \& Community Code (IAC3),  Universitat  de  les  Illes  Balears,  Palma  de  Mallorca,  Baleares  E-07122,  Spain}

\author{Ricard Aguilera-Miret}
\address{University of Hamburg, Hamburger Sternwarte, Gojenbergsweg 112, 21029, Hamburg, Germany}
\address{Institute of Applied Computing \& Community Code (IAC3),  Universitat  de  les  Illes  Balears,  Palma  de  Mallorca,  Baleares  E-07122,  Spain}

\author{Carles Bona}
\address{Departament  de  F\'{\i}sica,  Universitat  de  les  Illes  Balears,  Palma  de  Mallorca,  Baleares  E-07122,  Spain}
\address{Institute of Applied Computing \& Community Code (IAC3),  Universitat  de  les  Illes  Balears,  Palma  de  Mallorca,  Baleares  E-07122,  Spain}

\author{Juan Antonio Carretero}
\address{Departament  de  F\'{\i}sica,  Universitat  de  les  Illes  Balears,  Palma  de  Mallorca,  Baleares  E-07122,  Spain}
\address{Institute of Applied Computing \& Community Code (IAC3),  Universitat  de  les  Illes  Balears,  Palma  de  Mallorca,  Baleares  E-07122,  Spain}

\author{Joan Mass\'o}
\address{Departament  de  F\'{\i}sica,  Universitat  de  les  Illes  Balears,  Palma  de  Mallorca,  Baleares  E-07122,  Spain}
\address{Institute of Applied Computing \& Community Code (IAC3),  Universitat  de  les  Illes  Balears,  Palma  de  Mallorca,  Baleares  E-07122,  Spain}

\author{Matthew P. Smith}
\address{Nottingham Centre of Gravity,
	Nottingham NG7 2RD, United Kingdom}
\address{School of Mathematical Sciences, University of Nottingham,
	University Park, Nottingham NG7 2RD, United Kingdom}

\author{Kwabena Amponsah}
\address{Digital Research Service, University of Nottingham, University Park Campus, NG7 2QL, United Kingdom
}

\author{Kacper Kornet}
\address{Research Computing Services, University of Cambridge,  7 JJ Thomson Avenue, Cambridge, CB3 0RB, United Kingdom}

\author{Borja Mi\~nano}
\address{Institute of Applied Computing \& Community Code (IAC3),  Universitat  de  les  Illes  Balears,  Palma  de  Mallorca,  Baleares  E-07122,  Spain}

\author{Shrey Pareek}
\address{Institute of Applied Computing \& Community Code (IAC3),  Universitat  de  les  Illes  Balears,  Palma  de  Mallorca,  Baleares  E-07122,  Spain}

\author{Miren Radia}
\address{Research Computing Services, University of Cambridge,  7 JJ Thomson Avenue, Cambridge, CB3 0RB, United Kingdom}

\begin{abstract}
We present \mhduetX, an open source evolution
code for general relativistic magnetohydrodynamics with neutrino transport. The code solves the full set of Einstein equations coupled to a relativistic, magnetized fluid with an M1 neutrino radiation scheme using advanced techniques, including
adaptive mesh and large eddy simulation techniques, to achieve high accuracy. The Simflowny platform generates the code from a high-level specification of the computational system, producing code that runs with either the SAMRAI or AMReX infrastructure. The choice of AMReX enables compilation and execution on GPUs, running
an order of magnitude faster than on CPUs at the node level.
We validate the code against benchmark tests, reproducing previous results obtained with the SAMRAI infrastructure, and demonstrate its capabilities with simulations of neutron stars employing realistic tabulated equations of state. 
Resolution studies clearly demonstrate convergence faster than second order in the grid spacing. Scaling tests reveal excellent strong and weak scaling performance when running on GPUs.
The goal of the code is to provide a powerful tool for studying the dynamics of compact objects within multi-messenger astrophysics.
\end{abstract}

\maketitle
\tableofcontents

\section{Introduction}

The era of multi-messenger astronomy with gravitational waves and  electromagnetic counterparts was inaugurated with the detection of GW170817, consistent with the coalescence of a binary neutron star system~\cite{PhysRevLett.119.161101}. The LIGO-Virgo-KAGRA (LVK) collaboration reported the gravitational-wave signal, and, within 1.7~seconds of its peak, a short gamma-ray burst~(GRB) was observed~\cite{Abbott_2017}. From the same region of the sky, a kilonova (i.e., thermal emission powered by the radioactive decay of the neutron-rich ejecta expelled during the merger) was detected a few hours later~\cite{Balasubramanian:2021kny}. This single event has already provided invaluable insights, constraining the equation of state~(EoS) of matter at supranuclear densities~\cite{LIGOScientific:2018cki}, refining measurements of the Hubble constant~\cite{LIGOScientific:2017adf}, and ruling out several alternative theories of gravity~\cite{LIGOScientific:2018dkp}. However, to fully exploit the physical insights encoded from these observations, it is essential to develop theoretical models that can consistently connect the dynamics of matter, magnetic fields, and radiation in the strong-gravity regime to observable signatures.

In this context, general relativistic radiation magnetohydrodynamic (GRRMHD) simulations have become indispensable tools. They have revealed several key mechanisms for binary neutron star (BNS) and black hole–neutron star (BHNS) mergers.
On the one hand, magnetic fields are strongly amplified in the post-merger phase through various processes, some of them occurring in very small scales, and play a crucial role in launching relativistic jets that may power short gamma-ray bursts~(sGRBs) (see for instance \cite{Ciolfi2020b,2024arXiv240510081K} and references within).
On the other hand, matter ejection processes that contribute to kilonova emission rely critically on neutrino radiation transport, which is indispensable for accurately determining the composition and temperature of the ejecta, as both are strongly influenced by neutrino interactions (see for instance \cite{foucart_review,2024AnP...53600306R} and references within). 
These multi-physics processes occur across widely different scales, spanning several orders of magnitude in both space and time, which makes the simulations highly demanding and limits their ability to capture all relevant scales. This complexity, in turn, places stringent requirements not only on the underlying physical models but also on the computational methods and hardware used to evolve them.

Consequently, numerous codes have been developed over the past decades to address these problems with increasing accuracy and realism. While most existing GRRMHD codes are primarily designed for central processing unit (CPU) architectures, recent years have seen significant progress in porting them to the next level: efficient execution on graphical processing unit (GPU) architectures, with the associated gains in speed and scalability. Notable examples include AsterX~\cite{Kalinani:2024rbk}, Gram-X~\cite{Shankar:2022ful}, and AthenaK~\cite{Fields:2024pob}.

In this manuscript, we present the latest version of our open-source code, \mhduetX~\cite{mhduet}, describing in detail our efforts to adapt for high-performance GPU computing. In summary, \mhduet evolves magnetized fluids coupled to neutrinos within fully nonlinear general relativity. \mhduet achieves efficient performance on CPUs through both SAMRAI --a patch-based structured adaptive mesh refinement (AMR), developed over more than $20$ years at the Lawrence Livermore National Laboratory-- and AMReX --an AMR infrastructure developed as part of the DOE’s Exascale Computing Project--, and more importantly, also on GPUs via AMReX.
Unlike other codes, \mhduet is generated using Simflowny~\cite{Arbona:2013,Arbona:2017,Palenzuela_2021}, a simulation platform developed by the IAC3 group in Mallorca since 2008. Given a set of PDEs, Simflowny automatically produces the code required to numerically evolve these equations, greatly simplifying the use of HPC infrastructures for non computer-science specialists. This automatic code generation also enables rapid experimentation with different numerical techniques and physical models, a particularly valuable feature in emerging fields where models are still under active investigation.
The first version of the \mhduet code was introduced in Ref.~\cite{palenzuela18}, although without any specific name, and it already incorporated several of the features and capabilities presented here. These features have since been extended and enhanced to enable efficient simulations of fully relativistic astrophysical scenarios.
On the physics side, \mhduet includes fully nonlinear gravity by solving the conformal and covariant Z4, a strongly hyperbolic evolution formalism of Einstein equations; magnetohydrodynamics (MHD) with hyperbolic divergence cleaning; support for both tabulated, realistic finite-temperature equations of state (EoS) and an analytical hybrid EoS; and an M1 truncated moment scheme for neutrino transport. The code has also been extended to evolve scalar fields and explore theories beyond general relativity.
On the numerical side, the code incorporates high-resolution shock-capturing~(HRSC) methods; high-order adaptive mesh refinement~(AMR) with subcycling in time; Large Eddy Simulation (LES) techniques; implicit-explicit~(IMEX) Runge-Kutta time-integrators; and the ability to run efficiently on both CPUs and GPUs using the same codebase.

Different versions of \mhduet have been applied to a wide range of problems. The scalability of the new code and its comparison with our previous code HAD were studied in Ref.~\cite{Liebling:2020jlq}. High-density phase transitions in neutron star dynamics were explored in Ref.~\cite{Liebling:2020dhf}. The extension of Large-Eddy Simulation techniques with the gradient sub-grid-scale model to general relativistic MHD for neutron star merger simulations was developed in Refs.~\cite{vigano20,Aguilera-Miret:2020dhz}. It has also been used in to study binary neutron star mergers, focusing on the turbulent dynamo amplification of magnetic fields at merger and their effects on the remnant~\cite{Aguilera-Miret:2021fre,Palenzuela:2022kqk,Palenzuela:2021gdo,2023PhRvD.108j3001A,aguileramiret2024,aguileramiret2025}, with analogous studies carried out for black hole–neutron star mergers~\cite{Izquierdo_2024B,izquierdo2025}.
On more exotic fronts, \mhduet has been used to study gravitational wave emission from mergers of unequal-mass, highly compact boson stars~\cite{Bezares_2022b} and weakly interacting “dark” boson stars~\cite{bezpalen2}, as well as to investigate kinetic screening in binary neutron star mergers beyond General Relativity~\cite{Bezares_2022,Cayuso:2024ppe}.

Here we provide a comprehensive description of the \mhduet code, beginning with the equations of motion, followed by the numerical methods, and concluding with its high-performance computing implementation. We present tests demonstrating that the code reproduces results obtained with the SAMRAI infrastructure, confirming the validity of previous benchmarks. Additionally, we include tests involving neutron stars and using the newly incorporated tabulated EoS, showing convergence with increasing resolution and a reduction in constraint residuals. Finally, scaling results indicate that, when running on GPUs, the code achieves more than an order-of-magnitude speedup compared to CPUs on a node-to-node basis.

Unless otherwise stated, we adopt a unit system in which \(G = c = M_\odot = 1\). We follow the standard index convention, using the first Latin indices \(\{a, b, \dots\}\) to denote spacetime components and \(\{i, j, \dots\}\) to represent purely spatial components. The metric signature is chosen as \((-,+,+,+)\).

\section{Evolution formalism}

The spacetime deformation induced by a self-gravitating, magnetized fluid with neutrino transport is modeled by the covariant Einstein equations, in which the Einstein tensor $G_{ab}$ (encoding the curvature) is coupled to the stress-energy tensor $T_{ab}$ (describing the matter content). 
The stress-energy  is a combination of the contributions from the magnetized perfect fluid, $T_{a b}^{\text{mhd}}$ and from the neutrino radiation, $T_{a b}^{\text{rad}}$. The latter can be decomposed into three components, corresponding to three independent neutrino species; the electron neutrinos, electron anti-neutrinos, and heavy neutrinos, denoted by $ \left\{\nu_e, \bar{\nu}_e, \nu_x\right\}$. 
The Einstein equations then appear as
\begin{eqnarray}
	G_{ab} &=& 8\pi T_{a b} = 8\pi \left(T_{a b}^{\text{mhd}} + T_{ab}^{\text{rad}} \right) \,, \\
	T_{ab}^{\text{rad}} &=& \sum_{\nu_i} T_{ab}^{\nu_i}
	= T_{ab}^{\nu_e} + T_{ab}^{\bar{\nu}_e} + T_{ab}^{\nu_x} \,.
\end{eqnarray}

The dynamics of the matter are described by a series of covariant, quasi-conservation laws as:
\begin{itemize}
	\item The energy and momentum densities of each neutrino species,
	within the truncated moments formalism, obey the subsequent equations
	\begin{eqnarray}
		\nabla_{a} T^{a b}_{\nu_i} &=&  \mathcal{S}_{\nu_i}^{b} \,.
	\end{eqnarray}
	where the radiation force ${\cal S}^b_{\nu_i}$ accounts for the changes to the energy and momentum of the neutrinos due to their interaction with the fluid.	
	By using the $3+1$ decomposition,
	this equation can be written explicitly as a balance-law system almost equivalent to the one from general relativistic hydrodynamics.	
	\item The equations for a magnetized fluid are significantly simplified assuming the ideal MHD approximation $F_{ab}\,u^b = 0$, where $F_{ab}$ is the Faraday tensor describing the electromagnetic field and $u^b$ the fluid four-velocity. This condition ensures that the electric field vanishes in the frame comoving with the fluid, leading to the conservation of magnetic flux in the fluid. Therefore, the evolution equations follow from the  conservation of the total stress-energy tensor $T_{ab}$ (total energy and linear momentum conservation), and the conservation of the dual of the Faraday tensor ${}^* F^{ab}$ (conservation of magnetic flux),  namely
	\begin{eqnarray}
		\nabla_{a}T^{a b}_{\text{mhd}} &=& - \sum_{\nu_i} \mathcal{S}^{b}_{\nu_i} 
		= - \left(  \mathcal{S}^{b}_{\nu_e} + \mathcal{S}^{b}_{\bar{\nu}_e} + \mathcal{S}^{b}_{\nu_x} \right) \,, \\
		\nabla_{a}\text{*} F^{ab} &=& 0 \,,
	\end{eqnarray}
	By using the $3+1$ decomposition,
	these equation can be written explicitly as a balance-law system of general relativistic magneto-hydrodynamics with an extra source term arising from the interaction of the fluid with the neutrinos.
	\item The conservation of baryon number $n_b$, neutrino number $N_{\nu}^{a}$ and lepton number densities, can be expressed as
	\begin{eqnarray}
		\nabla_{a}(\rho u^{a}) &=& 0 \,, \\
		\nabla_{a} N_{\nu_i}^{a} &=& {\cal C}_{\nu_i} \,,		\\	
		\nabla_{a}(Y_{e}\rho u^{a}) &=& m_b \left( {\cal C}_{\bar{\nu}_e} - {\cal C}_{\nu_e} \right) \,,
	\end{eqnarray}
	where ${\cal C}_{\nu_i}$ accounts for changes to the neutrino densities due to their interaction with the fluid.
	Here we have defined the rest-mass density $\rho = m_b n_b = m_b (n_p + n_n)$, where $m_b$ is the reference baryon mass and $n_i$ is the number density of species $i$ in the fluid frame. The electron fraction $Y_e$ is defined as the net number of electrons per baryon. Due to charge neutrality in a neutral fluid, and ignoring heavier lepton families, the net electron number density (i.e, electrons minus positrons) match the number of protons $n_p = n_{e^-} - n_{e^+}$. 
	Therefore, we can finally write:
	\begin{equation}\label{Ye}
		Y_e = \frac{n_{e^-} - n_{e^+} }{n_p + n_n} \,.
	\end{equation}	
	Using the $3+1$ decomposition,
	these equations can also be written explicitly as balance-law equations.	
\end{itemize}

\subsection{Einstein equations}

The Einstein equations can be written as an explicit evolution system 
with a 3+1 decomposition,
which foliates the spacetime by a set of spacelike hypersurfaces. These hypersurfaces are  labeled by a time coordinate $t$ and
endowed with spatial coordinates $x^i$.  Within the 3+1 decomposition, the spacetime metric can be expressed as
\begin{eqnarray}
	ds^2 = -\alpha^2\,dt^2 
	+ \gamma_{ij}\left(dx^i + \beta^i\,dt\right)\left(dx^j + \beta^j\,dt\right),
	\nonumber
\end{eqnarray}
where $\alpha$ is the lapse function, $\beta^{i}$ is the shift vector, and $\gamma_{ij}$ is the induced 3-metric on each spatial slice.

We can define the time-like normal four-vector to the hypersurfaces $n_a = (-\alpha, 0)$ satisfying the normalization condition $n_a n^a = -1$. We can then construct the projector $\gamma_{ab} = g_{ab} + n_a n_b$, which projects any tensor onto the spacelike hypersurfaces. 
Finally, we can use the unit normal to define  the standard extrinsic curvature $K_{ij}$
as the Lie derivative of the metric	$K_{ij} = -\frac{1}{2}\, \mathcal{L}_n \gamma_{ij}$.

We consider the Z4 formalism of the Einstein equations that extend the original equations with the covariant derivatives of an auxiliary four-vector \(Z^a\), namely
\begin{eqnarray}
	R_{ab} &+& \nabla_a Z_b + \nabla_b Z_a = 8\pi\left(T_{ab} - \frac{1}{2} g_{ab} T\right) \nonumber \\
	&+& \kappa_{z} \, \left(  n_a Z_b + n_b Z_a - g_{ab}\, n^c Z_c \right),
	\label{eq:Z4_modified_Einstein}
\end{eqnarray}
such that the full set of dynamical variables  consists of the pair \(\{g_{ab}, Z^{a}\}\). 
The four-vector $Z_a$ can be interpreted as the time integral of the standard energy and momentum constraints, and thus it should vanish identically if such constraints are exactly satisfied. These physical constraints,
\begin{eqnarray}
	\Theta \equiv -n_a Z^a = 0 ~~,~~ Z_i \equiv \gamma^a_i Z_a =0  ~~,~~
\end{eqnarray}
are enforced dynamically by introducing linear terms in the evolution equations proportional to the damping coefficient $\kappa_z$ and to the constraints themselves. These damping terms ensure the exponential decay of constraint violations by efficiently controlling any deviation from \(Z^a = 0\). 

Our final system of evolution equations is based on the covariant conformal Z4 (CCZ4) formulation \cite{alic12, bezares17, palenzuela18}, which is just a conformal decomposition of the Z4 formalism supplemented by additional damping terms. The resulting evolution equations are expressed in terms of the conformal fields
\begin{eqnarray}
	& & 	\tilde{\gamma}_{ij} = \chi\,\gamma_{ij}
	\quad , ~
	{\tilde A}_{ij} = \chi \left( K_{ij} - \gamma_{ij} \frac{K}{3} \right) \nonumber \\
	& & {\hat \Gamma}^i = {\tilde \Gamma}^i + {2 \over \chi} Z^i 
	\quad,~
	{\hat K} \equiv K - 2\, \Theta \nonumber
\end{eqnarray}
where $\chi = (\det \gamma_{ij})^{-1/3}$ is the conformal factor and ${\tilde \Gamma}^i = {\tilde \gamma}^{ij} {\tilde \gamma}^{kl} \partial_l {\tilde \gamma}_{jk}$ the contracted Christoffel symbol of the conformal metric.
These new field definitions introduce some conformal constraints, namely
\begin{eqnarray}
  {\tilde \gamma} \equiv \det({\tilde \gamma}) =1  ~~,~~
  {\tilde A} \equiv \gamma^{ij} {\tilde A}_{ij} = 0 ~~.
\end{eqnarray}
They can also be enforced dynamically by adding linear terms in the evolution equations. These terms, proportional to the conformal constraints themselves and to the damping coefficient $\kappa_c$, ensure the exponential decay of the conformal constraint violations. 

The full set of evolution equations within this scheme, written as a function of the new conformal fields, and including all these damping terms, reads
\begin{widetext}
	\begin{eqnarray}
		\partial_t {\tilde \gamma}_{ij} 
		& =& \beta^k \partial_k {\tilde \gamma}_{ij} + {\tilde \gamma}_{ik} \, \partial_j \beta^k 
		+ {\tilde \gamma}_{kj} \partial_i \beta^k - {2\over3} \, {\tilde \gamma}_{ij} \partial_k \beta^k
		- 2 \alpha \Bigl( {\tilde A}_{ij}  - {1 \over 3} {\tilde \gamma}_{ij}\, {\tilde A} \Bigr) -  \frac{\alpha}{3}\kappa_{c}\tilde{\gamma}_{ij}\ln\tilde{\gamma} \label{system1}
		\\
		\partial_t {\tilde A}_{ij} 
		& =& \beta^k \partial_k{\tilde A}_{ij} + {\tilde A}_{ik} \partial_j \beta^k 
		+ {\tilde A}_{kj} \partial_i \beta^k - {2\over3} \, {\tilde A}_{ij} \partial_k \beta^k - \frac{\alpha}{3}\kappa_{c}\tilde{\gamma}_{ij}\tilde{A}
		\\
		& +& \chi \, \Bigl[ \, \alpha \, \bigl( {^{(3)\!}R}_{ij} + D_i Z_j + D_j Z_i 
		- 8 \pi G \, S_{ij} \bigr)  - D_i D_j \alpha \, \Bigr]^{\rm TF} 
		+ \alpha \, \Bigl( {\hat K} \, {\tilde A}_{ij} - 2 {\tilde A}_{ik} {\tilde A}^k{}_j \Bigr) 
		\nonumber \\
		\partial_t \chi & =& \beta^k \partial_k \chi 
		+ {2\over 3} \, \chi \, \bigl[ \alpha ({\hat K} + 2\, \Theta) - \partial_k \beta^k  \bigr] 
		\\
		\partial_t {\hat K} 
		& =&  \beta^k \partial_k {\hat K} 
		- D_i D^i \alpha
		+ \alpha \, \Bigl[ {1 \over 3} \bigl( {\hat K} + 2 \Theta \bigr)^2 
		+ {\tilde A}_{ij} {\tilde A}^{ij} + 4\pi G \bigl(\tau + S\bigr)
		+ \kappa_z (1 - \kappa_2) \Theta \Bigr]  
		\nonumber \\
		&+& 2\, Z^i \partial_i \alpha 
		\\
		\partial_t \Theta 
		& =&  \beta^k \partial_k \Theta + {\alpha \over 2} \Bigl[ {^{(3)\!}R} + 2 D_i Z^i
		+ {2\over3} \, {\hat K}^2 + {2\over3} \, \Theta \Bigl( {\hat K} - 2 \Theta \Bigr)
		- {\tilde A}_{ij} {\tilde A}^{ij}  \Bigr] - Z^i \partial_i \alpha 
		\nonumber \\
		&-& \alpha \, \Bigl[ 8\pi G \, \tau  + \kappa_z (2 + \kappa_2) \, \Theta \Bigr] 
		\\   
		\partial_t {\hat \Gamma}^i 
		& =& \beta^j \partial_j {\hat \Gamma}^i - {\hat \Gamma}^j \partial_j \beta^i 
		+ {2\over3} {\hat \Gamma}^i \partial_j \beta^j + {\tilde \gamma}^{jk} \partial_j \partial_k \beta^i
		+ {1\over3} \, {\tilde \gamma}^{ij} \partial_j \partial_k \beta^k 
		\\
		& -& 2 {\tilde A}^{ij} \partial_j \alpha + 2\alpha \, \Bigl[ {\tilde \Gamma}^i{}_{jk} {\tilde A}^{jk}
		- {3 \over 2 \chi} \, {\tilde A}^{ij} \partial_j \chi 
		- {2\over3} \, {\tilde \gamma}^{ij} \partial_j {\hat K} - 8\pi G \, {\tilde \gamma}^{ij} \, S_i \Bigr] 
		\\ 
		& +& 2 \alpha \, \Bigl[- {\tilde \gamma}^{ij} \bigl( {1 \over 3}\partial_j \Theta 
		+ {\Theta \over \alpha} \, \partial_j \alpha \bigr) 
		- {1 \over \chi} Z^i \bigl( \kappa_z + {2\over 3} \, ({\hat K} + 2 \Theta) \bigr) \Bigr]  
	\end{eqnarray}
\end{widetext}
where the expression $[\ldots]^{\rm TF}$ indicate the trace-free {(or trace-less)} component with respect to the metric $\tilde{\gamma}_{ij}.$ 
The Ricci terms can be written now as
\begin{eqnarray} 
	{^{(3)\!}R}_{ij} & +& 2 D_{(i} Z_{j)} 
	= {^{(3)\!}{\hat R}}_{ij} + {\hat R}^\chi_{ij} 
	\\   
	\chi {\hat R}^{\chi}_{ij} & =&  {1 \over 2} \, \partial_i \partial_j \chi 
	- {1 \over 2} \, {{\tilde \Gamma}^k}_{ij} \partial_k \chi 
	- {1 \over 4 \chi} \, \partial_i \chi \partial_j \chi 
	+ {2 \over \chi} Z^k {\tilde \gamma}_{k(i} \partial_{j)} \chi
	\nonumber \\ 
	& +& {1 \over 2 }{\tilde \gamma}_{ij} \, \Bigl[ {\tilde \gamma}^{km} \Bigl( {\partial}_k {\partial}_m \chi 
	-  {3\over 2 \chi} \, \partial_k \chi \partial_m \chi \Bigr)
	- {\hat \Gamma}^k \partial_k \chi \Bigr] 
	\\   
	{\hat R}_{ij} & =& - {1\over2} \, {\tilde \gamma}^{mn} \partial_m \partial_n {\tilde \gamma}_{ij}     
	+ {\tilde \gamma}_{k(i} \partial_{j)} {\hat \Gamma}^k + {\hat \Gamma }^k {\tilde \Gamma}_{(ij)k} 
	\nonumber \\        
	&+& {\tilde \gamma}^{mn} 
	\Bigl(  {{\tilde \Gamma}^k}_{mi} {\tilde \Gamma}_{jkn} 
	+ {{\tilde \Gamma}^k}_{mj} {\tilde \Gamma}_{ikn} + {\tilde \Gamma}^k{}_{mi} {\tilde \Gamma}_{knj}  \Bigr) 
\end{eqnarray} 
and the Laplacian of the lapse is just
\begin{equation} 
	D_i D^i \alpha 
    =\chi \, {\tilde \gamma}^{ij} \partial_i \partial_j \alpha - \chi {{\tilde \Gamma}^k} \partial_k \alpha
	- {1 \over 2} {\tilde \gamma}^{ij}\, \partial_i \alpha\, \partial_j \chi .
\end{equation}
Note that the following projections of the matter stress-energy 
appear in the CCZ4 equations
\begin{equation}
	\tau = n_a \, n_b \, T^{ab}  ~,~ 
	S_i = - n_a\, \gamma_{bi}\, T^{ab} ~,~
	S_{ij} = \gamma_{ai}\, \gamma_{bj}\, T^{ab} \,
	\nonumber
\end{equation}
corresponding to the energy density, the linear momentum density, and the stress tensor density. 

Einstein’s equations are covariant and hold in any coordinate system. However, to perform a numerical simulation, a specific choice of coordinates must be made. Within the 3+1 approach, this is achieved by prescribing evolution equations for the lapse and shift. These gauge conditions must be chosen carefully to ensure the well-posedness of the resulting system.
We use the 1+log slicing condition~\cite{BM} with a 
simplified version of the Gamma-freezing shift condition~\cite{alcub}, namely
\begin{eqnarray}
	\partial_t \alpha & =& \beta^i \partial_i \alpha 
	- 2 \,\alpha \, f_{\alpha}(\alpha) \, {\hat K} 
	\\ 
	\partial_t \beta^i & =& \beta^j \partial_j \beta^i + {3\over4} \, f_{\beta}(\alpha) \, {\hat \Gamma}^i - \eta (\beta^i - \beta^i_0)
	\label{system3}
\end{eqnarray}
where $\eta$ is a parameter scaling with the total mass such that $ \eta \approx 2/M$. In certain cases, the asymmetric ejection of mass and/or the emission of gravitational waves can impart a kick to the final remnant. To correct for this, we apply a small coordinate shift $\beta^i_0$ to realign the center of mass with the origin of our coordinate system~\cite{Hayashi:2022cdq}. Our standard gauge conditions employ $f_{\alpha}=f_{\beta}=1$ and $\beta^i_0=0$. The damping coefficients also scale with the total mass, and we usually set them to $\kappa_z \leq 1/M$ and $\kappa_c \leq 0.1/M$ in scenarios involving neutron stars.

\subsection{Magnetized perfect fluid}	
\label{sec:fluid}

The stress-energy tensor for a perfect fluid in the presence of an electromagnetic field
can be written as the combination of the two components, namely
\begin{eqnarray}\label{stress-energy-perfectfluid}
	&&T^\mathrm{mhd}_{ab} = T_{ab}^\mathrm{fluid} + T_{ab}^\mathrm{EM} 
	\\
	&=& \left[ \rho (1 + \epsilon) + p \right] u_{a} u_{b} + p g_{ab}
	\nonumber 
	+ {F_{a}}^{c} F_{bc}
	- \frac{1}{4} g_{ab} ~ F^{cd} F_{cd} ~~.
	\nonumber
\end{eqnarray}
Here we have introduced the following
fields to describe the state of a magnetized, perfect fluid, as measured by a co-moving observer: the rest-mass density of the fluid $\rho$; the  electron fraction $Y_e$, that measures the ratio of electrons per baryon; the internal energy $\epsilon$, which accounts 
for the thermal and binding energies; the fluid pressure $p$, which  contains information about the microphysical interaction of its constituents and is given by an equation of state~(EoS) relating the pressure to the other thermodynamic variables;
and the four-velocity $u^{a}$ that measures how the fluid moves with respect to Eulerian observers (i.e., ones moving along the normal $n^a$ to the space-like hypersurfaces). This four-velocity is commonly normalized $u^{a} u_{a} = -1$, and it can be decomposed into spatial and temporal components, that is, $u^{\mu} = W n^{\mu} + W v^{\mu}$. Here $v^{\mu}$ corresponds to the familiar three-dimensional velocities, while $W = - n_{a} u^{a} = (1 - v_i v^i)^{-1/2}$ is the standard Lorentz factor. 

The Faraday tensor $F_{ab}$ can be written in terms of the electric and magnetic fields $\left\{E^i, B^i\right\}$. In the ideal MHD approximation, the electric field in the co-moving frame vanishes and therefore it can be calculated in any other frame just using the magnetic field and the four-velocity. 
A common technique to 
dynamically control the solenoidal constraint, known as \textit{divergence cleaning}~\cite{2002JCoPh.175..645D}, introduces a new scalar field  $\phi$ in the Maxwell equations equivalent to the $Z^a$ four-vector in the Z4 formulation, namely
\begin{equation}\label{div_cleaning}
	  \nabla_{a}(  \text{*} F^{ac} + \hat{g}^{ac}\phi)  = \frac{\kappa_{b}%
	}{c_{b}^{2}}n^{c}\phi 
\end{equation}
with $\hat{g}^{ac} \equiv g^{ac}+(1- 1/c_{b}^{2}) n^{a}n^{c}$. This new field $\phi$, which propagates in a mode with a freely-specifiable speed $c_b$, can be interpreted as the time integral of the solenoidal constraint. Again, this constraint 
is damped exponentially by adding linear terms proportional to the damping parameter coefficient $\kappa_b$~\cite{Liebling_2010}. 
Therefore, the final set of physical or primitive MHD fields that describe the state of a magnetized perfect fluid is given by $(\rho,Y_e, \epsilon,p,v^{i}, B^i, \phi)$.

In order to capture properly the weak solutions of the non-linear general relativistic MHD equations 
in the presence of shocks, it is important to write the evolution system in local-conservation or balance law form~\cite{palenzuela15}. 
The conservation of baryon and lepton numbers, together with the conservation of total energy, linear momentum, and magnetic flux, provide a set of evolution equations for the conserved variables $ \left( \overline{D},\overline{D\!Y}_e, \overline{S}_i , \overline{\tau}, \overline{B}^i , \overline{\phi} \right) = \sqrt{\gamma} \left( D, D\!Y_e, \tau, S_i, B^i, \phi \right) $. The non-trivial relations between the conserved and   the primitive fields arise from their definitions:
\begin{eqnarray}\label{def_con2prim}
	D &=& \rho W  
	 \\
	D\!Y_e &=& \rho W Y_e 
	\nonumber \\   
	\tau &=&  h W^2 -p + B^2
	- \frac{1}{2} \bigg[ (B^k v_k)^2 + \frac{B^2}{W^2} \bigg] - D 
	\nonumber \\
	S_i &=&  (h W^2 + B^2) v_i - (B^k v_k) B_i  
	\nonumber 
\end{eqnarray}
where $h = \rho (1 + \epsilon) + p$ is the enthalpy of the fluid.

{As shown below,
the MHD equations are written in a mixture of conserved and primitive variables.} Consequently, the evolution of such system requires solving the primitive fields from the evolved conserved one at each time step. This algebraic (but transcendental) system of equations only becomes closed once the equation of state 
is provided, and can only be solved numerically in general.

There are two commonly used EoS types in the context of neutron star astrophysics. The first one is an analytical {\em hybrid EoS} $p=p(\rho, \epsilon)$, which splits the contributions to the pressure and the internal energy into a cold and a thermal part
\begin{eqnarray} 
	p = p_\mathrm{cold} (\rho) + p_\mathrm{th}  ~~,~~
	\epsilon = \epsilon_\mathrm{cold} (\rho) + \epsilon_\mathrm{th} ~~.
\end{eqnarray}
The thermal pressure is modeled by using the ideal gas EoS $p_\mathrm{th} = (\Gamma_\mathrm{th} - 1) \rho \epsilon$.
Here, $\Gamma_\mathrm{th}$ is the adiabatic index.
The cold contributions to pressure and the internal energy only depend on the density and can be modeled with a piece-wise polytrope via
\begin{eqnarray}\label{piecewise_p_eps_cold}
	p_\mathrm{cold} (\rho) = K_i \rho^{\Gamma_i} ~~,~~
	\epsilon_\mathrm{cold} (\rho) = a_i + \frac{K_i}{\Gamma_i -1} \rho^{\Gamma_i - 1} 
\end{eqnarray}
for $\rho_{i-1} \leq \rho \leq \rho_{i}$. 
Beginning with values for $K_0$ and $\Gamma_i$ and with $a_0=0$, the other constants $a_i$ and $K_{i}$ for $i > 0$
are found by imposing continuity.
In general, a good fit for most of the cold EoS can be obtained only with 4 different pieces.

The second popular, much more realistic EoS is given by a finite-temperature, {\em microphysical} EoS. This EoS is more relevant in relativistic astrophysical scenarios, especially those involving high temperatures and neutrino-rich environments such as core-collapse supernovae or neutron star mergers. In these cases, the pressure depends not only on the density \(\rho\) but also on additional variables such as the electron fraction \(Y_e\) and the temperature \(T\) (instead of the internal energy). Tabulated nuclear EoSs of the form $p=p(\rho,T,Y_e)$ incorporate finite-temperature nuclear physics, composition changes, and nuclear statistical equilibrium in a self-consistent manner.
Notice that, during the procedure to obtain the primitive fields, one needs to compute also $\epsilon = \epsilon(\rho, T, Y_e)$ in order to calculate the right-hand-side of the MHD evolution equations. 
The specific procedure to convert from the set of conserved variables to the set of primitive variables follows from the algorithm described in Ref.~\cite{kastaun20,10.1093/mnras/stab2606}, which is summarized in Appendix \ref{reprimand}.

The resulting system of balance law equations is extended by employing \emph{Large Eddy Simulation} (LES) techniques to model the influence of unresolved sub-grid scales through additional terms ${\cal G}$ in the evolution equations~\cite{vigano20,Palenzuela:2022kqk}, which are given explicitly in Appendix \ref{appendix:LES}. This is achieved by applying a spatial filter to the governing balance laws, which decomposes each field into a resolved component and a \emph{subgrid-scale} (SGS) component. The filtering operation introduces extra terms that represent the momentum and energy fluxes of small-scale turbulence. In particular, the SGS \emph{gradient model} approximates these subgrid contributions in terms of the gradients of the resolved fields. The inclusion of these SGS terms in the equations allows for the recovery, at least partially, of the effects
that the unresolved sub-grid dynamics induce over the resolved scales. While LES introduces a modest computational overhead in standard GRMHD simulations, it potentially extends the modeling of physical processes (such as turbulence) to scales smaller
than those captured by conventional numerical methods,
without the prohibitive cost of fully resolving those extended scales.

The MHD evolution equations, with the neutrino source contributions and the sub-grid-scale tensors, can be written as a function of the CCZ4 evolution fields and the densitized MHD conserved fields as follows
\begin{widetext}
	\begin{eqnarray}
		\partial_t {\bar D} 
		&+& \partial_k [(- \beta^k + \alpha v^k) {\bar D} 
		- \alpha {\cal G}_D^{k})] = 0 \label{eq:mhd_system} \\
		\partial_t \overline{D\!Y}_e 
		&+& \partial_k [ (- \beta^k + \alpha v^k) \overline{DY}_e  
		- \alpha {\cal G}_Y^{k} ] 
		= - \alpha \sqrt{\gamma} m_{b} ({\cal C}_{\nu_e}  - {\cal C}_{\bar{\nu}_e}) 
		\nonumber \\		
		\partial_t {\bar \tau}
		&+& \partial_k [- \beta^k {\bar \tau} + \alpha ({\bar S}^k - {\bar D} v^k) 
		- \alpha {\cal G}_{\tau}^{k}] 
		= \frac{\alpha}{\chi} {\bar S}^{ij} {\tilde A}_{ij}
		+ \frac{\alpha}{3} tr{\bar S}\, ({\hat K} + 2 \Theta)
		- {\bar S}^j \partial_j \alpha
		- \alpha \sqrt{\gamma} \left( {\cal S}_n^{\nu_e} + {\cal S}_n^{\bar{\nu}_e} + {\cal S}_n^{\nu_x} \right)
		 \nonumber \\
		\partial_t {\bar S}_i
		&+& \partial_k [- \beta^k {\bar S}_i + \alpha {{\bar S}^k}_i 
		- \alpha {\cal G_S}^{k}_i ] 
		= \frac{\alpha}{2 \chi} \left({\bar S}^{jk} {\partial_{i}} {\tilde \gamma}_{jk} - tr{\bar S} \partial_i \chi \right)
		+ {\bar S}_j \partial_i \beta^j - ({\bar \tau} + {\bar D}) \partial_i \alpha 
		- \alpha \sqrt{\gamma} \left( {\cal S}_i^{\nu_e} + {\cal S}_i^{\bar{\nu}_e} + {\cal S}_i^{\nu_x} \right) 
		\nonumber \\
		\partial_t {\bar B}^{i} &+& 
		\partial_k [{\bar B}^i (\alpha v^k - \beta^k) 
		- {\bar B}^k (\alpha v^i - \beta^i) + \alpha \chi {\tilde \gamma}^{ki} {\bar \phi} 
		- \alpha {\cal G}_B^{ik}] 
		= {\bar \phi} [ - \alpha (  \chi {\hat \Gamma}^i - 2 Z^i ) 
		+
		{\tilde \gamma}^{ki} (- \frac{\alpha}{2} \partial_k \chi +\frac{\chi}{c_b^{2}}  \partial_k \alpha 	
		)] \nonumber \\ 
		\partial_t {\bar \phi} &+& \partial_k [- \beta^k {\bar \phi} + \alpha\, c_b^2 {\bar B}^k] = 
		- \alpha\, c_b^{2} {\bar \phi}\, ({\hat K} + 2 \Theta) + c_b^2 {\bar B}^k (\partial_k \alpha)  -\alpha \kappa_b {\bar \phi}  ~~. \nonumber
	\end{eqnarray}
\end{widetext}
We have defined 
${\bar S}_{ij} =  \sqrt{\gamma} S_{ij}$ and $tr {\bar S} = \gamma_{jk} {\bar S}^{jk}$, being 
\begin{eqnarray}
	S_{ij} &=& \frac{1}{2} \left(v_i S_j + v_j S_i \right) 
	+ \gamma_{ij} p 
	- \frac{1}{2 W^2} \bigg[ 2 B_i B_j - \gamma_{ij} B^2  \bigg] \nonumber \\    
	&-& \frac{1}{2} (B^k v_k) \bigg[ B_i v_j + B_j v_i - \gamma_{ij} (B^m v_m)  \bigg].
	\label{def_cont} \nonumber
\end{eqnarray}
{Notice that we have introduced the source terms accounting for the interaction with the neutrinos $\{ {\cal C}_{\nu_e}, {\cal S}_n^{\nu_e}, {\cal S}_i^{\nu_e}\}$ which will be written explicitly in the next subsection.}

\subsection{Neutrinos: truncated moments (M1)}
\label{sec:M1}

Neutrino radiation transport is fundamentally described macroscopically by the neutrino distribution function \(f_\nu(x^a, p^a)\) that depends on the spacetime coordinates $x^a = (t, x^i )$ 
and the 4-momentum $p^a$ of the neutrinos. This distribution function  evolves according to the fully relativistic Boltzmann equation, which includes a collision term accounting for the weak interactions---such as neutrino emission, absorption, and scattering---that couple neutrinos to matter. Solving Boltzmann’s equation explicitly requires the time evolution of a 6-dimensional function over both physical and momentum space, a prohibitive computational challenge, and so instead various approximations have been adopted~\cite{foucart_review}. 

A common and efficient approach is the \textit{truncated moment} formalism~\cite{thorne81,shi11}, in which only the lowest moments of the distribution function in momentum space are evolved. Further simplifications can be obtained within the gray approximation, that is, 
considering energy-integrated moments. In this approach, it is standard to consider three independent neutrino species: the electron neutrinos $\nu_e$, the electron anti-neutrinos $\bar{\nu}_e$, and the
heavy-lepton neutrinos $\nu_x$ that combines 4 species 
$\left\{\nu_{\mu} , \bar{\nu}_{\mu}, \nu_{\tau} , \bar{\nu}_{\tau}\right\}$. 
This merging is justified because, at the low temperatures and neutrino energies reached in our simulations, relative to those in relativistic heavy-ion collisions, the interactions that distinguish between individual species are subdominant relative to their common neutral-current interactions. 

In the gray approximation and considering only the first two moments of the distribution function, we evolve for each species the projections of the stress-energy tensor $T^{ab}_{\nu}$ for each neutrino species. A convenient decomposition of this tensor can be obtained by considering an observer co-moving with the fluid, such that
\begin{equation}
	T^{ab}_{\nu_i} = J u^a u^b + H^a u^b + H^b u^a + Q^{ab}
\end{equation}
{where $J$, $H^a$, and $Q^{ab}$ are the energy density, flux or momentum density, and pressure tensor of the neutrinos as measured by a fluid co-moving observer,
satisfying $H^a u_a = Q^{ab} u_b = 0$}. Notice that we have dropped the index $\nu_i$ in the projections, assuming that the following calculations apply for each neutrino species.
For these observers, the interaction between the neutrino radiation and the fluid can be expressed as
\begin{equation}
	{\cal S}^a = (\eta - \kappa_a J) u^a - (\kappa_a + \kappa_s) H^a ~~.
\end{equation}
The functions $(\eta,\kappa_a,\kappa_s)$ are, respectively, the (energy-averaged) neutrino emissivity, absorption opacity, and elastic scattering opacity, to be calculated from the fluid state using the EoS tabulated information~(see Appendix~\ref{app:neutrino}). Scattering is assumed to be isotropic and elastic, although inelastic scattering could in principle be treated within this formalism as absorption events immediately followed by emission.

The conservation of energy and linear momentum of each neutrino species leads to a system of balance law equations when the decomposition of $T^{ab}_{\nu}$ is performed by an inertial observer
\begin{equation}
	T^{ab}_\mathrm{rad} = E n^a n^b + F^a n^b + F^b n^a + P^{ab}
\end{equation}
where we can interpret $E$, $F^a$, and $P^{ab}$ as the energy density,
flux density, and pressure tensor of the neutrinos as measured by an inertial observer, satisfying $F^a n_a = P^{ab} n_b = 0$. The resulting evolution equations for the neutrino energy and the linear momentum densities resembles closely the ones obtained for the fluid counterparts.

Although weak reactions conserve the total lepton number of the system, they can alter the electron fraction of the matter. 
For this reason, it is convenient to also evolve the number density of neutrinos. Following  the phenomenological approach proposed in Ref.~\cite{Foucart_2016,rad2022}, we introduce a neutrino number current for each species $N^a_{\nu_i}$ following a conservation equation
\begin{equation}
	\nabla_a N^a = {\cal C} =  \eta^0 - \kappa_a^0 n
\label{eq:balance}
\end{equation}
where $n = -N^a u_a$ is the neutrino density in the fluid frame and $(\kappa_a^0, \eta^0)$ are the neutrino number absorption and emission coefficients, also to be computed from the fluid state and the EoS tables.
Assuming that the neutrino number density and the radiation flux are aligned (i.e., which is a reasonable assumption but not true in general), one can define the closure relation 
\begin{equation}
	N^a = n \left( u^a + \frac{H^a}{J}  \right) ~.
\end{equation}
With this choice, Eq.~\eqref{eq:balance} defines a balance-law evolution equation for  $N=-N^a n_a$, with all terms explicitly specified as functions of the evolved neutrino fields~\cite{rad2022}.

Following a procedure equivalent to the one performed for the MHD equations, we can define densitized conserved quantities $({\bar N}, {\bar E},{\bar F}_i) = \sqrt{\gamma}(N, E, F_i)$.
The evolution equations for each neutrino species, in terms of the CCZ4 evolved fields and 
these conserved neutrino fields, can be written as~\cite{Izquierdo_2023,Izquierdo_2024}
\begin{widetext}
	\begin{eqnarray}
		\partial_t {\bar N} 
		&+& \partial_k [(- W \beta^k + \alpha W v^k + \alpha \frac{H^k}{J}) \frac{{\bar N}}{\Gamma} ] = 
		\alpha \sqrt{\gamma} {\cal C}
		\label{eq:Nneutrino} \\
		\partial_t {\bar E}
		&+& \partial_k [- \beta^k {\bar E} + \alpha {\bar F}^k ] 
		= \frac{\alpha}{\chi} {\bar P}^{ij} {\tilde A}_{ij}
		+ \frac{\alpha}{3} tr{\bar P}\, ({\hat K} + 2 \Theta)
		- {\bar F}^j \partial_j \alpha + \alpha \sqrt{\gamma} {\cal S}_n  
		\label{eq:Eneutrino} \\
		\partial_t {\bar F}_i
		&+& \partial_k [- \beta^k {\bar F}_i + \alpha {{\bar P}^k}_i] 
		= \frac{\alpha}{2 \chi} \left({\bar P}^{jk} {\partial_{i}} {\tilde \gamma}_{jk} - tr{\bar P} \partial_i \chi \right)
		+ {\bar F}_j \partial_i \beta^j - {\bar E} \partial_i \alpha
		+ \alpha \sqrt{\gamma} {\cal S}_i	
		\label{eq:Fneutrino}
	\end{eqnarray}
\end{widetext}
where we have defined $\Gamma \equiv W \left( E - F_i v^i\right)/J$
and used the shortcuts ${\bar P}_{ij}=\sqrt{\gamma} P_{ij}$ and $tr {\bar P} = \gamma_{jk} {\bar P}^{jk}$.
The source terms in these equations, namely 
\begin{eqnarray}
	&& {\cal C}=\eta^0 - \kappa_a^0 n = \eta^0 - \kappa_a^0 \frac{N}{\Gamma} \\
	&&{\cal S}_n =
	- n_a {\cal S}^a =
	W \left[ (\eta + \kappa_s J) - (\kappa_a + \kappa_s) (E - F_i v^i) \right] \nonumber \\
	&&{\cal S}^i = 
	\perp^i_b {\cal S}^b = 
	W (\eta - \kappa_a J) v^i - (\kappa_a + \kappa_s) \gamma_{a}^i H^{a} 
	\nonumber ~~,
\end{eqnarray}
are the same that appear in the MHD equations as a consequence of total conservation of energy, momentum, and lepton number. These expressions are obtained using the useful relations between the projections in the fluid and Eulerian frames
\begin{eqnarray}
	J = W^2 (E -2F^iv_i + P^{ij}v_i v_j) \label{shibataJ} \\
	\label{shibataHn}
	H^a n_a = W (J - E  + F^k v_k) \\
	\label{shibataH}
	\gamma_{ia}H^{a} = W ( F_i - P^k_i v_k - J v_i) .
\end{eqnarray}
Note that, using the generic relations (\ref{shibataHn},\ref{shibataH}), it is straightforward to reconstruct
$H^a = - (H^b n_b) n^a + \gamma_b^a H^b$. 

It is important to remark that, at this point, these equations are still exact. However, they are not in a closed form, since the evolution equation for the second moment $P^{ij}$ will depend in general on the third moment. The main idea of the M1 scheme is to truncate the moment expansion by providing an approximate analytic closure for these equations $P^{ij}=P^{ij}(E,F^k)$.
Since the second moment generically depends on the global geometry of the radiation field, such a closure can not be exact in general. 
As is common when using the M1 scheme, we adopt the so-called \textit{Minerbo} closure, which is exact in both the optically thick limit (with matter and radiation in thermodynamic equilibrium) and in the propagation of radiation in a transparent medium from a single point source.

In the optically thin limit, we assume that radiation is streaming at the speed of light in the direction of the radiation flux, leading to the closure {for free-streaming neutrinos given by}
\begin{equation}\label{Pthin}
	P^\mathrm{thin}_{ij} = \frac{F_i F_j}{F^k F_k} E ~~.
\end{equation}
In the optically thick limit, the neutrinos are trapped and reach thermal equilibrium with the matter and become nearly isotropic in momentum space. This means that the neutrino pressure tensor is approximately isotropic in the frame comoving with the fluid, just like for a radiation gas in equilibrium.
For Eulerian observers, the projections of the neutrino stress-energy tensor can be written as
\begin{eqnarray}
	&&P_{ij}^\mathrm{thick} = \frac{4}{3} J_\mathrm{thick} W^2 v_i v_j \nonumber \\
	&&~~~~	+ W (\gamma_{ia}H_\mathrm{thick}^a v_j + \gamma_{ja} H_\mathrm{thick}^a v_i) + \frac{1}{3} J_\mathrm{thick} \gamma_{ij}
    \label{Pthick} \\
	&&J_\mathrm{thick} = \frac{3}{2 W^2 + 1} \left[ (2 W^2 - 1) E - 2 W^2 F^k v_k \right] \\
	&&\gamma^i_a H^a_\mathrm{thick} =
	 \frac{F^i}{W} + \frac{W v^i}{2 W^2 + 1} 
	\left[(4 W^2 + 1) F^k v_k - 4 W^2 E \right] ~~ .\nonumber 
\end{eqnarray}

Finally, we combine both limits to construct the {\em Minerbo closure}
\begin{equation}\label{minerbo}
	P_{ij} = \frac{3 \chi - 1}{2} P^\mathrm{thin}_{ij}
	+ \frac{3 (1- \chi)}{2} P^\mathrm{thick}_{ij}
\end{equation}
where $\chi \in [\frac{1}{3},1]$ is the Eddington factor
\begin{equation}\label{eddington}
	\chi (\xi)= \frac{1}{3} + \xi^2 \left(
	\frac{6 - 2\xi + 6 \xi^2}{15} \right)
	~~,~~ \xi^2 = \frac{H_a H^a}{J^2} ~~.
\end{equation}
By construction, both limits of the neutrino pressure tensor are correctly recovered. In the optically thick regions $H^a  \approx 0$, implying that $\chi \approx 1/3$, leading to $P_{ij} \approx P^\mathrm{thick}_{ij}$. On the other hand, in  optically thin regions $|H^a|  \approx J$, implying that $\chi \approx 1$ and then $P_{ij} \approx P^\mathrm{thin}_{ij}$.

The calculation of the pressure neutrino tensor involves the following steps to be performed at each point: first, calculate the primitive fields for the fluid, which will determine the fluid velocity vector.
Then, one can compute $P_{ij}$ combining the optically thick and thin limits with the Minerbo closure (\ref{minerbo}) and the evolved neutrino fields $\left\{N,E,F^i\right\}$. Notice however that, since $\chi\left(\xi (J, H^a)\right)$ is described by equation (\ref{eddington}), and those fields also depend on  $P_{ij}$ through equations (\ref{shibataJ}-\ref{shibataH}), we obtain a non-linear equation 
\begin{equation}\label{minerbo_transcendental}
	R = \frac{\xi^2 J^2 - H_a H^a}{E^2} = 0
\end{equation}
which needs to be solved numerically for $\xi$, similar to solving for the primitive fluid variables as discussed in Section~\ref{sec:fluid}.

\section{Numerical methods}

The governing equations to be solved form a first order in time  evolution system of partial differential equations, which
can be written generically as
\begin{eqnarray}\label{PDEequation}
	\partial_t \textbf{u} = {\cal L}(\textbf{u}) 
\end{eqnarray}
where $\textbf{u}$ denotes the set of evolved fields and ${\cal L}({\textbf{u}})$ is an operator that depends on the fields and its spatial derivatives. 
This continuum problem can be transformed into a semi-discrete one by 
discretizing the space with a grid spacing $\Delta x$ along each coordinate direction.
In the semi-discrete problem, the continuum solution $\textbf{u}$ is converted into $\textbf{U}$, representing a quantity defined discretely over the spatial grid. Analogously, the continuum operator ${\cal L}(\textbf{u})$ is substituted with the discrete $L(\textbf{U})$, which approximates the continuum spatial derivatives on the discrete grid up to a certain accuracy order. The resulting equations at each grid point have a  right-hand-side that depends only on that point, such that they constitute ODEs in time. Within this \textit{method of lines}, the equations can be evolved at each point using a time integrator to solve for the next time step.

In this section we describe in detail the main numerical schemes that we usually employ in our simulations. The time integrator is given by a generic Implicit-Explicit Runge-Kutta (IMEX) method, which reduces to the standard explicit Runge-Kutta (RK) when there are no potentially stiff terms in the equations. Then we describe the finite-difference spatial discretization for smooth and non-smooth solutions, namely centered discretizations based on Taylor expansions and High-Resolution Shock Capturing (HRSC) schemes. Finally, we discuss the special case of HRSC schemes for advection-diffusion equations.

\subsection{The Implicit-Explicit Runge-Kutta time integrator}
\label{sect:IMEX}

It is useful to decompose the system of equations with potentially stiff terms in the following way
\begin{eqnarray}\label{eq_stiff}
	\partial_t \textbf{U} &=&  \mathcal{F} (\textbf{U})  + \frac{1}{\epsilon} \mathcal{R} (\textbf{U}) \,,
\end{eqnarray}
where $\epsilon$ is the relaxation time, $\mathcal{R} (\textbf{U})$ accounts for the stiff part (i.e., the neutrino-fluid interaction terms in the neutrino radiation transport equations) while $\mathcal{F}(\textbf{U})$ accounts for all the other, non-stiff terms which can be treated explicitly. It is assumed that in the limit $\epsilon \rightarrow \infty$ the system is hyperbolic with a spectral radius $c_{h}$ (i.e., the absolute value of the maximum eigenvalues). At the other limit $\epsilon \rightarrow 0$, the system is clearly stiff since the time scale $\epsilon$ of the relaxation (or stiff term) $\mathcal{R} (\textbf{U})$ is very different from the speeds $c_h$ of the hyperbolic (or non-stiff) part $\mathcal{F} (\textbf{U})$. In the stiff limit ($\epsilon \rightarrow 0$) the stability of an explicit time evolution scheme  is only achieved with a time step size $\Delta t \leq \epsilon$. This restriction is much stronger than the one given by the Courant-Friedrichs-Lewy (CFL) condition $\Delta t \leq \Delta x/c_h$. 

An IMEX Runge-Kutta scheme consists of applying an implicit discretization to the stiff terms and an explicit one to the non-stiff ones (see e.g.~\cite{Pareschi:2005} and references within). When applied to the system (\ref{eq_stiff}), it takes the form
\begin{eqnarray}\label{eq_IMEX}
	{\bf U}^{(i)} = {\bf U}^n &+& \Delta t \sum_{j=1}^{i-1} {\tilde{a}}_{ij} \mathcal{F}({\bf U}^{(j)}) 
	\nonumber \\
	&+& \Delta t  \sum_{j=1}^{i} a_{ij} \frac{1}{\epsilon} \mathcal{R}({\bf U}^{(j)}) \,, \\
	{\bf U}^{n+1} = {\bf U}^n &+& \Delta t \sum_{i=1}^{q} {\tilde{\omega}}_{i} \mathcal{F}({\bf U}^{(i)})
	+ \Delta t  \sum_{i=1}^{q} \omega_{i} \frac{1}{\epsilon} \mathcal{R}({\bf U}^{(i)}) \,,
	\nonumber
\end{eqnarray}
where ${\bf U}^{(i)}$ are the auxiliary intermediate values of the Runge-Kutta. The matrices $\tilde{A}= (\tilde{a}_{ij})$, $\tilde{a}_{ij} = 0$ for $j \geq i$ and $A= (a_{ij})$ are $q \times q$ matrices such that the resulting scheme is explicit in $\mathcal{F}$ and implicit in $\mathcal{R}$. An IMEX Runge-Kutta is characterized by these two matrices and the coefficient vectors $\tilde{\omega}_i$ and $\omega_i$. Since the simplicity and efficiency of solving the implicit part at each step is of great importance, it is natural to consider diagonally-implicit Runge-Kutta (DIRK) schemes ($a_{ij}=0$ for $j > i$) for the stiff terms.
IMEX RK schemes can be represented by a double \emph{tableau} in the usual Butcher notation~\cite{Butcher:2008}
\begin{equation}
	\begin{minipage}{1in}
		\begin{tabular} {c c c}
			${\tilde c}$  & \vline & ${\tilde A}$  \\
			\hline 
			& \vline & ${\tilde \omega}^T$ \\
		\end{tabular}
	\end{minipage}
	\begin{minipage}{1in}
		\begin{tabular} {c c c}
			${c}$  &  \vline & ${A}$  \\
			\hline 
			&  \vline & ${\omega}^T$  \\
		\end{tabular}
	\end{minipage}
	\label{butcher_tableau}
\end{equation}
to characterize the explicit and implicit RK schemes, respectively. The coefficients $\tilde{c}$ and $c$ used for the treatment of non-autonomous systems are given by the following relations
\begin{eqnarray}\label{eq_definition_cs}
	{\tilde c}_{i} = \sum_{j=1}^{i-1}~ {\tilde{a}}_{ij} \,, ~~~~~~
	{c}_{i} = \sum_{j=1}^{i}~ {a}_{ij}.
\end{eqnarray}

The IMEX RK schemes might be denoted as RK($s, \sigma, p$), which is characterized by the number $s$ of stages of the implicit scheme, the number $\sigma$ of stages of the explicit scheme, and the order $p$ of the IMEX scheme. Explicit RK schemes might be characterized using only the doublet ($\sigma, p$), while the implicit ones requires only the doublet ($s, p$).

These RK schemes need to satisfy certain mixed order conditions to achieve high order accuracy. For a given explicit tableau, it is straightforward to show that not all the ordering conditions can generically be satisfied without decreasing the CFL coefficient. Here we present two IMEX schemes that satisfy these conditions and are constructed upon the explicit part of the two most commonly used Runge-Kutta methods: the standard fourth-order~\cite{originalRK4}, RK4($4,4$), and the third-order Strong-Stability-Preserving (SSP) scheme of Shu-Osher~\cite{shu_osher}, SSP-RK($3,3$). Notice that RK4($4,4$) is close to being SSP (see, e.g., \cite{gottlieb98,gottlieb01}).

The Butcher \emph{tableau} for the implicit RK that couples with the explicit standard RK4($4,4$) scheme is shown in Table \ref{RK4}. 
The most convenient option, keeping the large $\text{CFL}=2$ and the $L$-stability, is to achieve at least a second order implicit scheme denoted as I42L($4,2$). The resulting IMEX($4,4,2$), in addition to being of high order, have a large $\text{CFL}_{\text{eff}} \equiv \text{CFL}/\sigma=1/2$. 
\begin{table}[h]
	\caption{\emph{Tableau} for the IMEX($4,4,2$), composed by the explicit RK4($4,4$) scheme combined with the $L$-stable I42L($4,2$). \label{RK4}}
	\begin{minipage}{1.4in}
		\begin{tabular} {c| c c c c}
			$0$ &  $0$ & $0$ & $0$ & $0$\\
			$1/2$ & $1/2$ & $0$ & $0$ & $0$ \\  
			$1/2$ & $0$& $1/2$ &$0$& $0$\\  
			$1$ & $0$ & $0$ & $1$ & $0$ \\  
			\hline \\[-1.0em] 
			& $1/6$& $1/3$ & $1/3$ & $1/6$ \\
		\end{tabular}
	\end{minipage}
	\begin{minipage}{1.4in}
	\begin{tabular} {c | c c c c}
		$0$ &  $0$ & $0$ & $0$ & $0$\\
		$1/2$ & $1/4$ & $1/4$ & $0$ & $0$ \\  
		$1/2$ & $0$& $1/6$ &$1/3$& $0$\\  
		$1$ & $1/6$ & $1/3$ & $1/3$ & $1/6$ \\  
		\hline \\[-1.0em] 
		& $1/6$& $1/3$ & $1/3$ & $1/6$ \\
	\end{tabular}
\end{minipage}
\end{table}

The implicit RK method, denoted as I32L($3,2$), that couples with the explicit Shu-Osher SSP-RK(3,3) is presented in Table~\ref{RK3SSP}. The resulting IMEX($3,3,2$) also has a large $\text{CFL}_{\text{eff}} = 1/3$ while satisfying the L-stability condition.

\begin{table}[h]
	\caption{\emph{Tableau} for the IMEX($3,3,2$), composed by the  explicit SSP-RK(3,3) scheme combined with the $L$-stable I32L($3,2$). \label{RK3SSP}}
	\begin{minipage}{1.4in}
		\begin{tabular} {c | c c c }
			$0$ &  $0$ & $0$ & $0$ \\
			$1$ & $1$ & $0$ & $0$\\  
			$1/2$ & $1/4$ & $1/4$ & $0$  \\  
			\hline \\[-1.0em] 
			& $1/6$& $1/6$ & $2/3$ \\
		\end{tabular}
	\end{minipage}
	\begin{minipage}{1.4in}
	\begin{tabular} {c | c c c }
		$0$ &  $0$ & $0$ & $0$ \\
		$1$ & $1/2$ & $1/2$ & $0$ \\  
		$1/2$ & $1/6$ & $1/6$ & $2/3$\\  
		\hline \\[-1.0em] 
		& $1/6$& $1/6$ & $2/3$\\
	\end{tabular}
\end{minipage}

\end{table}

\noindent{\bf Dense output interpolator}.
The boundary conditions on refined grids when using sub-cycling, as discussed in Sec.~\ref{sec:subcycling}, require an interpolation of the solution between different time levels.
A continuous, high accurate interpolation of the numerical solution between time steps can be constructed using the evaluation of the solution at the different RK sub-steps~\cite{Hairer:1987}. Its generic form is given by
\begin{equation}
	U^{n + \theta} = U^{n} + \sum_{j=1}^{s} {b}_{j}(\theta) {k}_j ~~~,~~~
	\theta = \frac{t - t^n}{t^{n+1} - t^{n}}
\end{equation}
where $b_i(\theta)$ are the coefficients to build
the interpolator for a given RK scheme. Notice that
the $m-$derivative can also be computed from this dense
output interpolator
\begin{equation}\label{denseoutput_derivative}
	\frac{d^{m}}{dt^{m}} U (t^n + \theta \Delta t) = 
	\frac{1}{h^m} \sum_{j=1}^{s} {k}_j 
	\frac{d^{m}}{d\theta^{m}} {b}_{j}(\theta)  + O(h^{4-m})~~~.
\end{equation}

For the standard RK$(4,4)$, it can be shown that there is a unique third order interpolator that can be written as
\begin{eqnarray}
	b_{1}(\theta) &=& \theta - \frac{3}{2} \theta^2
	+ \frac{2}{3} \theta^3
	~~,~~
	b_{2}(\theta) = b_3(\theta) = \theta^2 -
	\frac{2}{3} \theta^3
	~~,~~
	\nonumber \\
	b_{4}(\theta) &=& -\frac{1}{2}\theta^2 +
	\frac{2}{3} \theta^3                 
\end{eqnarray}

For the SSP-RK$(3,3)$, a second order interpolator
which also satisfies the SSP condition can be written as
\begin{equation}
	b_{1}(\theta) = \theta - \frac{5}{6} \theta^2
	~~,~~
	b_{2}(\theta) =  \frac{1}{6} \theta^2
	~~,~~
	b_{3}(\theta) = \frac{4}{6}\theta^2.
\end{equation}

\subsection{Spatial Discretization for smooth solutions}

High-order discrete derivative operators can be found using a Taylor expansion of the smooth solution around a specific position $x_i$ of the discrete grid. By default, we employ standard fourth-order centered finite differences, such that continuum derivatives are approximated as $\partial_i u = D_i U + {\cal O}({\Delta x}^4)$. The first-order derivative operators along the x-direction have the form
\begin{eqnarray}
	D_x U_{i,j,k} &=& \frac{1}{12 \Delta x}
	\left( U_{i-2,j,k} - 8\, U_{i-1,j,k} 
	\nonumber \right. \\
	&+& \left. 8\, U_{i+1,j,k} - U_{i+2,j,k}    \right)   + {\cal O}(\Delta x^4).
\end{eqnarray}

Second-order derivative operators can be constructed by applying the first-order operator twice. This is a convenient choice for the (commutative) cross-derivatives
\begin{eqnarray}
	D_{xy} U_{i,j}  = D_{yx} U_{i,j} &=& D_y \left( D_x U_{i,j} \right) ~~.
\end{eqnarray}
However, the stencil of the second order derivative along a single coordinate direction (i.e., $xx$) would be twice as wide as that of the cross-derivatives. Therefore, with scalability in mind, it is preferable to change to a more compact fourth order operator which keeps the original stencil width, namely
\begin{eqnarray}
	D_{xx} U_{i,j,k} &=& \frac{1}{12 \Delta x^2}
	\left( -U_{i-2,j,k} + 16 \, U_{i-1,j,k} -30 \, U_{i,j,k} 
	\right. \nonumber \\
	&+& \left. 16\, U_{i+1,j,k} - U_{i+2,j,k}       \right) 
	+ {\cal O}(\Delta x^4).
\end{eqnarray}

We use centered derivative operators for all the derivative terms except for the advection terms, which are generically proportional to a vector $\beta^i$. In such cases, we use the following one-sided derivative schemes
\begin{eqnarray}
	D_x U_{i,j,k} &=& \frac{1}{12 \Delta x}
	\left( - U_{i-3,j,k} +  6\, U_{i-2,j,k} - 18\, U_{i-1,j,k}
	\right. \nonumber \\
	&+& \left. 10\, U_{i,j,k} + 3 \,U_{i+1,j,k}  \right) + {\cal O}(\Delta x^4) ~,~ \mathrm{if}~ \beta^{x} < 0 \nonumber \\
	D_x U_{i,j,k} &=& \frac{1}{12 \Delta x}
	\left( U_{i+3,j,k} -  6\, U_{i+2,j,k} + 18\, U_{i+1,j,k}
	\right. \nonumber \\
	&-& \left. 10\, U_{i,j,k} - 3 \,U_{i-1,j,k}  \right) + {\cal O}(\Delta x^4) ~,~ \mathrm{if}~ \beta^{x} > 0. \nonumber
\end{eqnarray}

\noindent{\bf Kreiss-Oliger dissipation}.
Discrete numerical solutions generically contain high-frequency unphysical modes, with a wavelength smaller than the grid size $\Delta x$, that can grow rapidly and spoil the physical solution. These modes can be suppressed by including small, artificial Kreiss-Oliger~(KO) dissipation along each coordinate direction~\cite{Calabrese_2004}. For instance, along the $x$-direction, the KO dissipation operator suitable for our fourth-order scheme can be written as
\begin{eqnarray}
	Q^x_d U_{i,j,k} 
	&=& \frac{\sigma}{64 \Delta x}
	\left(U_{i-3,j,k}  -6 \, U_{i-2,j,k} + 15 \, U_{i-1,j,k} 
	\nonumber \right. \\
	&-& \left.20 \, U_{i,j,k} 
	+ 15\, U_{i+1,j,k} - 6 \, U_{i+2,j,k} + U_{i+3,j,k}   \right) \nonumber
\end{eqnarray}
where $\sigma$ is a positive, adjustable parameter that controls the small amount of dissipation (typically $\sigma \lesssim 0.1$) added to the right-hand side of the discretized equations. 

\subsection{Spatial Discretization for non-smooth solutions}
\label{sec:nonsmooth}

Equations which are intrinsically non-linear might develop non-smooth features during the evolution, even when initialized with smooth initial data. In such cases it is advisable to use High-Resolution-Shock-Capturing~(HRSC) methods, which are specifically designed to handle shocks and discontinuities 
by taking advantage of the integrated or weak-form of the equations~\cite{toro97,Leveque98}.

A system of equations in balance law form can be written as
\begin{eqnarray}\label{PDEequationdecomposed}
	\partial_t {\bf U} + \partial_k {\cal F}^k({\bf U}) = S({\bf U})
\end{eqnarray}
where ${\bf U}$ is the list of evolved fields, and ${\cal F}^k({\bf U})$ and  $S({\bf U})$ correspond to their fluxes and  sources, which might be non-linear but depend only on the fields and not on their derivatives. 
The main difficulty is to provide a stable and accurate, non-oscillatory, numerical approximation to $\partial_k {\cal F}^k({\bf U})$. Let us consider again for simplicity only the $x$-direction and denote $F_i \equiv F^x(U_{i, j,k})$. 
A straightforward differentiation yields the conservative approximation~\cite{shu98} 
\begin{equation}
	\left. \frac{\partial F}{\partial x}\right|_{x_i}
	= \frac{1}{\Delta x} \left(\hat{F}_{i+ 1/2} - \hat{F}_{i-1/2}\right) ~~,
\end{equation}
{where $\hat{F}_{i\pm 1/2}$ are high-order reconstructions of the numerical fluxes, evaluated at the interfaces $x_i \pm \frac{\Delta x}{2}$, computed from $F_{i \pm 1/2} \equiv F(U_{i \pm 1/2})$}. 
Stated in this form, the problem consists of finding a high-order approximation to the interface values of $\hat{F}_{i+ 1/2}$ from the fluxes evaluated at neighboring points. Thus one can set $\hat{F}_{i+ 1/2} = R (F_{[s]})$, where $R()$ is a highly accurate reconstruction scheme providing a stable interface flux value from point-wise values,
while the index $[s]$ spans through the interpolation stencil.
The crucial issue in HRSC methods is how to approximately solve the Riemann problem, by reconstructing the fluxes at the interfaces such that no spurious oscillations appear in the solutions. 

The non-oscillatory condition 
can be enforced by considering the Lax-Friedrichs splitting at each grid point, yielding the following combination of fluxes and fields:
\begin{eqnarray}\label{flux_decomposition}
	F^{\pm}_{i} = \frac{1}{2} \left( F_i \pm \lambda U_i \right) 
\end{eqnarray}
where $\lambda$ is the maximum propagation speed of the system among the neighboring points~\cite{shu98}. Then, from the neighboring nodes $\{x_{i-n},..,x_{i+1+n}\}$, we reconstruct the fluxes at the left and right of each interface as
\begin{equation}
	F^L_{i+1/2} = R(\{F^+_{[s]}\}) ~~,~~
	F^R_{i+1/2} = R(\{F^-_{[s]}\}) ~~.
\end{equation}
The number $2(n+1)$ of such neighbors used in the reconstruction procedure depends on the order of the method. Our schemes incorporate some commonly used reconstructions, such as the piecewise parabolic method (PPM)~\cite{colella84}, the Weighted-Essentially-Non-Oscillatory~(WENO) reconstructions~\cite{jiang96,shu98}, the monotonicity preserving (MP5) scheme~\cite{suresh97}, as well as other implementations such as the FDOC families~\cite{bona09}.
Our default choice is the MP5 scheme, whose reconstruction can be formally written as
\begin{eqnarray}
	F^L_{i+1/2} &=& R_{\text{MP5}}\left( F^+_{i-2}, F^+_{i-1}, F^+_{i}, F^+_{i+1},F^+_{i+2}  \right) \\
	F^R_{i+1/2} &=& R_{\text{MP5}}\left( F^-_{i+3}, F^-_{i+2}, F^-_{i+1}, F^-_{i},F^-_{i-1}  \right) 
\end{eqnarray}
with more details regarding $R_{\text{MP5}}$ presented below.
Then, we use a simple flux formula to compute the final flux at each interface, namely 
\begin{equation}\label{LLF2}
	\hat{F}_{i+1/2} = F^L_{i+1/2} + F^R_{i+1/2}.
\end{equation}
Notice that these methods do not require the characteristic decomposition of the system of equations, making them efficient and readily adaptable to a variety of balance law systems.

~\\
\noindent{\bf Monotonic Preserving 5th-order reconstruction}.
The monotonicity preserving fifth-order~(MP5) schemes of Suresh \& Huynh~\cite{suresh97} achieve high-order interface reconstruction by performing a two-step procedure. First, it provides an accurate polynomial interpolation, and then, it limits the resulting value to preserve both monotonicity near discontinuities and accuracy in smooth regions. 
Here we employ the fifth-order accurate scheme based on the (unlimited) interface value given by
\begin{equation}\label{MP5_unal}
	F_{i+ 1/2}  = \frac{1}{60} \left(2 F_{i-2} - 13 F_{i-1} + 47 F_{i} + 27 F_{i+1} - 3 F_{i+2}\right)
\end{equation}
which relies on the five point values $F_{i-2},..,F_{i+2}$. Together with this interpolation, we also define the monotonicity-preserving bound
\begin{eqnarray}
	&& f^\mathrm{MP} = F_i + \Minmod \left(\Delta_{i+ 1/2}, \alpha \Delta_{i-1/2} \right) 
\end{eqnarray}
where we have introduced the undivided difference
$\Delta_{i+ 1/2} = F_{i+1} - F_i$ and 
\begin{eqnarray}	
	&& \Minmod(a, b) = \frac{\sgn(a) + \sgn(b)}{2}
	\min (|a|, |b|) \nonumber \\
	&& \Median(a, b, c) = a + \Minmod(b - a, c - a) ~~. \nonumber
\end{eqnarray}
The parameter
$\alpha \geq 2$ controls the maximum steepness of the left sided slope and preserves monotonicity during a single Runge-Kutta
stage, provided the CFL number is smaller than $1/(1 + \alpha)$. In practice, setting $\alpha = 4$ allows large CFL values of approximately $0.4$ while still preserving the non-oscillatory behavior~\cite{suresh97}. 

The final reconstruction can be written as
\begin{equation}
	R (F_{[s]})) =
	\begin{cases}
		& F_{i+1/2}  ~~ \text{if}~~   ( F_{i+ 1/2} - F_i)( F_{i+ 1/2} - f^\mathrm{MP}) < 0 , \\
		& \Median ( f^\mathrm{min}, F_{i+ 1/2}, f^\mathrm{max})  ~~~\text{otherwise}
	\end{cases}
\end{equation}
where the values $f^\mathrm{min}$ and $f^\mathrm{max}$, required to preserve accuracy near smooth extrema and provide monotone profile close to discontinuous data, are calculated as
\begin{eqnarray}\label{MP5_bounds}
	f^\mathrm{min} &=& \max [\min ( F_i, F_{i+1}, f^\mathrm{MD}) , \min ( F_i, f^\mathrm{UL}, f^\mathrm{LC})] , \nonumber \\
	f^\mathrm{max} &=& \min [\max ( F_i, F_{i+1}, f^\mathrm{MD}) , \max ( F_i, f^\mathrm{UL}, f^\mathrm{LC})] . \nonumber
\end{eqnarray}
These quantities require some additional bounds. By defining $d_i = \Delta_{i+1/2} - \Delta_{i-1/2}$, these intermediate bounds are given by
\begin{eqnarray}\label{MP5_bounds2}
	f^\mathrm{UL} &=& F_i + \alpha \Delta_{i-1/2} , \\
	f^\mathrm{MD} &=& \frac{F_i + F_{i+1}}{2} - \frac{1}{2} d^{M4}_{i+ 1/2} , \\
	f^\mathrm{LC} &=& F_{i} + \frac{1}{2} \Delta_{i-1/2} + \frac{4}{3} d^{M4}_{i- 1/2} 
\end{eqnarray}
where we have estimated the local curvature as
\begin{equation}\label{MP5_dm4}
	d^{M4}_{i+ 1/2}
	= \Minmod (4d_i - d_{i+1}, 4 d_{i+1} - d_i, d_i, d_{i+1})~~.
\end{equation}
The reconstruction illustrated preserves monotonicity and does not degenerate to first-order in proximity of smooth extrema.

\subsection{Linear reconstruction for advection-diffusion equations}

The truncated moments approach (M1) employed to model the neutrino transport exhibits an hyperbolic character in optically thin regions, where neutrinos free-stream, but becomes parabolic in optically thick regions. Such systems can be generically classified as advection–diffusion equations. 
Although these systems can also be expressed in balance law form, it is not advisable to use directly the methods for non-smooth solutions described in the previous subsection. 
The main reason is that the choice of the spatial discretization might be constrained by the \emph{ill-posedness} of the diffusion equation\cite{hiscock85,andersson11}.

Here we follow the same strategy introduced in Ref.~\cite{rad2022}, such that the fluxes are computed using a flux-splitting approach~\cite{Leveque98},
namely
\begin{equation}
	\hat{F}_{i+1/2} = F^{\text{HO}}_{i+1/2} - A_{i+1/2}\left[1 - \Phi_{i+1/2}\right] \left(F^{\text{HO}}_{i+1/2} - F^{\text{LO}}_{i+1/2}\right) \,.
\end{equation}
The reconstructed flux is obtained as a linear combination of a non-diffusive high order (second order) flux $F^{\text{HO}}$, that works well in smooth regions, and a diffusive low order Lax-Friedrichs (first order) flux $F^{\text{LO}}$, that behaves well near discontinuities. 
The flux limiter, $\Phi$, switches between the high and low order flux. Additionally, we define a coefficient depending on the opacities and the grid spacing, $A(\Delta x_i , \kappa_a , \kappa_s )$, which allow us to switch off the diffusive correction at high optical depth. All these quantities are given explicitly by~\cite{rad2022}
\begin{eqnarray} 
	F^{\text{HO}}_{i+1/2} &=& \frac{1}{2} \left(F_i + F_{i+1} \right) \,,\\ 
	F^{\text{LO}}_{i+1/2} &=& \frac{1}{2} \left(F_i + F_{i+1} \right) - \frac{\lambda_{i+1/2}}{2} \left(u_i + u_{i+1}\right) \,, \nonumber \\
	\Phi_{i+1/2} &=& \text{max}\left[0, \text{min}\left(1, 2 \frac{u_i - u_{i-1}}{u_{i+1}-u_i}, \nonumber 2 \frac{u_{i+2}-u_{i+1}}{u_{i+1}-u_i}  \right)   \right] \,, \\ \nonumber
	A_{i+1/2} &=& \text{tanh}\left[\frac{1}{\kappa_{i+1/2} \Delta x}\right] \,. \nonumber
\end{eqnarray}
where an averaged opacity coefficient is defined as
\begin{equation}
	\kappa_{i+1/2} = \frac{1}{2} \left[  (\kappa_{a})_i + (\kappa_{a})_{i+1} + (\kappa_{s})_i + (\kappa_{s})_{i+1} \right] \,.
	\nonumber
\end{equation}

Notice that $A_{i+1/2}=1$ in optically thin regions (i.e., advection limit), while $A_{i+1/2} \approx 0$
at high optical depths (i.e., the diffusion limit). Therefore, in the diffusion limit $\hat{F}_{i+1/2} \approx F^{\text{HO}}$, such that the scheme reduces to a centered second-order finite-difference scheme. This discretization has been shown to be asymptotic preserving~\cite{rider02}, and therefore avoids the ill-posedness of the diffusion equation discussed earlier.

In some cases, there might appear odd-even oscillations due to a decoupling of consecutive grid points in the spatial discretization scheme. A necessary and sufficient condition for this problem to appear is that~\cite{rad2022,odd_even}
\begin{equation}
	(u_i  - u_{i-1}) (u_{i+1}  - u_{i}) < 0   ~~\text{and}~~
	(u_{i+1}  - u_{i}) (u_{i+2}  - u_{i+1}) < 0  ~.
	\nonumber
\end{equation}
When the above two conditions are satisfied simultaneously, we set $A_{i+1/2}=1$, which suffices to resolve the issue. 

\section{High Performance Computing strategies}

This section outlines the automatic code generation process for \mhduetX, which can be configured to target either the SAMRAI or AMReX infrastructures. We also describe the strategies employed for mesh refinement as well as the sub-cycling algorithms.
Note that currently \mhduet adopts only vertex-centered discretizations, as these provide simpler AMR interfaces and reduce the number of required computational kernels. This typically leads to improved efficiency on modern HPC architectures, particularly when employing high-order accurate schemes.

\subsection{Automatic code generation with Simflowny for AMReX \& SAMRAI infrastructures} 
\label{sec:automatic}
\noindent{\bf Simflowny}. The code presented here has been generated using Simflowny~\cite{arbona13,arbona18,Palenzuela_2021} together with the infrastructures SAMRAI~\citep{hornung02} and AMReX~\cite{AMReX_JOSS}. Simflowny is an open-source and user-friendly platform continually developed by the IAC3 group since 2008 to facilitate the use of HPC infrastructures by non-specialist scientists. 
Using a Domain Specific Language along with a web-based integrated development environment~(IDE), Simflowny assists researchers in implementing scientific models by
automatically generating efficient parallel code with adaptive mesh refinement.

Simflowny splits the physical models and problems from the numerical techniques. The problems are instantiations of the models with specific initial data and boundary conditions. Simflowny provides an arsenal of discretization techniques that include all the ones that have been described in the previous section, and more. The user selects a model of equations, a specific problem and a discretization policy to generate a code. The automatic generation of the simulation code allows the parallelization and adaptive mesh refinement features to be incorporated consistently.

Although Simflowny has a graphical user interface for introducing the equations, the system of equations in MHDuet, due to their complexity and tensorial nature, is not amenable to coding by hand. Furthermore, Simflowny is not a computer algebra system and it does not provide symbolic capabilities to manipulate the equations and its tensors. Instead, we have developed a Maple~\cite{Maple} script that creates a MathML fully expanded representation of the system of equations presented above, and this is injected automatically to Simflowny through a built-in YAML reader~\cite{Arbona_2025}.

Simflowny has traditionally produced code relying on the SAMRAI infrastructure. Alternatively, beginning this year~\cite{Arbona_2025}, it can also produce code relying on the AMReX infrastructure.  The combination of Simflowny with either SAMRAI or AMReX provides a final code with a good balance of speed, accuracy, scalability, ability to switch physical models (flexibility), and the capability to run on different architectures (portability). It should be stressed that all the interfacing between Simflowny with either SAMRAI or AMReX is performed internally, through an XSD schema file that serves as a skeleton for the automatically generated code, which contains all the details necessary for the parallelization, GPU handling (if supported) and even mesh refinement, as explained in the following sections.  
 
~\\
\noindent{\bf SAMRAI}. The infrastructure SAMRAI (Structured Adaptive Mesh Refinement Application Infrastructure) \cite{hornung02}\footnote{See also the website\\{\tt https://computation.llnl.gov/project/SAMRAI/}} offers patch based, structured AMR developed over more than 25 years by the Center for Applied Scientific Computing at the Lawrence Livermore National Laboratory. The latest upgrades to the AMR algorithms improve the performance and scale well up to 1.5M cores and 2M MPI tasks~\cite{gunney16}, at least for certain problems. We have extensive experience with SAMRAI, which has proven to be both reliable and efficient~\cite{palenzuela18}. Unfortunately, it does not support GPUs natively, although it should be noted that it has external support for GPUs through the libraries RAJA\cite{RAJA} and Umpire\cite{UMPIRE}. We have found that the modern native implementation of GPU support within AMReX provides a more natural path to exascale. 

~\\
\noindent{\bf AMReX}. The development of AMReX (Adaptive Mesh Refinement framework for Exascale computing)\cite{AMReX_JOSS}\footnote{See also the website \\{\tt https://amrex-codes.github.io/amrex/}} is more recent than that of SAMRAI. The project started around 2017 at Lawrence Berkeley National Laboratory, as an evolution of an earlier mature framework called BoxLib. AMReX was specifically designed with exascale computing in mind, focusing on performance portability across emerging architectures including GPUs. AMReX has demonstrated strong scaling up to approximately 2 million CPU cores on supercomputers such as Summit at Oak Ridge National Laboratory, and its GPU implementations have shown excellent performance on machines with NVIDIA, AMD, and Intel GPUs, as it can achieve up to 2 orders of magnitude speedup for certain applications when using GPU acceleration compared to CPU-only implementations~\cite{Zhang_2021}. In this paper, we will focus on producing Simflowny code for AMReX, due to its native support for GPU architectures. When running AMReX on a CPU system, the parallelization strategy is based on a combination of MPI~\cite{mpi41} and OpenMP~\cite{OpenMP} using tiling, also known as cache blocking.

\subsection{Mesh Refinement: refinement criteria, prolongation and restriction}

Mesh refinement introduces additional sub-grids within a base (coarse) grid in regions where increased resolution can significantly improve the solution accuracy. In the adaptive case, these sub-grids can dynamically track the regions in which higher resolution is required, and be removed when no longer needed. Mesh refinement enables a more efficient use of computational resources, particularly in problems where the dynamics are concentrated in localized regions of the computational domain.

There are several strategies to decide which regions need further refinement or which sub-grids can be removed, subject to
some \textbf{refinement criteria}. In particular, there are two refinement tagging strategies integrated in Simflowny.:
\begin{itemize}
	\item Fixed Mesh Refinement (FMR): a set of boxes prescribed statically at the beginning of the simulation. These boxes are specified for each level subject to the condition that each box is nested within a box on the next coarser level.
	\item Adaptive Mesh Refinement (AMR): the cells to be refined are calculated dynamically by setting a criteria (i.e., a measurement of the error in the solution, or a function of the fields surpassing certain threshold). 
\end{itemize}

Notice that fixed and dynamical tagging strategies (i.e., FMR and AMR) can be combined in the same simulation, such that some levels are specified statically while others are refined dynamically. As the simulation evolves, the evaluation of the AMR tagging criteria will generally change, requiring the refinement of new regions as older ones might be removed. This re-meshing procedure is performed periodically according to a user parameter. 

If a new refinement level is added dynamically during the simulation (i.e., or the region of a given level increases due to the dynamical AMR criteria), the domain of that grid increases with respect to the coarser level. The new grid points on the fine level are set by the {\bf prolongation} procedure,
interpolating the solution from the coarse grid into the fine one. This spatial interpolation must be at least as accurate as the spatial derivative operators in order to prevent degrading the accuracy of the scheme. One of the simplest and most efficient options is to use Lagrange interpolating functions. Given a solution $U_i$ at grid-points $x_i$, one can construct a Lagrangian polynomial function of order $k$ passing through a set of $k+1$ of points $\{(x_1,U_1), (x_2,U_2),...(x_k,U_k),(x_{k+1},U_{k+1})\}$, namely
\begin{equation}
	U(x) = \sum\limits_{j=1}^{k+1} l_j(x) U_j ~~~,~~~
	l_j(x) = \prod_{\substack{m=1 \\ m \neq j}}^{k+1} \frac{x - x_m}{x_j - x_m}
\end{equation}
where $x$ is the location where the interpolant is sought, satisfying $x_1 < x < x_{k+1}$. To construct a symmetric Lagrangian polynomial of fifth-order, suitable for our fourth-order spatial scheme, six points are required (i.e, three at each side of the point to be interpolated). Such Lagrangian polynomial interpolation can be simplified to define the point $x_{i+1/2}$ in the new refined (child) mesh, reducing simply to
\begin{eqnarray}
	U_{i+1/2} &=& \frac{1}{256}  \biggl[ 150 (U_{i} + U_{i+1}) 
	- 25 (U_{i-1} + U_{i+2}) \nonumber \\
	&&  ~~~~~  + 3 (U_{i-2}+U_{i+3}) \biggr].
\end{eqnarray}
In structured grids it is common to choose refined grids such that the points of the coarse grid also exist in the fine grid (i.e., the ratio between their resolutions is a factor of two).
Therefore, this interpolation is the only one required.
A comparison between the Lagrangian and a more sophisticated WENO interpolation indicates that the simple and efficient Lagrange interpolation, combined with HRSC finite difference schemes, suffices to retain high-order of accuracy and essentially non-oscillatory properties even for strong shocks in mesh refinement scenarios~\cite{Sebastian:2003}.
Finally, note that similar Lagrangian polynomial interpolation is  employed to extrapolate to the points in the ghost zones, outside the computational domain, defined for convenience to impose different types of boundary conditions.

The information obtained in a fine child grid needs to be communicated to its parent grid. In particular, the region of a parent grid that overlaps with a fine grid is replaced by the data in the fine level via the \textbf{restriction} operation. Although one could imagine computing a spatial average, for efficiency reasons we choose a direct copy as our restriction operator, made possible by using only integer refinement ratios such that the location of overlapping parent grid-points will always correspond to fine grid-points.

Finally, we note that refluxing, the flux-correction procedure required to enforce global conservation when sub-cycling is used on refined levels, is not currently implemented in \mhduetX. Consequently, strict conservation of the baryon number is not guaranteed.

\subsection {Sub-cycling in time}
\label{sec:subcycling}

For explicit schemes applied to hyperbolic systems, stability requires that the time step must satisfy the CFL condition $\Delta t \le \lambda_{\rm CFL} \Delta x$, where $\lambda_{\rm CFL}$ depends on the problem's dimensionality and the time integrator. In the presence of multiple refinement levels $l$ ranging from 0 to $L$, stability can be ensured by enforcing the CFL condition for the finest grid resolution ${\Delta x}_{L}$. However, this approach is generally quite inefficient, as coarser grids are evolved with unnecessarily small time steps
A more efficient approach is to evolve finer levels using multiple time steps per coarse-level step, a technique commonly known as \textbf{sub-cycling} in time. This allows each level to use a time step consistent with its own CFL condition. When integer refinement ratios are used, levels remain time-aligned after a fixed number of steps. 

Sub-cycling introduces the challenge of providing boundary data for the fine grids at intermediate times between coarse-level steps. Several approaches have been proposed, including the original Berger–Oliger (BO) algorithm~\cite{BERGER1984484} and the tapering scheme~\cite{lehner}, which offers higher accuracy at the cost of increased computational expense. In this work, we adopt the \emph{Berger–Oliger without order reduction} (BOR) algorithm~\cite{McCorquodale:2011,Mongwane:2015}, which provides a good balance of speed, efficiency, and accuracy

\begin{figure}[h]
	\includegraphics[width=1\columnwidth]{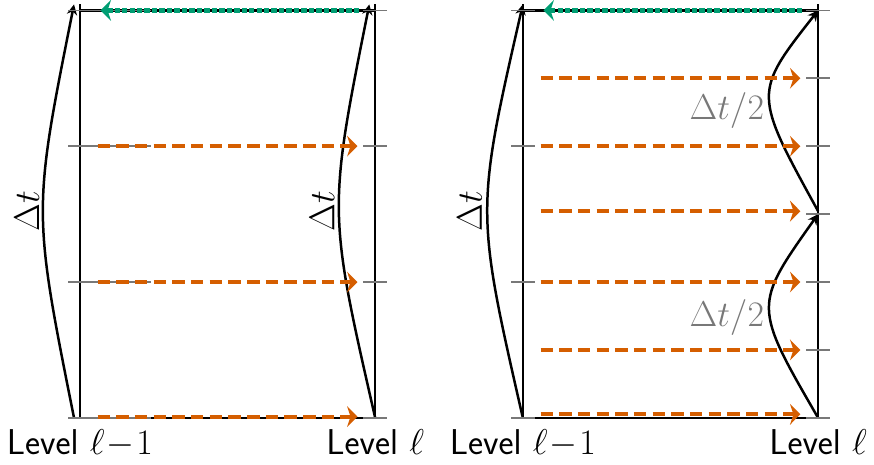}
	\caption{Synchronization between two levels with no sub-cycling (left) and with any of the Berger-Oliger schemes (right), using a Runge-Kutta with three substeps. Prolongation operations between levels is denoted with red arrows, while that restriction is marked with green ones.}
	\label{figure:subcycling}
\end{figure}

Let us introduce the convention ${}^lU_i^n$ to represent $U(x_i,t^n)$ on level $l$, such that its child is ${}^{l+1}U_i^{n}$ and the time-advanced value 
of that same point is ${}^{l+1}U_{i}^{n+1}$.
The standard BO algorithm advances the coarse level ${}^{l}U^n_i$ one
time step to ${}^{l}U^{n+1}_i$. One then interpolates this data linearly
in time to set the fine grid points ${}^{l+1}U^{n+1}_i$ that align with the coarse grid.
Spatial interpolation sets those points that do not so align with the parent.
Although this scheme is very easy to implement, it is only first order accurate in time.

The BOR algorithm offers a major improvement over the BO one by incorporating all Runge–Kutta sub-steps
$\{U^n,U^{(k)},U^{n+1}\}$ to construct a dense output interpolator of order $q= p-1$, only one order lower than the RK scheme.
In the second step, this interpolator is used for computing all the time derivatives of the solution on the fine grid~\cite{McCorquodale:2011,Mongwane:2015}. 
This is achieved by noticing that a direct Taylor expansion of the solution at $t=t^n$ leads to
\begin{equation}
	U_{n+1} = U_{n} + \Delta t \, U'_{n}
	+ \frac{1}{2} \Delta t^2 \, U''_{n}
	+ \frac{1}{6} \Delta t^3 \, U'''_{n}
	+ O(\Delta t^4),
\end{equation}
where a prime indicates derivative with respect to time. Performing a similar expansion for the solution at the different steps of the RK scheme (i.e., $U^{(k)}$) results into a system of equations relating the time derivatives $(U',U'',U''')$ and $U^{(k)}$. The detailed process would be as follows: (i) one can compute the time derivatives of the solution on the coarse grid from the dense output interpolator, (ii) with these time derivatives one can calculate the solution $^l U^{(k)}$ corresponding to the RK steps of the coarse grid. Moreover, one can also compute the solution $^{l+1} U^{(k)}$, corresponding to the RK steps of the fine grid, by changing $\Delta t \rightarrow \Delta t/2$ in the previous system of equations. From here it is straightforward to calculate the solution, at any space grid point at for the different RK steps, required for the evolution of the boundary points of the fine grid. The final scheme is at least accurate at order $q$ in time.
The prolongation and restriction operations involved in each sub-cycling scheme are schematically displayed in Figure~\ref{figure:subcycling}.

A detailed implementation for two commonly-used RK schemes can be found in~\cite{palenzuela18}, where we 
extended the algorithm to allow arbitrary resolution ratios between consecutive AMR grid.


\begin{figure}[h]
	\includegraphics[width=1\columnwidth]{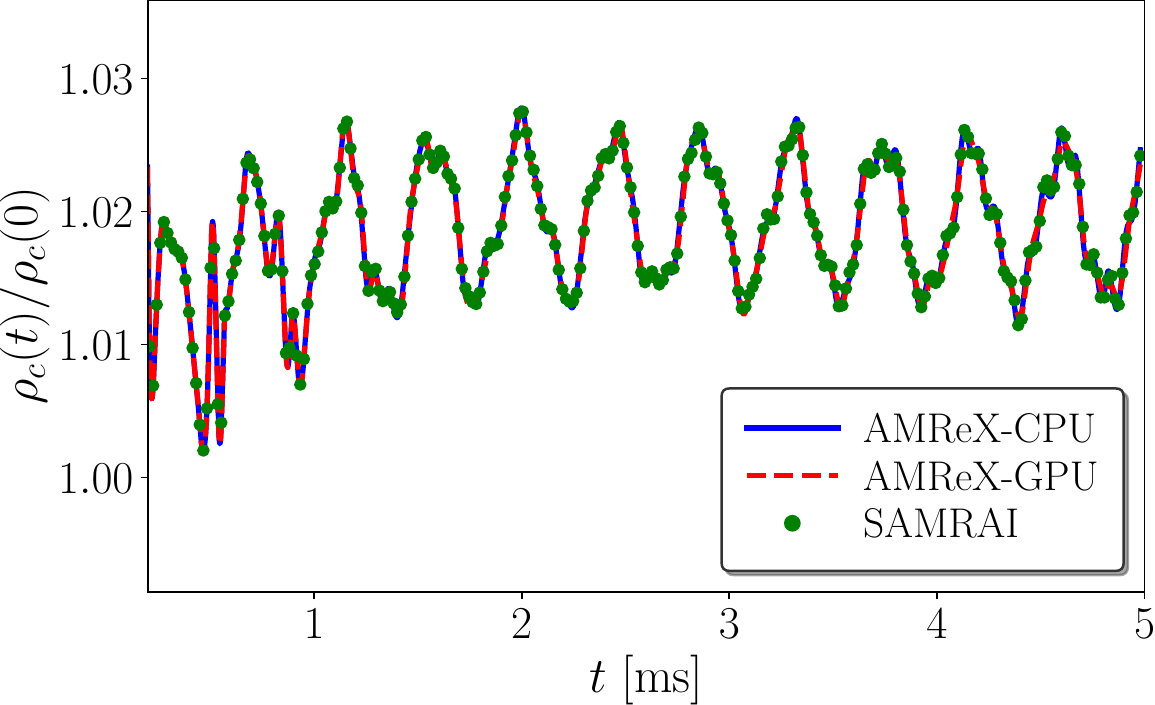}\\
	\includegraphics[width=1\columnwidth]{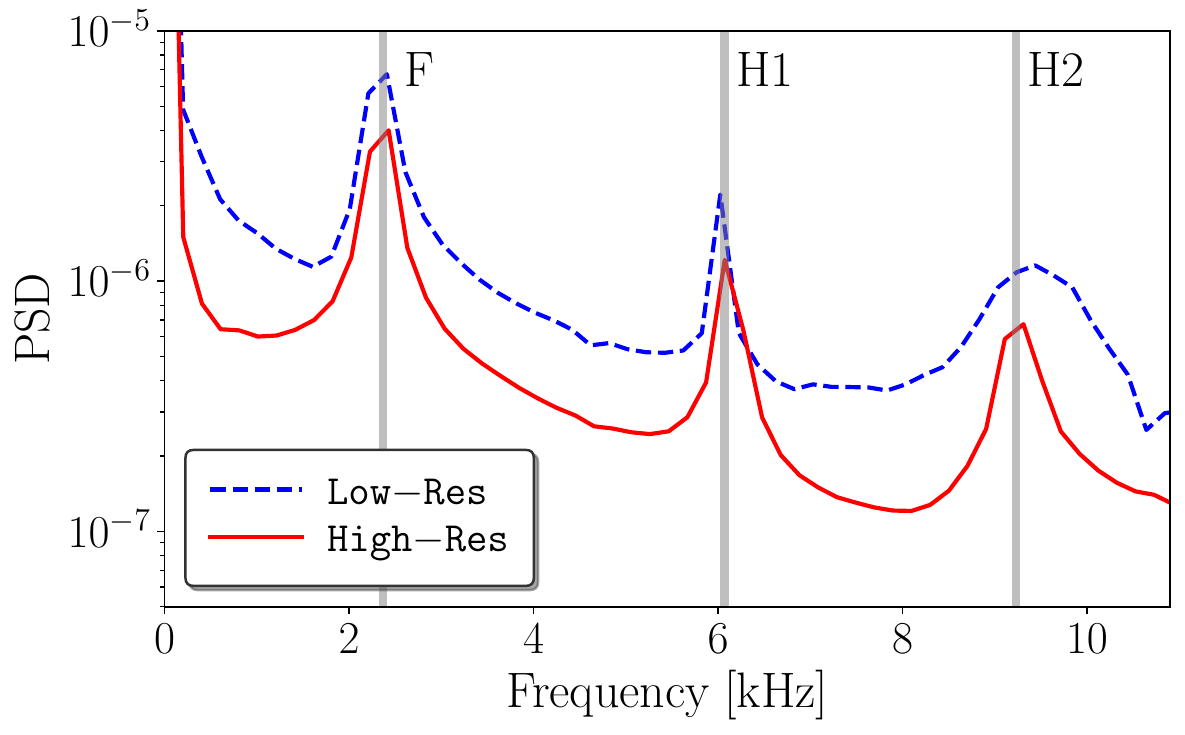}
	\caption{{\em Single NS with the LS-K220 EoS.} 
     \textbf{Top panel:} 
     The central rest-mass density, defined as 
     $\rho_{c}(t)\equiv \rho(r\!=\!0,t)$, is normalized by its initial value and plotted as a function of time. The different curves represent solutions obtained with versions of the code running on SAMRAI, AMReX-CPU, and AMReX-GPU infrastructures. 
     The agreement demonstrates that the SAMRAI and AMReX versions of the code generate the same results.
     \textbf{Bottom panel:} The power spectral density of the central density for the low and high resolutions evolved with AMReX-GPU. 
     Vertical lines indicate frequencies obtained by previous codes as discussed in Section~\ref{sec:coldstar}.
     }
	\label{figure:singleNS_LSS220}
\end{figure}

\section{Numerical Tests}

Here we present a set of tests involving the evolution of neutron stars employing realistic, temperature-dependent, tabulated equations of state. The recovery from the conserved to the primitive fields is handled through an internal implementation of the RePrimAnd library following~\cite{kastaun20,10.1093/mnras/stab2606}. In these tests, we successfully reproduce results reported in previous works, including binary neutron star simulations and the use of tabulated EoS with Large-Eddy Simulation techniques~\cite{Liebling:2020jlq,Liebling:2020dhf,Palenzuela:2022kqk}.
Note that the code can be generated for either the SAMRAI or AMReX infrastructures (see Section~\ref{sec:automatic}). The AMReX version can be compiled to run on both CPU and GPU architectures. Finally, we emphasize that the open-source release of the code~\cite{mhduet_webpage} includes initial data and representative parameter files.

\subsection{Cold magnetized star with tabulated LS-K220 EoS}
\label{sec:coldstar}

We evolve a cold, non-rotating, single NS with the tabulated LS-K220 EoS on a dynamic spacetime, duplicating a test from Ref.~\cite{Palenzuela:2022kqk}.
The star has a gravitational mass $1.72~M_\odot$ and radius $11.95$~km, constructed assuming $\beta$-equilibrium with a temperature $T=0.01$~MeV. We perturb the star by increasing the initial temperature to $T=0.05$~MeV and by adding a purely poloidal magnetic field with maximum magnitude $10^{15}$~G.
The boundaries of the grid are located at $\pm 1892$~km, employing 8 levels of refinement to achieve a finest resolution of $\Delta x = 230$~m covering the whole star. We also perform a high resolution simulation with another level of refinement, doubling the highest resolution to $\Delta x = 115$~m.

We perform three simulations corresponding to SAMRAI (on CPU), AMReX-CPU, and AMReX-GPU. The central density as a function of time is displayed in the top panel of Fig.~\ref{figure:singleNS_LSS220}. The different cases match almost identically, indicating that the AMReX and SAMRAI versions of the code generate the same evolutions.

We display the Fourier transform of the central density for both
resolutions in Fig.~\ref{figure:singleNS_LSS220}.
The first three peaks in the power spectral density agree
very well with the quasi-normal modes~(QNM) obtained with different codes~\cite{Palenzuela:2022kqk,neilsen2014magnetized,2013PhRvD..88f4009G}, shown with vertical lines.
The agreement improves with higher resolution.

\begin{figure*}[t]
	\includegraphics[width=\linewidth]{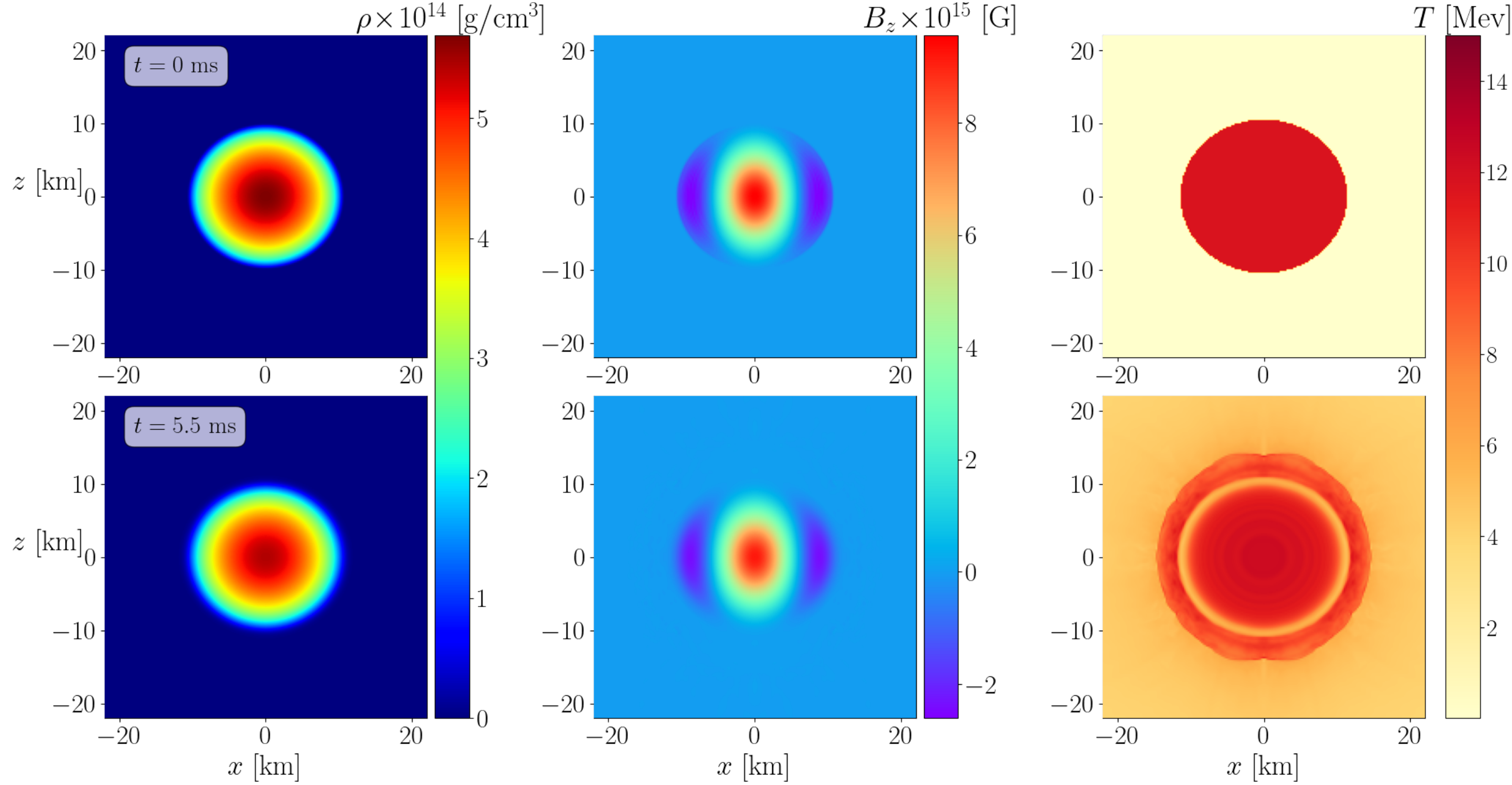}
	\caption{{\em Rotating, hot, magnetized star using the DD2 tabulated EoS.}
	Snapshots on a meridional plane: from left to right, density, magnetic field and temperature, at the initial (top) and final time (bottom) of our simulation. The agreement within the stellar interior for the two times indicates a stationary solution is achieved as discussed in Section~\ref{sec:hotstar}. }
	\label{figure:rotNS_DD2_A}
\end{figure*}

\begin{figure}[h]
	\includegraphics[width=1\columnwidth]{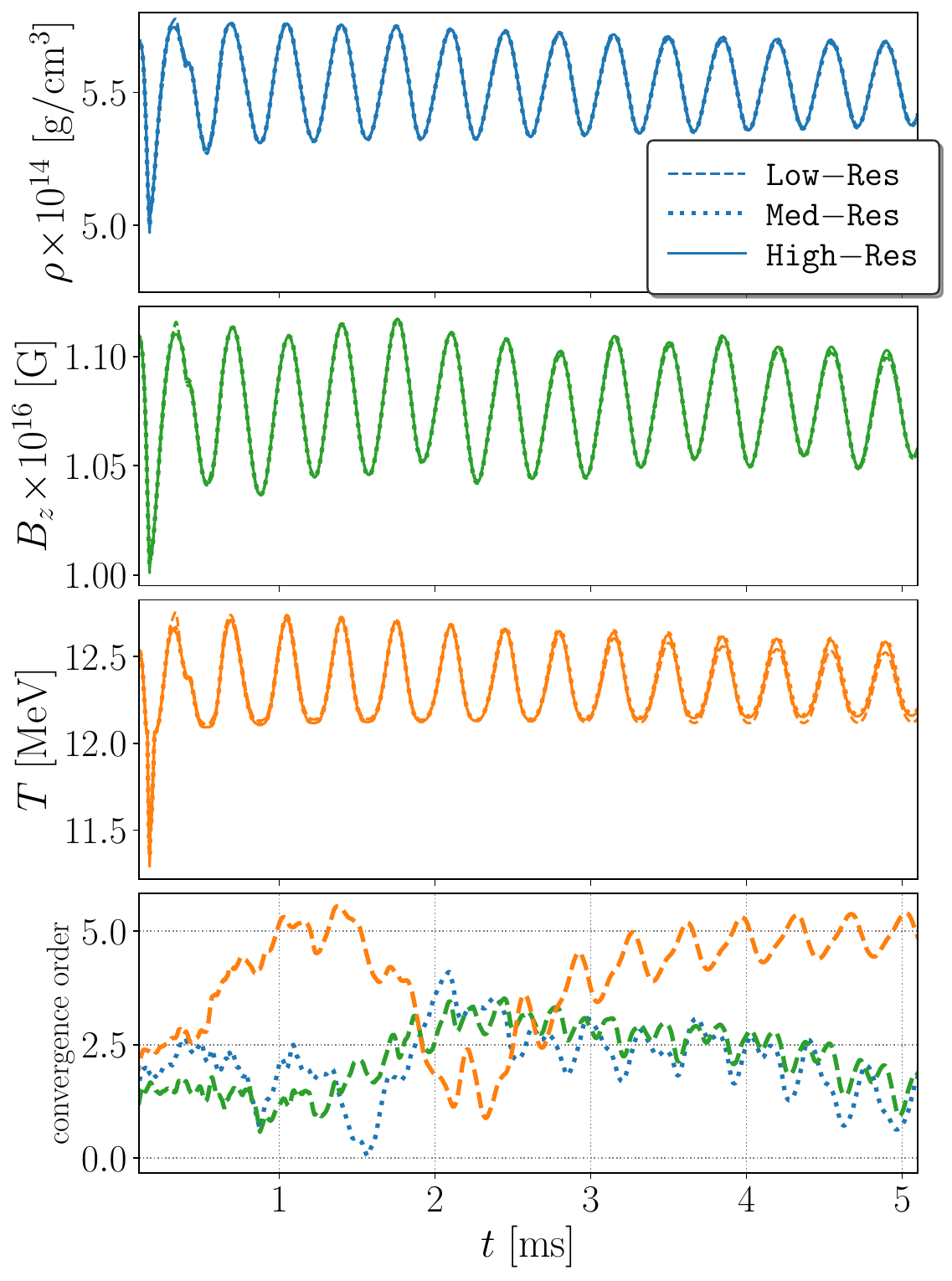}
	\caption{{\em Rotating, hot magnetized NS using the DD2 tabulated EoS.} 
           Stellar properties at the center: from top to bottom,  density, 
           $z$-component of the magnetic field, and temperature.
           The evolution oscillates about a unique, stationary
           solution with an amplitude slowly decreasing with time. 
			Tiny differences between the high and low resolution runs become apparent only at late times. The density and magnetic field converge between second and third order, whereas the temperature converges even faster.
           }
           \label{figure:rotNS_DD2_B}

\end{figure}

\subsection{Hot, rotating, magnetized star with tabulated DD2 EoS}
\label{sec:hotstar}

We evolve a hot, rotating, single NS with the tabulated DD2 EoS on a dynamic spacetime. The star has gravitational mass $1.4~M_\odot$, radius $13.66$~km, and a rotational period of $2.2$~ms. It is constructed assuming $\beta$-equilibrium and a temperature of $T=12$~MeV. Here, we perturb the star by: (i)~constructing the initial data with the tabulated EoS matched to a polytrope below an intermediate energy density, and (ii)~setting a strong poloidal magnetic field with maximum strength $B=1.2 \times 10^{16}$~G.
The boundaries are located at $\pm 1892$ km, employing 8 levels of refinement to achieve a finest resolution of $\Delta x = 230$~m (low resolution), $\Delta x = 168$~m (medium resolution) and $\Delta x = 115$~m (high resolution) covering the whole star. 

We perform the simulations using the AMReX code run on both CPU and GPU, again finding almost perfect agreement. Therefore, we present only one set of results. 
First, we display the density, the $z$-component of the magnetic field, and the temperature on a meridional plane in Fig.~\ref{figure:rotNS_DD2_A} at the initial and final time $t=5$~ms. 
The star exhibits significant oscillations, during which some matter is ejected into the atmosphere, although the stellar interior remains largely unaffected by the dynamics near the surface

We examine the density, magnetic field strength, and temperature at the stellar center in Fig.~\ref{figure:rotNS_DD2_B}.
The star is perturbed away from equilibrium mainly by numerical discretization and by our addition of initial perturbations. Nevertheless, the star oscillates
around a unique solution with an amplitude that slowly decreases with time. This behavior suggests that the code is accurately evolving the perturbed solution.
In the bottom panel we compute the convergence order of these quantities, and find that most converge between second and third order, whereas the temperature converges even faster.

\begin{figure}[h]
	\includegraphics[width=1\columnwidth]{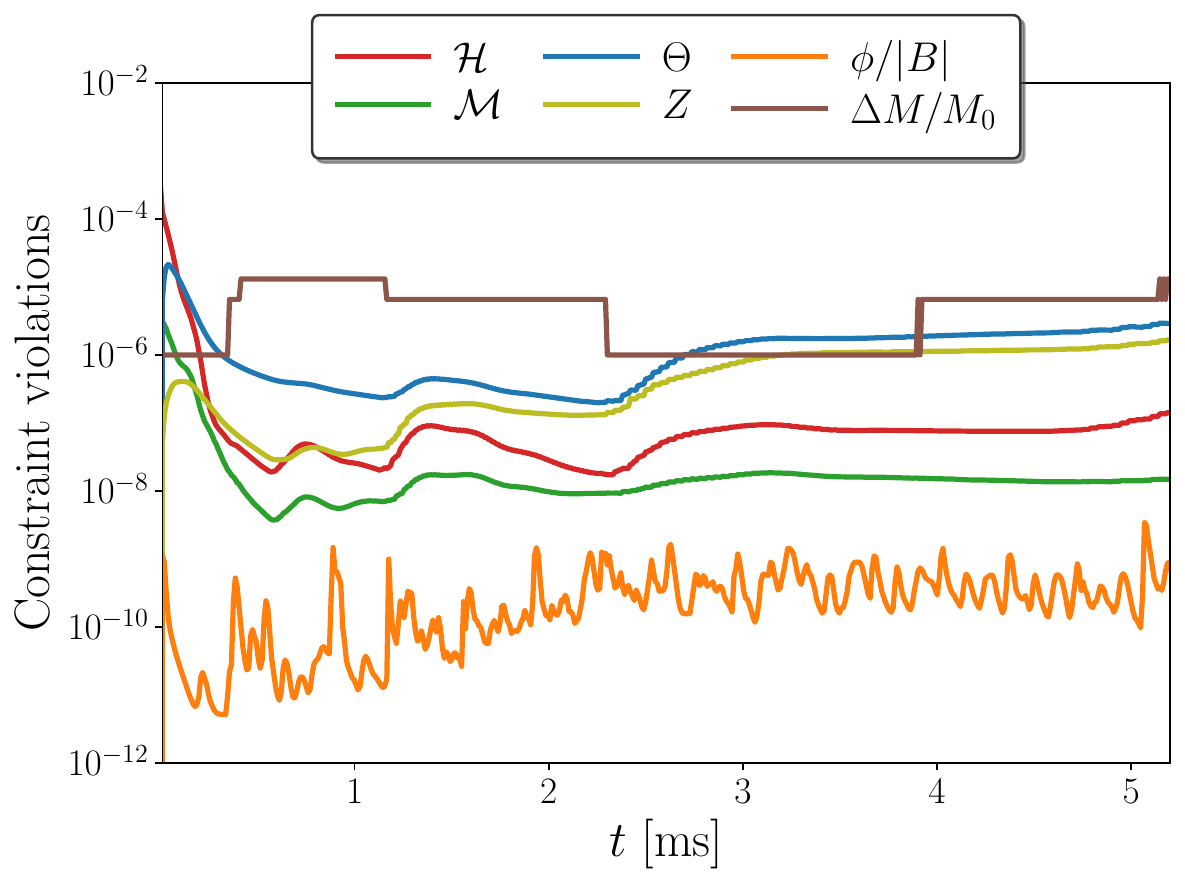}
	\caption{{\em Rotating, hot magnetized NS using the DD2 tabulated EoS.} Estimates of the accuracy of the solution for the high-resolution simulation. The fractional change in baryonic mass of the star $\Delta M/ M_0$ (i.e., which ideally should remain constant during the evolution) and the L2-norm of the energy and momentum constraint violations. Also displayed are the time integrals of the energy constraint residual, $\theta$, and the solenoidal constraint violation, properly normalized (i.e., $\phi/|B|$).
    }\label{figure:rotNS_DD2_C}
\end{figure}

We also consider certain global constraints to assess the accuracy of the solution. We display the variation of the total baryonic mass (i.e., computed as the integral of the conserved density field) for the highest resolution simulation in Fig.~\ref{figure:rotNS_DD2_C}. 
Because our fluid evolution is a conservative scheme, this mass should in principle remain constant, but, in practice, a number of avenues allow for its non-conservation (outer and AMR boundaries, flooring of the atmosphere, etc).
We find that its value varies by roughly $0.001\%$ over the
duration of our simulation.\footnote{Note that the step-like appearance of the data is due to the limited number of significant digits in the integral output.} We also check the L2-norm of the energy and momentum constraint residuals, which remain very small throughout the simulation, as well as the time integrals of the energy-momentum and the solenoidal constraints. These results suggest that the code converges to a consistent, stationary stellar solution.

\begin{figure*}[t]
	\includegraphics[width=\linewidth]{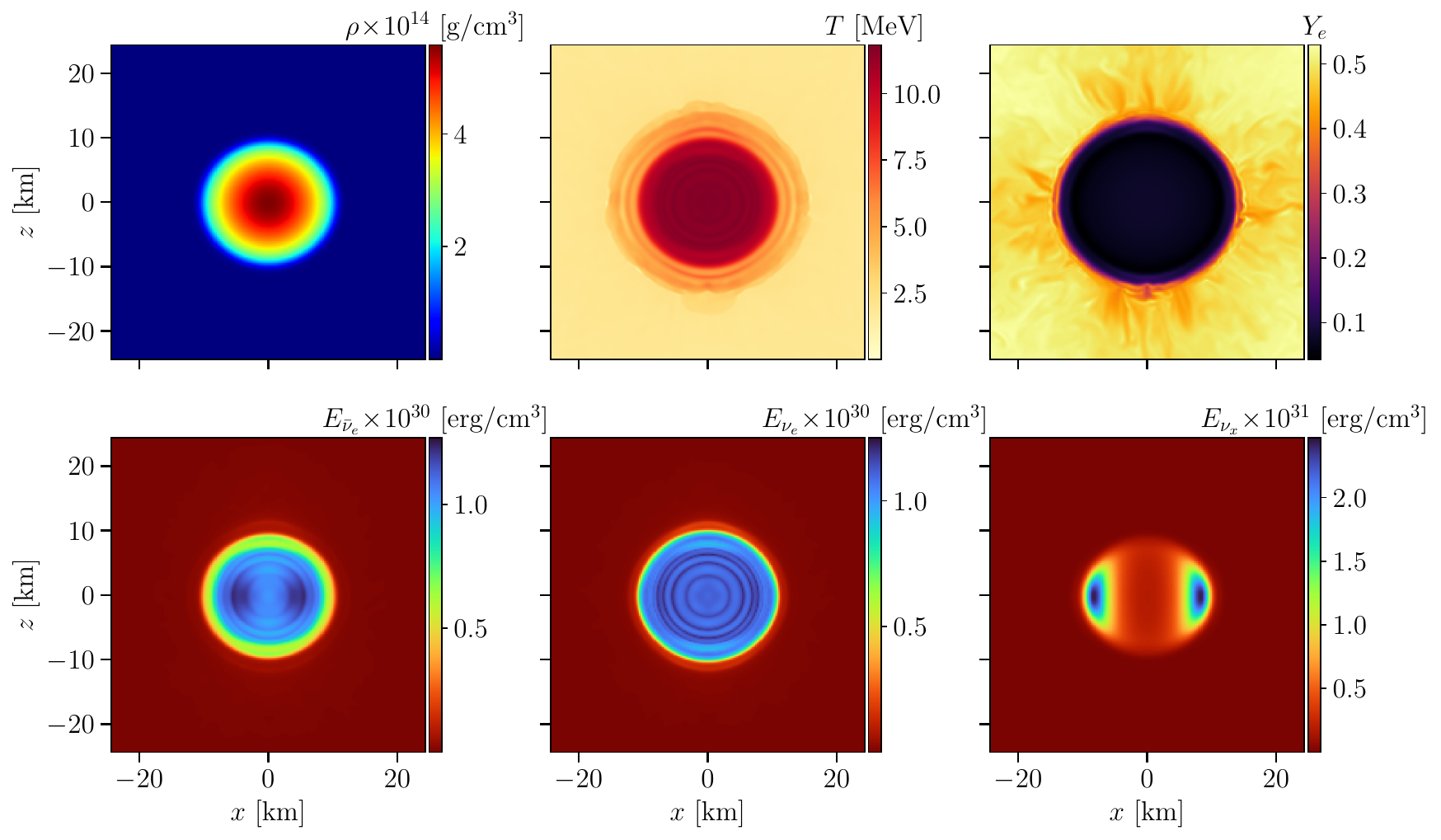}
	\caption{{\em Rotating, hot, magnetized star using the DD2 tabulated EoS and including neutrinos.}
		Snapshots on a meridional plane, from left to right: (top)~the rest-mass density, temperature, and the  electron fraction; (bottom)~the neutrino energy densities for the three species  at the final time (i.e., $t=10$~ms) of our simulation. 
	} 
	\label{figure:rotNS_DD2_D}
\end{figure*}

Finally, we demonstrate our implementation of the M1 scheme 
by repeating the evolution of the same hot, rotating star but including neutrino transport. We evolve until $t = 10$~ms, and show snapshots of the final state for the high-resolution run with $\Delta x = 115$~m in      Fig.~\ref{figure:rotNS_DD2_D}. The evolution is similar to that without neutrinos, but neutrino transport slowly cools the star. 

The neutrino energy densities shown in the bottom row of Fig.~\ref{figure:rotNS_DD2_D} demonstrate that the configuration has  not yet reached equilibrium for the heavy-lepton neutrinos, which take longer to equilibrate than the other species due to the smaller absorption rates. Their characteristic equilibration time is of the order of several tens of milliseconds for this model. The toroidal configuration of $E_{\nu_x}$ (bottom right panel) follows the pattern of the Lorentz factor, which reaches maximum values of $W \simeq2$ in the equatorial outer regions of the star. This leads the interaction rates in the Eulerian frame to be roughly two times bigger than in the center of the star, causing a faster equilibration. This toroidal shape, indeed, becomes more evident as time evolves, but will eventually disappear on longer timescales when the entire star reaches the equilibrium configuration. Some traces of this toroidal shape are also visible in $E_{\overline{\nu}_e}$ for the same reason, but are absent in $E_{\nu_e}$ because electron neutrinos have the highest absorption rates. 
For this particular system, the electron neutrinos in the stellar interior rapidly reach equilibrium with the fluid. Consequently, as expected, their distribution in the dense region follows the same pattern as the fluid temperature, $T$.
These energy densities match our physical expectations and suggest that the neutrino transport scheme is working correctly.

\begin{figure}[t]
	\includegraphics[width=1\columnwidth]{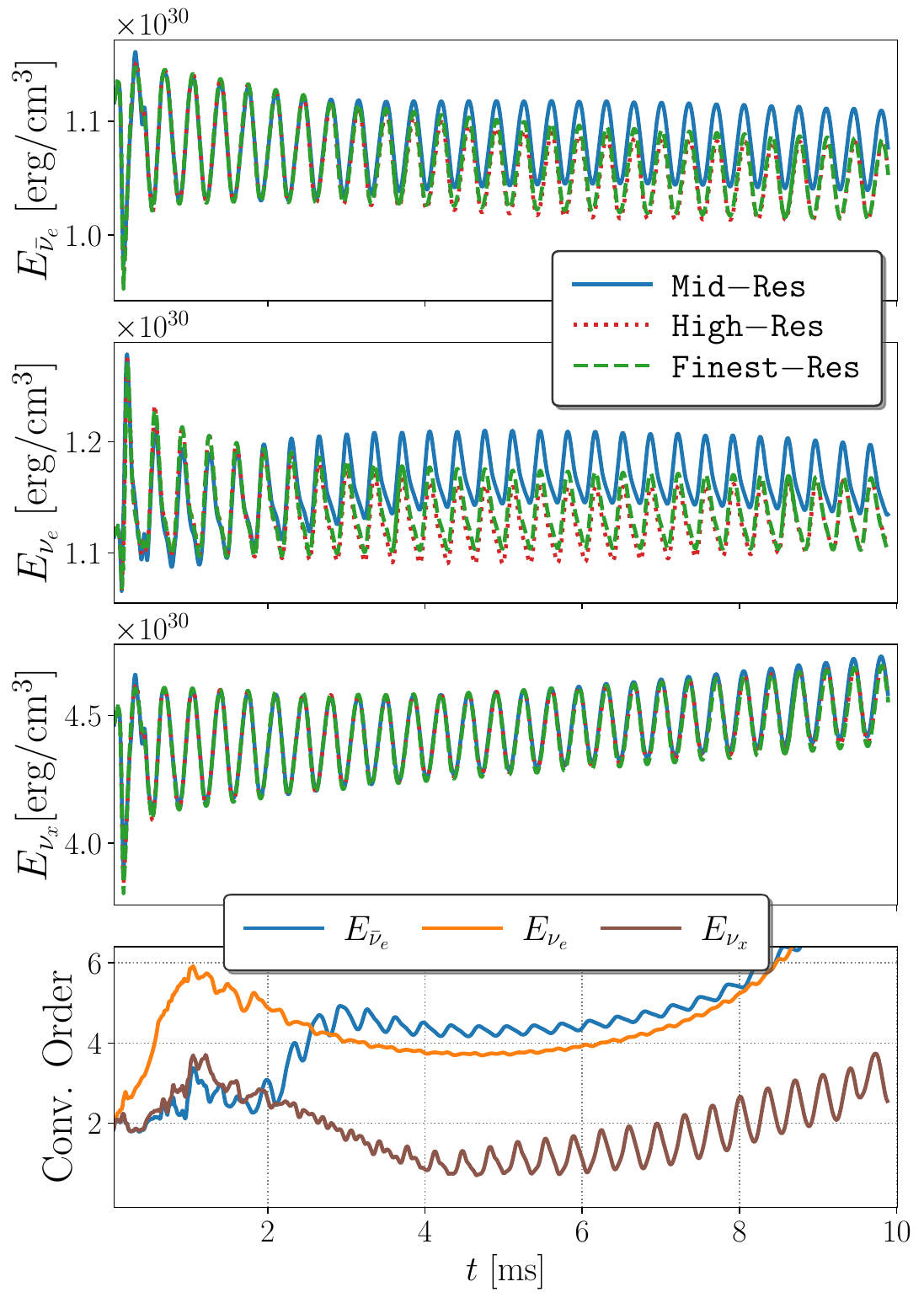}
	\caption{{\em Rotating, hot, magnetized star using the DD2 tabulated EoS and including neutrinos.}
	The neutrino energy densities at different resolutions ($\Delta x = 168$~m, $115$~m, and $82$~m) as functions of time, along with (bottom) the order of convergence for each of them.
}
	\label{figure:rotNS_DD2_E}
\end{figure}

We also assess the convergence of the solution by examining the neutrino energy densities at the center, shown in Fig.~\ref{figure:rotNS_DD2_E} for three different resolutions, including a very high-resolution  case with $\Delta x = 82$~m. The simulations converge faster than second order for the three highest resolutions. These simulations indicate that a  minimum resolution of $\Delta x \lesssim {\cal O}(100\mathrm{m})$ is required to capture       accurately the neutrino dynamics. More extensive benchmarks and results, including binary neutron star mergers with neutrinos and large-eddy simulations, will be presented in a forthcoming dedicated publication.

\subsection{Binary neutron star with tabulated DD2 EoS}

We evolve an equal-mass BNS in a close quasi-circular-orbit with the DD2 tabulated EoS. The binary has a gravitational ADM mass of $2.52~M_\odot$ with an initial separation of $38.40$~km, which corresponds to an angular velocity of $2155$~rad/s. It is constructed assuming 
cold stars in  $\beta$-equilibrium. As in the previous tests above, the boundaries are located at $\pm 1892$~km, employing 8 levels of refinement to achieve a finest resolution of $\Delta x = 185$~m (low resolution), $\Delta x = 154$~m (medium resolution) and $\Delta x = 132$~m (high resolution) covering the stars. 

\begin{figure*}[t]
	\includegraphics[width=\linewidth]{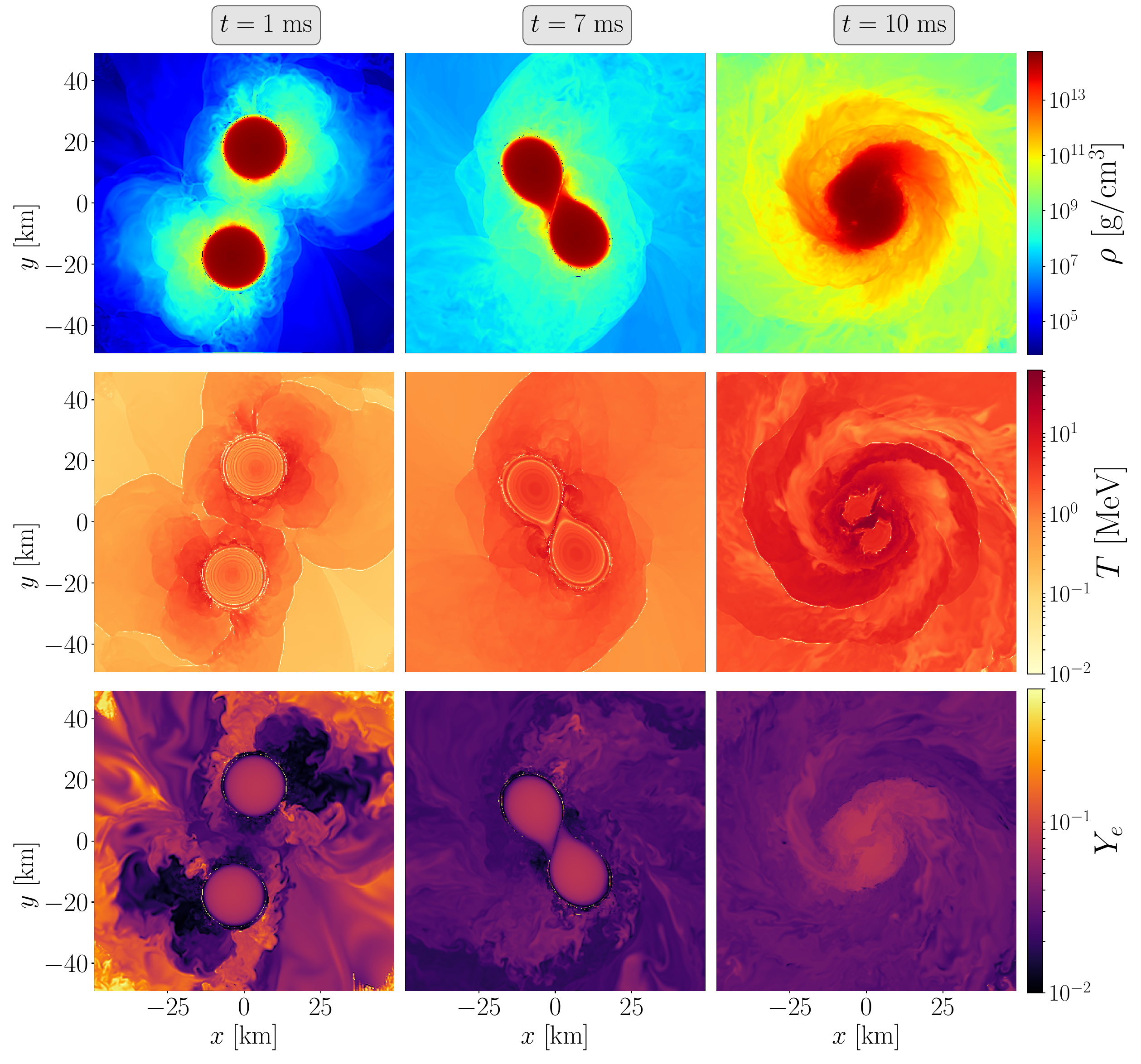}
	\caption{
        {\em Binary neutron star using the DD2 tabulated EoS.} 
		The coalescence dynamics is illustrated across the three stages (inspiral, merger, and post-merger) with snapshots on the equatorial plane showing the rest-mass density, temperature, and electron fraction at representative times of the simulation.
	}
	\label{figure:binNS_DD2_A}
\end{figure*}

Fig.~\ref{figure:binNS_DD2_A} depicts the binary dynamics with snapshots of the rest-mass density in the orbital plane at representative stages of the coalescence. After completing roughly two orbits, the neutron stars collide and merge, giving rise to a massive, differentially rotating remnant surrounded by complex spiral structures and shock-heated matter.

\begin{figure}[h]
	\includegraphics[width=1\columnwidth]{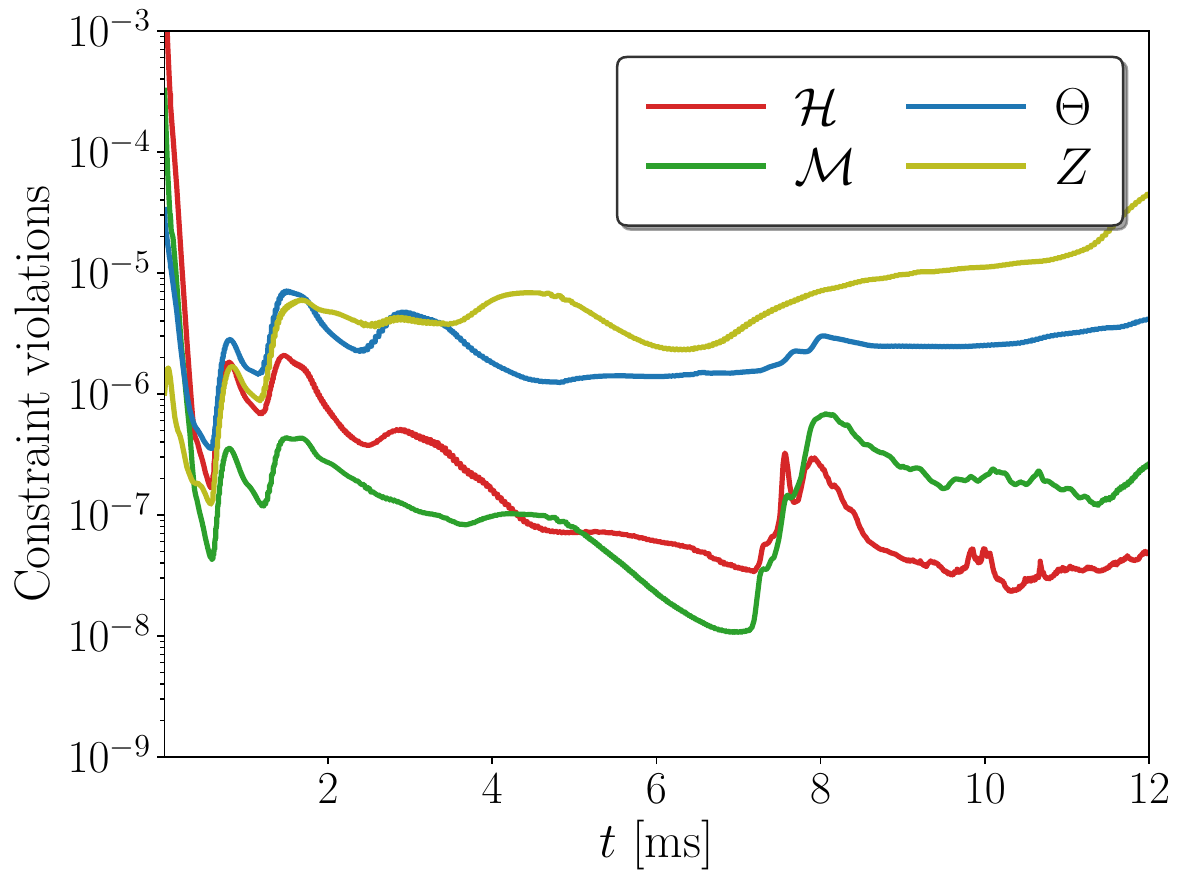}
	\caption{ {\em Binary neutron star using the DD2 tabulated EoS.} 
	Estimates of the solution’s accuracy, showing the L2-norm of the energy and momentum constraint violations, along with their time integrals $\theta$ and $|Z|$.
	}\label{figure:binNS_DD2_B}
\end{figure}

To assess the stability of the evolution, Fig.~\ref{figure:binNS_DD2_B} shows estimates of the simulation errors. All plotted constraints (energy, momentum, and their time integrals) remain stable and within narrow bounds. This behavior persists until the end of our simulation, with no visual indications of any imminent violations

\begin{figure}[h]
	\includegraphics[width=1\columnwidth]{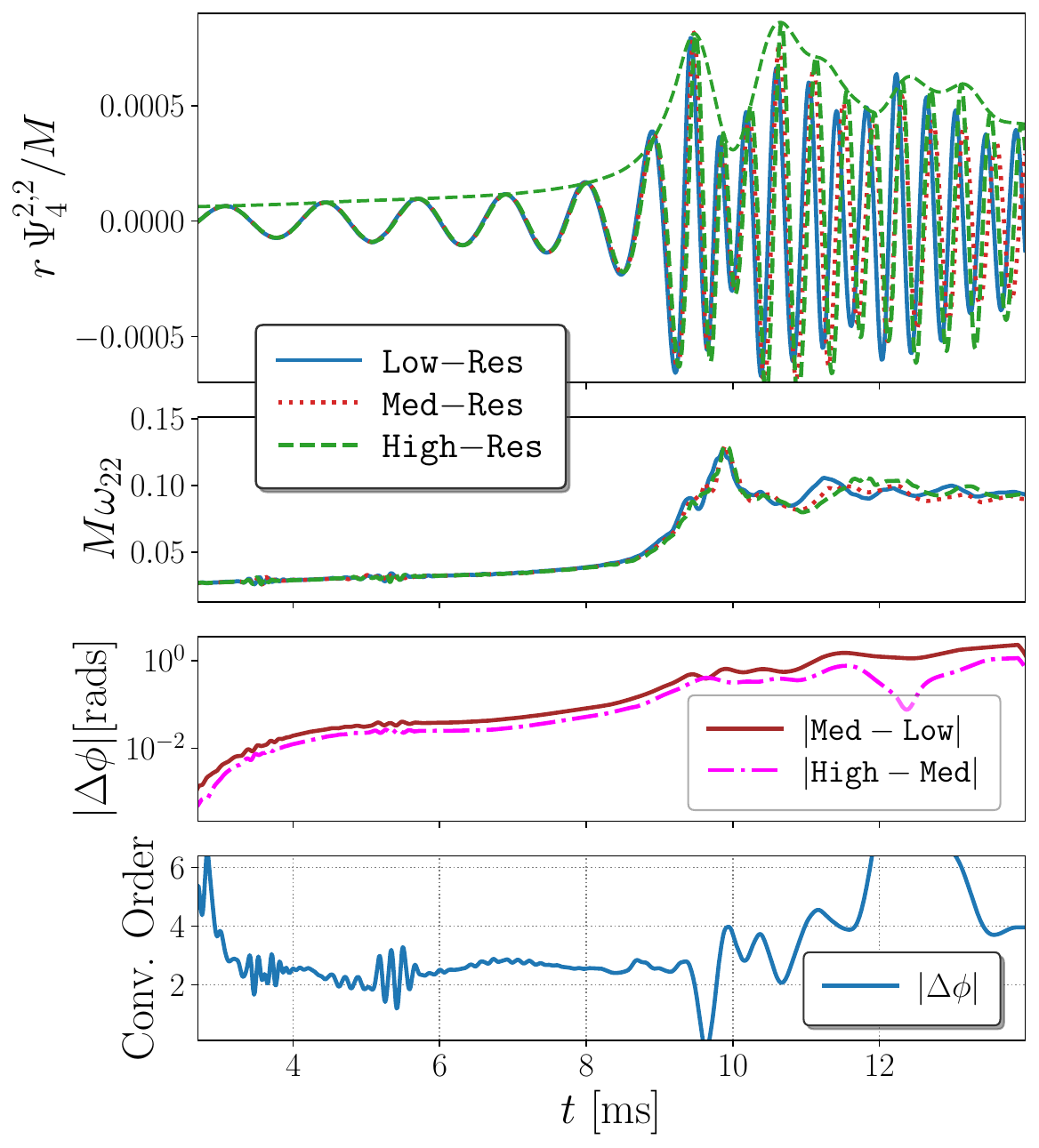}
	\caption{{\em Binary neutron star using the DD2 tabulated EoS.} 
	Gravitational waveform emitted during the coalescence. Shown are the real part of the dominant mode ($l=m=2$) of the Newman–Penrose scalar $\Psi_4$ (top) and its instantaneous angular frequency (upper middle). 
    The phase error~(lower middle) and respective convergence order~(bottom)
    indicate convergence better than second order. 
    }\label{figure:binNS_DD2_C}	
\end{figure}

{Finally, we present in Fig.~\ref{figure:binNS_DD2_C} the gravitational waveform produced during the binary coalescence. In particular, the real part and the angular frequency of the dominant $l=m=2$ mode are displayed for three different resolutions in the two upper panels. This mode is obtained from the spin-weighted spherical harmonic decomposition of the Newman–Penrose scalar 
$r \Psi_4$. The phase differences between resolutions and the corresponding convergence order are shown in the two lower panels, indicating a convergence order between 2 and 3. 
}

High order convergence of the phase error in a BNS merger is a demanding and
important test as discussed within the context of future gravitational wave
detectors in Ref.~\cite{2025arXiv250810981K}. A uniform convergence order
is not expected here, in part, because of variations among the dynamic refinement among the three resolutions. Our use of a tabulated EoS also introduces
interpolation error. 
This convergence order is comparable to previous results in the literature; see, for instance, Fig.~6 in Ref.~\cite{2025arXiv250810981K}, Fig.~19 in Ref.~\cite{2025ApJS..277....3C}, and Fig.~3 in Ref.~\cite{Habib:2025bkb}.

\section{Performance}

Testing the speed and scaling of a code such as this can be quite extensive because of the variety of regimes and scales involved with the various physics solvers. For example, a binary in quasi-circular orbit involves significant re-gridding associated with moving the grids as they track the two compact objects. Evolutions with a tabulated EoS and neutrino radiation involve loading large tables which test the memory bandwidth of machines.

We first present a strong scaling (fixed problem size run on an increasing number
of nodes) test of \mhduet using the hot rotating star from Section~\ref{sec:hotstar}. We run the simulation for 8 coarse time steps on MareNostrum~(MN5) and Tursa clusters. In particular, the general purpose partition of MareNostrum 5 has 6408 CPU-nodes, each one with two Intel Sapphire Rapids~8480+ at 2~GHz (112~cores per node) with 256~GB of main memory, using DDR5. The accelerated partition of MN5 has 1120~GPU-nodes, each with with four  Nvidia Hopper GPUs with 64~GB of HBM2 memory. Tursa has 64~nodes with four Nvidia Ampere A100-80 GPUs in each and also 114~nodes with four Ampere A100-40 GPUs in each. 

\begin{figure}[h]
	\includegraphics[width=1\columnwidth]{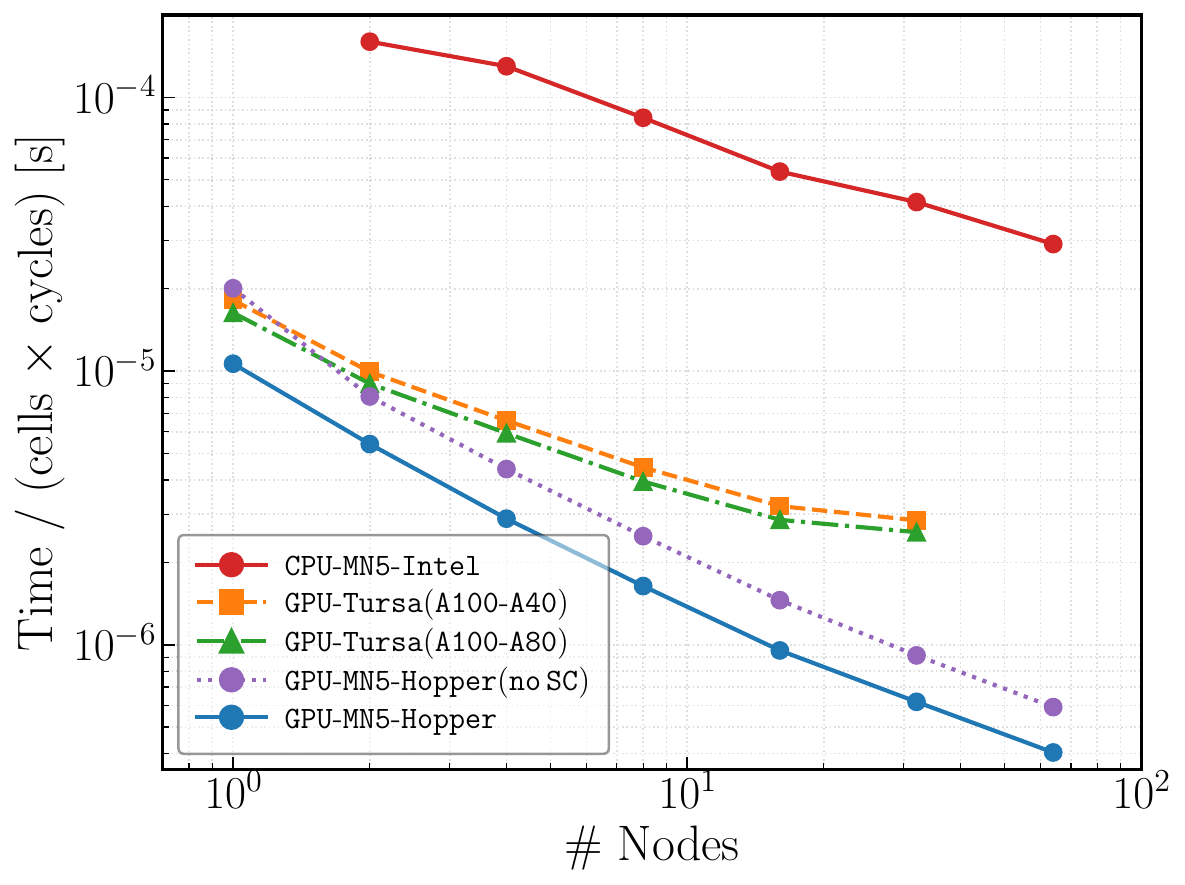}
	\caption{{\em Strong scaling with the single, rotating, hot NS using the DD2 tabulated EoS.} The CPU implementation does not scale well because a large fraction of the available memory is employed saving the tabulated EoS. The GPU-MN5 is at least 20 times faster than CPU-MN5.
    }
	\label{figure:scaling_rotatingNS}
\end{figure}

The strong scaling results are displayed in Fig.~\ref{figure:scaling_rotatingNS}.
The scaling suffers on CPUs because a large fraction of RAM is dedicated to holding the EoS table.
On the other hand, the scaling achieved on GPUs is quite good, up to a factor of 16 with respect to the minimum number of nodes required for this problem, especially for GPU-MN5. It is worth emphasizing that GPU-MN5 is approximately 20 times faster than its CPU counterpart in a node-to-node comparison.  We consider that comparing performance on a node-to-node basis provides the most meaningful measure, as the cost of a GPU node is of the same order as that of a CPU node (i.e., the GPU node is typically 2–4 times more expensive, though in some cases the cost can be up to eight times higher). 
In contrast, a GPU-MN5 node contains 4 GPUs while a CPU-MN5 node has 112 cores, and so one could compare a single GPU against a single CPU core. Such a comparison would yield a speedup of roughly 500.

In the same figure, we also compare the efficiency and scalability of the GPU-MN5 code with and without sub-cycling in time. While the performance gain depends on the specific problem, for this test we observe a speed-up of roughly 1.5–2 when sub-cycling is enabled, with no noticeable loss in scalability.

It is also interesting to compare the scaling on Hopper nodes with those of A100 nodes. While the performance on one node of each is similar (i.e., Hopper is approximately twice as fast as A100), the scaling on Hopper nodes continues quite well up to 64 nodes (i.e., 256 GPUs) for this test in contrast to the flattened scaling achieved on A100 nodes.
How much the better scaling of the Hopper nodes can be attributed to this new GPU over the A100 is not so clear, given that the nodes also differ in their interconnect, CPU, and amount of RAM.

\begin{figure}[h]
	\includegraphics[width=1\columnwidth]{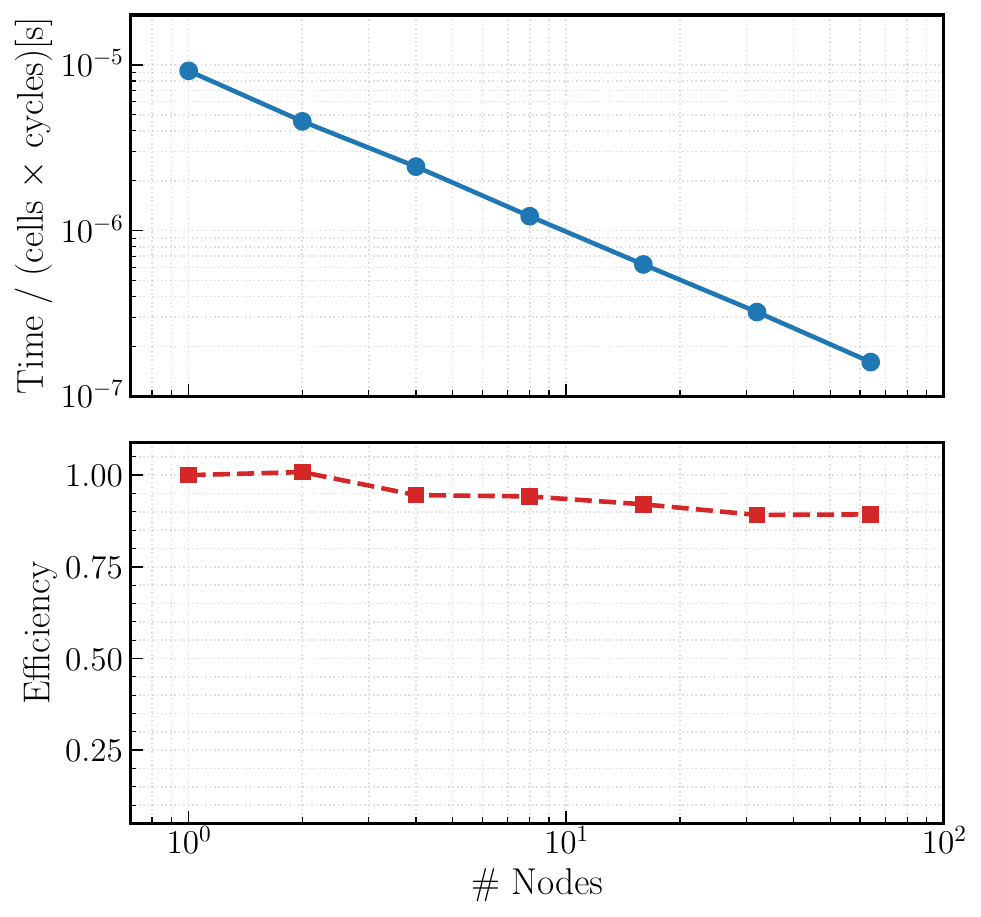}
	\caption{{\em Weak scaling with the single, rotating, hot NS using the DD2 tabulated EoS.} The code exhibits excellent scaling on GPU nodes, achieving an efficiency of about 90\% at the maximum nodes available on GPU-MN5. 
	}
	\label{figure:weak_scaling_rotatingNS}
\end{figure}

Although strong scaling generally suggests good weak scaling, we compute it explicitly for the same rotating, hot neutron star. The results are shown in Fig.~\ref{figure:weak_scaling_rotatingNS}. As expected, the weak scaling performance on GPU-MN5 is excellent, reaching approximately 90\% efficiency at the maximum number of GPU nodes available on the cluster. This test further demonstrates the strong scalability of the GPU implementation of the \mhduet code in realistic problems.

\section{Conclusion}

We have presented the \mhduet code, highlighting its flexibility, high-performance capabilities, and applicability to fully relativistic, astrophysical scenarios. The code has been rigorously tested on single magnetized neutron stars, successfully recovering the quasi-normal modes from previous studies. We have also evolved hot, magnetized stars, demonstrating accurate handling of both thermal and magnetic effects. Furthermore, we have applied \mhduet to fully relativistic, magnetized binary neutron star mergers, capturing the post-merger dynamics. Convergence studies of these scenarios consistently demonstrate that the solutions converge faster than second order in the grid spacing.

In all tests, the results obtained with the SAMRAI infrastructure are fully consistent with those from AMReX, validating the robustness and portability of the code across different frameworks. Performance benchmarks reveal that AMReX-GPU reaches more than an order-of-magnitude speedup compared to AMReX-CPU when comparing GPU and CPU nodes, highlighting the potential of GPU acceleration to dramatically reduce computational costs for demanding simulations. Together, these results demonstrate that \mhduet provides a reliable and efficient platform for high-fidelity simulations of compact objects, enabling detailed studies of their multi-messenger signatures and opening new avenues for large-scale parameter surveys previously inaccessible due to computational limitations.

\section*{Acknowledgements}

This work was supported by the project PID2022-138963NB-I00, funded by the Spanish Ministry of Science, Innovation and Universities (MCIN/AEI/10.13039/501100011033). JAC acknowledges support from a predoctoral fellowship (FPI) associated with this project (reference PREP2022-000480).
MB acknowledge partial support from the STFC Consolidated Grant no. ST/Z000424/1 and Gravitational Waves consolidated grant no. APP46676. 
This work was also supported by the National Science Foundation via grant PHY-2409407.
This research used computational resources provided by
the Advanced Cyberinfrastructure Coordination Ecosystem: Services \& Support (ACCESS) program, which is supported by National Science Foundation grants 2138259, 2138286, 2138307, 2137603, and 2138296,
and by
the Unity platform at the Massachusetts Green High Performance Computing Center (MGHPCC) provided by the University of Rhode Island.
The author thankfully acknowledges the computer resources at MareNostrum and the technical support provided by Barcelona Supercomputing Center (RES-AECT-2025-2-0007 and RES-AECT-2025-2-0024).
This work used the DiRAC Extreme Scaling service (Tursa) at the University of Edinburgh, managed by the EPCC on behalf of the \href{https://www.dirac.ac.uk}{STFC DiRAC HPC Facility}. The DiRAC service at Edinburgh was funded by BEIS, UKRI and STFC capital funding and STFC operations grants. DiRAC is part of the UKRI Digital Research Infrastructure.
\appendix

\section{Large Eddy Simulation techniques}
\label{appendix:LES}

The main idea of Large-Eddy Simulation (LES) is to resolve the large-scale dynamics explicitly, while modeling the influence of the unresolved subgrid scales through additional terms in the evolution equations. These techniques are particularly relevant for problems in which small-scale dynamics play a critical role, such as turbulence.
The effect of finite resolution in a numerical simulation can be thought as equivalent to a low-pass spatial filter applied to the governing evolution equations. Applying the filter to the continuous set of fields \(\mathbf{u}(t,\mathbf{x})\) effectively decomposes the fields into a resolved component \(\overline{\mathbf{u}}\) and an unresolved SGS contribution \(\mathbf{u}' = \mathbf{u} - \overline{\mathbf{u}}\). 

The filtered (resolved) field is formally computed as
\begin{equation}
	\overline{\mathbf{u}}(t,\mathbf{x}) = \int_{-\infty}^{+\infty} G(\mathbf{x} - \mathbf{x}')\, \mathbf{u}(t,\mathbf{x}')\, d\mathbf{x}'~,
	\label{eq:filter_def_bold}
\end{equation}
where \(G\) is the filter kernel with a dependence on the filter width, $\Delta_f$, which is typically chosen to be comparable to the grid cell size, $\Delta$.  An homogeneous, isotropic, low-pass filter is independent of the direction (i.e., $G({\bf x}-{\bf x'})=G(|{\bf x}-{\bf x'}|)$) and only smooths out fluctuations on length scales smaller than the filter size, leaving unchanged the variations of the solution at larger length scales. In addition, the filter operator is linear and commutes with spatial derivatives. Generically, it can be written for any dimension $D$ as
\begin{eqnarray}
	G(|{\bf x}-{\bf x'}|) = \prod_{i=1}^{D} G_i(|x_i-x'_i|) ~,
	\label{eq:kernel_dim}
\end{eqnarray}
where $G_i(|x_i-x'_i|)$ is just the one-dimensional kernel function.
The simplest low-pass filter is the mean value in a cubic domain with size $\Delta_f$ in each Cartesian direction $\{x_i\}$, described by the normalized kernel
\begin{eqnarray}
	G_i(|x_i-x'_i|) = 
	\begin{cases}
		1/\Delta_f ~~~~ {\rm if} ~~|x_i-x'_i| \le \Delta_f/2\\
		0 ~~~~~~~~~ {\rm otherwise}
	\end{cases}
	\label{eq:kernel_space_box}
\end{eqnarray}
Despite the appealing simplicity of the box filter, which makes it very useful to perform numerical calculations, we will see below that it is not suitable for analytical calculations involving its derivatives, since they are not continuous. Therefore, at a formal level, it is more practical to introduce the normalized Gaussian kernel, which in the space domain can be written as
\begin{eqnarray}
	G_i(|x_i-x'_i|) =  \left( \frac{1}{4 \pi \xi } \right)^{1/2}
	\exp\left(\frac{-|x_i-x'_i|^2}{4 \xi}\right) ~,
	\label{eq:kernel_space_gaussian}
\end{eqnarray}
where $\xi$ defines the effective filtering width. Besides having the same zeroth and first moments, Gaussian and box filters have the same second moment if we set $\xi=\Delta_f^2/24$.

When the filtering operator is applied to a nonlinear balance law,
\begin{equation}
	\partial_t \mathbf{u} + \partial_k \mathbf{F}^k(\mathbf{u}) = \mathbf{0}~,
	\label{eq:generic_balance_bold}
\end{equation}
the linearity of differentiation allows us to write
\begin{equation}
	\partial_t \overline{\mathbf{u}} + \partial_k \overline{\mathbf{F}^k(\mathbf{u})} = \mathbf{0}~.
	\label{eq:filtered_balance_bold}
\end{equation}
However, because \(\mathbf{F}^k(\mathbf{u})\) is generally a nonlinear function of \(\mathbf{u}\), it is not true that \(\overline{\mathbf{F}^k(\mathbf{u})} = \mathbf{F}^k(\overline{\mathbf{u}})\). We therefore define the SGS residual, or subfilter-scale tensor, as
\begin{equation}
	\overline{\boldsymbol{\tau}}^k_F = \mathbf{F}^k(\overline{\mathbf{u}}) - \overline{\mathbf{F}^k(\mathbf{u})}~.
	\label{eq:SFS_def_bold}
\end{equation}
With this definition, the filtered balance law becomes
\begin{equation}
	\partial_t \overline{\mathbf{u}} + \partial_k \mathbf{F}^k(\overline{\mathbf{u}}) = \partial_k \overline{\boldsymbol{\tau}}^k_F~.
	\label{eq:filtered_balance_SGS_bold}
\end{equation}
The right-hand side represents the influence of the unresolved scales on the resolved dynamics and requires a closure model to express \(\overline{\boldsymbol{\tau}}^k_F\) in terms of the resolved fields.

We adopt the gradient model as our SGS closure, which is derived by performing a Taylor expansion of the filter kernel and expressing the subgrid-scale contributions in terms of spatial derivatives of the resolved fields~\cite{leonard75,muller02a}. For example, when filtering the product of two fields, \(f\) and \(g\), a first-order expansion in the filter width parameter \(\xi \propto \Delta_f^2\) yields
\[
\overline{fg} \approx \overline{f}\,\overline{g} + 2\,\xi\,\nabla\overline{f}\cdot\nabla\overline{g}~.
\]
More generally, for a nonlinear flux \(\mathbf{F}^k(\mathbf{u})\), the gradient model approximates the SGS tensor as
\[
\tau^k_F \approx -\xi\,\nabla\left(\frac{d\mathbf{F}^k}{d\mathbf{u}}\Big|_{\overline{\mathbf{u}}}\right) \cdot \nabla\overline{\mathbf{u}}~,
\]
where the derivative \(\frac{d\mathbf{F}^k}{d\mathbf{u}}\) is evaluated with the resolved field \(\overline{\mathbf{u}}\). This expression captures the leading-order effects of the unresolved fluctuations on the fluxes.

The relativistic MHD system is more complex, as it evolves a set of conserved variables $\mathbf{C}$, whose fluxes $\mathbf{F}^{k}(\mathbf{P})$
depend on a distinct vector of primitive fields 
$\mathbf{P}$. In practice, the numerical scheme evolves the filtered conserved variables $\overline{\mathbf{C}}$.  Since the nonlinear fluxes depend on the primitive variables, we first reconstruct the primitives from  $\overline{\mathbf{C}}$, yielding a set of reconstructed fields, $\widetilde{\mathbf{P}}$.
The filtered evolution equations can then be written as
\begin{equation}
	\partial_t \overline{\mathbf{C}} + \partial_k \mathbf{F}^{k}(\widetilde{\mathbf{P}}) = \partial_k \overline{\boldsymbol{\tau}}^{k}_F~,
	\label{eq:filtered_conservation_final_bold}
\end{equation}
where the SGS residual in the fluxes is defined as
\begin{equation}
	\overline{\boldsymbol{\tau}}^{k}_F = \mathbf{F}^{k}(\widetilde{\mathbf{P}}) - \overline{\mathbf{F}^{k}(\mathbf{P})}~.
	\label{eq:SFS_flux_final_bold}
\end{equation}
Here, \(\mathbf{F}^{k}(\widetilde{\mathbf{P}})\) represents the flux evaluated using the reconstructed primitive fields, while \(\overline{\mathbf{F}^{k}(\mathbf{P})}\) is the filtered flux computed from the full primitive fields.

By substituting our gradient model approximation for the SGS term into the filtered conservation equations, we obtain a closed set of evolution equations that incorporate the effects of subgrid-scale turbulence through additional derivative terms. For the GRMHD equations, this derivation follows straightforwardly, though with extensive algebra, and its detailed procedure is discussed in Refs.~\cite{carrasco19,vigano20,Palenzuela:2022kqk}. We emphasize that the gradient model is conceptually similar to a reconstruction numerical scheme, without relying on any a priori physical assumptions.

Notice that we hereafter have simplified the notation by removing the tildes and bars from the filtered fields and fluxes, for the sake of clarity. All fields in the equations are implicitly meant to be the filtered values (i.e., simply resolved by the discretized equations, as in any simulation).

Connecting with the notation employed in the equations (\ref{eq:mhd_system}), the additional SGS terms are given by:
\begin{eqnarray}
	{\cal G}_{D}^{k} &=& -~{\cal C_N}~\xi \sqrt{\gamma} \, H_{N}^k ~~, \nonumber \\
	{\cal G}_{Y}^{k} &=& -~{\cal C_Y}~\xi \sqrt{\gamma} \, H_{N_Y}^k ~~, \nonumber \\	
	{\cal G}_{S}^{ki} &=& -~{\cal C_T}~\xi \sqrt{\gamma} \, H_{T}^{ki} ~~, \nonumber \\
	{\cal G}_{B}^{ki} &=& -~{\cal C_M}~\xi \sqrt{\gamma} \, H_{M}^{ki} ~~. \label{eq:sgs_gradient}
\end{eqnarray}
The coefficient $\xi= \gamma^{1/3} \Delta^2/24$ has the proportionality to the spatial grid spacing squared, which is typical of SGS models and ensures by construction the convergence to the continuous limit (vanishing SGS terms for an infinite resolution). 
Importantly, for each equation there is a coefficient ${\cal C}_i$, which is meant to be of order one for a numerical scheme having a mathematically ideal Gaussian filter kernel and neglecting higher-order corrections. 
Although ideally the coefficient is of order one~\cite{carrasco19,vigano20}, because numerical 
dissipation and dispersion decrease the effect of the subgrid term, in practice one finds that increasing the coefficient can mitigate this decrease.
Generally, we set the coefficient associated with the magnetic field ${\cal C}_M = 8$, which has been shown
to reproduce the magnetic field amplification more accurately~\cite{vigano20,aguilera2020} with our current numerical scheme. The remaining coefficients are set to zero, as the fluid dynamics appear to be adequately captured without the need of subgrid terms.

The cumbersome expressions of the tensors $H$ have been obtained in detail for the special \cite{carrasco19} and general relativistic \cite{vigano20} cases. 
Notice that here we have extended the expressions reported in the Appendix A of Ref.~\cite{aguilera2020} to accommodate for the additional variables $Y_e$ and $D_Y$ (primitive and conserved, respectively). 

The final expressions for $(H_{N}^k, H_{N_Y}^k , H_{T}^{ki} , H_{M}^{ki})$, as a function of auxiliary fields $\Psi$ and other $H$-s, are given by:
\begin{widetext}
	\begin{eqnarray}
		{\Psi}_{v}^k &=& \frac{2}{ {\Theta}} \left\lbrace \nabla ( {v}\cdot  {B}) \cdot \nabla  {B}^k  - \nabla  {\Theta} \cdot \nabla  {v}^k   
		+ \frac{ {B}^k}{ {\mathcal{E}}} \left[   {\Theta} \nabla  {B}^j \cdot \nabla  {v}_j +  {B}_j \nabla  {B}^j \cdot \nabla ( {v}\cdot  {B}) -  {B}^j \nabla  {v}_j \cdot \nabla  {\Theta} \right]  \label{hTauv}  \right\rbrace ~, \nonumber  \\
		{\Psi}^{ki}_{M} &=& \frac{4}{ {\Theta}} \left[   {\Theta} \, \nabla  {B}^{[i} \cdot \nabla  {v}^{k]} +   {B}^{[i} \nabla  {B}^{k]} \cdot \nabla ( {v}\cdot  {B}) -  {B}^{[i} \nabla  {v}^{k]} \cdot \nabla  {\Theta} \right] ~,
		\nonumber \\
		{\Psi}_{\Theta} &=&  \frac{ {\Theta}}{ {\Theta} - {E}^2} \left\lbrace \nabla  {B}_{j} \cdot \nabla  {B}^{j} - \nabla  {E}_{j} \cdot \nabla  {E}^{j} -  {B}_{[i} {v}_{k]} \,  {\Psi}^{ki}_{M} \right\rbrace ~~,~~
		{\Psi}_{A}  =  {W}^2 \left(  {p} \, \frac{d {p}}{d {\epsilon}} +  {\rho}^2 \, \frac{d {p}}{d {\rho}} \right) ~, \nonumber \\
		H_{\rm p} &=&  \frac{ {\mathcal{E}} \,  {W}^2 ({ {\Theta}-  {E}^2 })}{( {\rho} \,  {\mathcal{E}} -  {\Psi}_{A})( {\Theta} -  {E}^2 )  {W}^2 +  {\Psi}_{A} \,  {\Theta}} \left\lbrace  {\rho} \left( \nabla \frac{d {p}}{d {\rho}} \cdot \nabla  {\rho} + \nabla \frac{d {p}}{d {\epsilon}} \cdot \nabla  {\epsilon} \right)  - 2 \frac{d {p}}{d {\epsilon}} \, \nabla  {\rho} \cdot \nabla  {\epsilon}  \right. \nonumber\\
		&-&  \left.  \left( {\mathcal{E}} \frac{d {p}}{d {\epsilon}} -  {\Psi}_{A}\right) \left[ \frac{ {W}^2}{4} \nabla  {W}^{-2} \cdot \nabla  {W}^{-2} + \nabla  {W}^{-2} \cdot \nabla (\ln  {\rho}) \right]  -  \frac{2}{ {W}^2}\frac{d {p}}{d {\epsilon}} \left[   \nabla  {B}_j \cdot \nabla  {B}^j -   {W}^{4} \nabla  {W}^{-2} \cdot \nabla  {h} \right]  \right. \label{tau_p}\\ 
		&-&  \left.  \left( {\mathcal{E}} \frac{d {p}}{d {\epsilon}} +  {\Psi}_{A}\right) \left[  {v}_j  {\Psi}_{v}^j +  \nabla  {v}_{j} \cdot \nabla  {v}^{j} +  {W}^2 \, \nabla  {W}^{-2} \cdot \nabla  {W}^{-2} \right]  +   \frac{ {\Psi}_{\Theta}}{ {\mathcal{E}}  {\Theta}} \left[ \left( {\mathcal{E}} \frac{d {p}}{d {\epsilon}} +  {\Psi}_{A}\right)( {\Theta}-  {E}^2 ) - \frac{ {\Psi}_{A} \,  {\Theta}}{ {W}^2}   \right]   \right\rbrace 
		\nonumber \\
		&+& \nabla \frac{d {p}}{d {Y_e}}  \cdot \nabla Y_e - \frac{2}{D} \frac{d {p}}{d {Y_e}} \nabla Y_e \cdot \nabla D  \nonumber\\
		H_{\Theta} &=&  {\Psi}_{\Theta} + \frac{ {\Theta}}{ {\Theta} - {E}^2} H_p ~, \label{tau_Theta} \\ 
		H_{v}^k &:=&  {\Psi}_{v}^k - \left(  {v}^k + \frac{ {v}\cdot  {B}}{ {\mathcal{E}}}  {B}^k \right)  \frac{H_{\Theta}}{ {\Theta}} ~, \label{Tauv} \\
		H^{k}_{N} &=&  2 \, \nabla  {D} \cdot \nabla  {v}^k +  {D} \, H^{k}_v ~~~~,~~~~
		H^{k}_{N_Y} =  2 \, \nabla  {D\!Y_e} \cdot \nabla  {v}^k +  {D\!Y_e} \, H^{k}_v ~~,~~ \\
		H^{ki}_{M} &=&  2  {B}^{[i} H_{v}^{k]} + 4 \, \nabla  {B}^{[i} \cdot \nabla  {v}^{k]} ~~\rightarrow~~ 
		H_{E}^i = \frac{1}{2} \epsilon^{i}_{\phantom ijk } H_{M}^{jk} ~,
		\label{HNME} \\
		H^{ki}_{T} &=& 2 \left[ \nabla  {\mathcal{E}} \cdot \nabla ( {v}^k  {v}^i ) +  {\mathcal{E}} \left(  {v}^{(k} H_{v}^{i)} +  \nabla  {v}^{k} \cdot \nabla  {v}^{i} \right)  +   {v}^k  {v}^i H_{p} \right] 
		- 2\left[ \nabla  {B}^{k} \cdot \nabla  {B}^{i} + \nabla  {E}^{k} \cdot \nabla  {E}^{i} +  {E}^{(k} H_{E}^{i)}   \right]  \nonumber \\
		&+& (\gamma^{ki} -  {v}^k  {v}^i)  \left[ H_p + \nabla  {B}_{j} \cdot \nabla  {B}^{j} + \nabla  {E}_{j} \cdot \nabla  {E}^{j} +  {E}_{j} H_{E}^{j} \right]~, \label{HT}     
	\end{eqnarray}
\end{widetext}

Notice that we need the derivatives $(dp/d\rho, dp/d\epsilon, dp/dY_e)$ in order to compute the gradient SGS terms. They can be computed analytically for hybrid EoSs, but only numerically for the tabulated ones.

\section{Characteristic structure}

We examine the characteristic structure of the equations because this information is essential for the HRSC methods employed to evolve the fluid and neutrinos.

\noindent{\bf Ideal MHD}.
The hyperbolic structure of the ideal MHD equations allows for seven
physical waves: two Alfv\'en, two fast and two slow magnetosonic, and one entropy wave. The characteristic structure of these equations in the fully relativistic case was studied by Anile \cite{1989cup..book.....A}. It was found that only the entropic and the Alfvén waves can be explicitly
written in closed form, while the other four velocities are found by solving a quartic polynomial.

The seven waves can be ordered as follows
\begin{eqnarray}
	\lambda^+_\mathrm{fast} \ge \lambda^+_\mathrm{alfven} \ge \lambda^+_\mathrm{slow} \ge \lambda_\mathrm{entropic}
	\ge \lambda^-_\mathrm{slow} \ge \lambda^-_\mathrm{alfven} \ge \lambda^-_\mathrm{fast}.
	\label{seven_waves} \nonumber
\end{eqnarray}

A very useful upper bound for fast waves (which have the maximum speed)
can be found by considering the degenerate case of normal propagation~\cite{2003ApJ...589..444G,Del_Zanna_2007}. In that case there is an analytical expression for the
two fast magnetosonic waves. Considering only the $x$-direction, it can be written as
\begin{widetext}
\begin{eqnarray}\label{fast_waves}
	{\lambda^x}^{\pm}_\mathrm{fast} &=& -\beta^x + \frac{\alpha}{1 - v^2 a^2}
	\left[ (1-a^2) v^x \pm \sqrt{ a^2 (1-v^2) [(1-v^2 a^2)\gamma^{xx}
		- (1-a^2)(v^x)^2] }
	\,\right]
\end{eqnarray}
\end{widetext}
where
\begin{eqnarray}
	a^2 = c_s^2 + c_a^2 - c_s^2 c_a^2.
\end{eqnarray}
Here $c_s$ is the sound speed (which for an ideal EoS is just $c_s^2 = (\Gamma p)/h$ c) and $c_a$ is the Alfv\'en speed, which can be written in terms of the comoving magnetic four-vector $b^{\mu}$ as follows
\begin{eqnarray}
	c_a^2 = \frac{b^2}{h + b^2} ~~~~,~~~ b^2 = B^2/W^2 + (v_k B^k)^2 ~~.
\end{eqnarray}

\noindent{\bf M1 neutrino}.
The hyperbolic structure of the M1 formalism is straightforward in the limiting regimes, either when the neutrinos are free-streaming (i.e., optically thin) or when they are trapped in the fluid (i.e., optically thick). The velocities can be calculated both in the thin and thick regimes, and then interpolated using the Minerbo closure
\begin{equation}\label{minerbo_velocity}
	\lambda^i_\mathrm{M1} = \frac{3 \chi - 1}{2} \lambda_\mathrm{thin}^{i}
	+ \frac{3 (1- \chi)}{2} \lambda_\mathrm{thick}^{i}
\end{equation}
where
\begin{widetext}
	\begin{eqnarray}
		\lambda_\mathrm{thin}^{i} &=& |\beta^i| + \alpha \max \left[ \frac{|F^i|}{\sqrt{F_k F^k}} , \frac{E\, |F^i|}{F_k F^k}  \right]\\
		\lambda_\mathrm{thick}^{i} &=& |\beta^i| + \alpha \max \left[ \frac{2 W^2 |v^i| + \sqrt{(2 W^2 + 1) \gamma^{ii} - 2 W^2 v^i v^i}}{2 W^2 + 1}, |v^i| \right] ~.\nonumber
	\end{eqnarray}
\end{widetext}


\section{Time integration of equation with stiff source}

Note that we can split the vector of all fields $\bf{U}$ into two parts, one 
containing those fields for which their evolution equations contain stiff terms
and the other for which all terms can be treated explicitly, respectively $\bf{U} = (\bf{V},\bf{W})$.
The evolution equations for both types of fields can then  be written generally as
\begin{eqnarray}\label{split}
	\partial_t {\bf W} &=& \mathcal{F}_W({\bf V},{\bf W}) \\
	\partial_t {\bf V} &=& \mathcal{F}_V({\bf V},{\bf W}) + \frac{1}{\epsilon({\bf W})} \mathcal{R}_V({\bf V},{\bf W}) ~~.
\end{eqnarray}
where we have considered that the relaxation parameter $\epsilon$ can depend also on the ${\bf W}$ fields. 
Here we have used the same notation as in Section~\ref{sect:IMEX},
where $\mathcal{R}$ accounts for the stiff terms while $\mathcal{F}$ accounts for the non-stiff ones.
Therefore, the flux terms that can be treated explicitly are $\mathcal{F}_W$ and $\mathcal{F}_V$ while the stiff terms constitute $\mathcal{R}_V$.
The evolution procedure to compute each step $U^{(i)}$
can be split in two substeps
\begin{enumerate}
	\item compute the explicit intermediate values $\{\bf{V^*},\bf{W^*}\}$, that is,
	\begin{eqnarray}\label{first_step}
		{\bf W}^{*} = {\bf W}^n &+& \Delta t~ \sum_{j=1}^{i-1}~ {\tilde{a}}_{ij} \mathcal{F}_W({\bf U}^{(j)}) 
		\nonumber \\
		{\bf V}^{*} = {\bf V}^n &+& \Delta t~ \sum_{j=1}^{i-1}~ {\tilde{a}}_{ij} \mathcal{F}_V({\bf U}^{(j)}) 
		\nonumber \\
		&+& \Delta t~ \sum_{j=1}^{i-1}~ a_{ij} \frac{1}{\epsilon^{(j)}} \mathcal{R}_V({\bf U}^{(j)})  
	\end{eqnarray} 
	where we have defined $ \epsilon^{(j)} = \epsilon({\bf W}^{(j)})$ 
	and the coefficients $\tilde{a}_{ij}$ and $a_{ij}$ form the Butcher tableau for the explicit and implicit RK schemes, respectively.
	\item compute the implicit part, involving only ${\bf V}$, by solving the implicit equation
	\begin{eqnarray}\label{second_step}
		{\bf V}^{(i)} &=& {\bf V^*} 
		+ a_{ii}~\frac{\Delta t}{\epsilon^{(i)}}~\mathcal{R}_V({\bf V}^{(i)},{\bf W}^{(i)})
		\nonumber  \\
		{\bf W}^{(i)} &=& {\bf W}^{*} .
	\end{eqnarray}
\end{enumerate}

There are different ways to solve the equation in the second step (\ref{second_step}) depending on the nature of the implicit part.
If the source terms depend linearly on the evolution fields, then the stiff part can be written in the following way
\begin{eqnarray}\label{stiff_part}
	\mathcal{R}_V({\bf V},{\bf W}) = A({\bf W}) {\bf V} + S(\bf{W}).
\end{eqnarray}  
In this way, the implicit equation can be solved just by inverting the matrix, namely
\begin{eqnarray}\label{invert_matrix}
	{\bf V}^{(i)} &=& [I - a_{ii}~\frac{\Delta t} {\epsilon^{(i)}} A({\bf W^*})]^{-1} ~
	( {\bf V^*} + a_{ii}~\frac{\Delta t}{\epsilon^{(i)}}~S({\bf W^*}) )
	\nonumber  \\
	{\bf W}^{(i)} &=& {\bf W}^{*} .
\end{eqnarray}  

If instead the sources depend non-linearly on the evolution fields,
then one needs to find the zeros of the full non-linear implicit problem
\begin{eqnarray}\label{nonlinearstiff}
	{\bf G} &=& {\bf V}^{(i)} - {\bf V^*} 
	- a_{ii}~\frac{\Delta t}{\epsilon^{(i)}}~\mathcal{R}_V({\bf V}^{(i)},{\bf W}^{(i)}).
\end{eqnarray}

Another option would be to linearize the stiff term around $\{ {\bf V}'\}$ (assuming ${\bf W^{(i)}}$ is known), namely
\begin{eqnarray}\label{linearization2}
	\mathcal{R}_V({\bf V}^{(i)},{\bf W}^{(i)}) &\approx& \mathcal{R}_V({\bf V}',{\bf W}^{(i)}) \\
	&+& \left( \frac{\partial \mathcal{R}_V}{\partial {\bf V}} \right)_{{\bf V}',{\bf W}^{(i)}}
	({\bf V}^{(i)} - {\bf V}'). \nonumber 
\end{eqnarray}

By substituting the previous expansion (\ref{linearization2}) in (\ref{second_step}), 
adding and subtracting ${\bf V}'$ on the right-hand-side, and rearranging the terms, it is obtained an expression that can
be solved explicitly, namely
\begin{eqnarray}\label{invert_matrix2}
 M &\equiv& \left[I - a_{ii}~\frac{\Delta t} {\epsilon^{(i)}} \left( \frac{\partial \mathcal{R}_V}{\partial {\bf V}}
 \right)_{{\bf V}',{\bf W}^{(i)}} \right]^{-1} ~,
 \\
	{\bf V}^{(i)} &=& {\bf V}' + M [ {\bf V}^* - {\bf V}' +
	a_{ii}~\frac{\Delta t} {\epsilon^{(i)}} ~\mathcal{R}_V({\bf V}',{\bf W}^{(i)}) ] ~.
\nonumber	
\end{eqnarray}  

~\\
\noindent{\bf Stiff terms in the neutrino transport}. 
Following the approach outlined explicitly in Ref.~\cite{rad2022}, we describe here the prescription for handling the stiff terms in the neutrino equations. 

The evolution for the neutrino number density, given by  Eq.~\eqref{eq:Nneutrino}, has a linear stiff term of the type given in~\eqref{invert_matrix}. This allows for an analytic solution, namely 
\begin{eqnarray}
	{\bar N} = {\bar N}^* 
	+ a_{ii} \Delta t \alpha \sqrt{\gamma} \left(\eta^0 - \kappa_a^0  \frac{N}{\Gamma} \right)  .
\end{eqnarray}
By dividing by $\sqrt{\gamma}$ and defining
${N}^* \equiv {\bar N}^* /\sqrt{\gamma}$, this relation
can be inverted directly as
\begin{equation}
	N = \frac{N^* 
		+ a_{ii} \Delta t \alpha \eta^0}{
		1 + a_{ii} \Delta t \alpha \kappa_a^0 /\Gamma } .
\end{equation}
Because $\Gamma$ depends on $(E,F_i)$ as specified in Eq.~(\ref{Jminerbo}), namely
\begin{eqnarray}
	\Gamma = 
	W \left( \frac{E - F_i v^i}{J} \right), \label{eq:Gamma_nu_number}
\end{eqnarray}
this equation for $N$ will be solved only after these fields have been updated.

The M1 equations~(\ref{eq:Eneutrino}-\ref{eq:Fneutrino}),  governing the neutrino energy density and momentum,  are coupled and cannot be solved independently. Together, they constitute the full nonlinear system of the type given in \eqref{nonlinearstiff}. This results in a four-dimensional system that must be solved simultaneously using an iterative method. Again, for the undensitized variables (i.e., dividing by $\sqrt{\gamma}$), the equations to be solved are
\begin{eqnarray}
	G &=& E - E^* \\
	&-& a_{ii} \Delta t \alpha W \left[ (\eta + \kappa_s J) - (\kappa_a + \kappa_s) (E - F_i v^i) \right] \nonumber \\
	G_i &=& F_i - F_i^* \\
	&-& a_{ii} \Delta t \alpha \left[ W (\eta - \kappa_a J) v_i - (\kappa_a + \kappa_s) H_i  \right]. \nonumber
\end{eqnarray}
The multidimensional Newton-Raphson solver
involves the inverse of the Jacobian of ${\bf G}$ with respect to ${\bf V} =(E, {F}_i)$. The solution at the iteration $n$ can be calculated from the solution at the iteration $n-1$ as follows
\begin{equation}
	{\bf V}^{n}  = {\bf V}^{n-1} - \left(\frac{\partial {\bf G}  }{\partial {\bf V}}\right)^{-1} {\bf G}
\end{equation}
and where the Jacobian of $G$ can be written in terms of the Jacobian of $\mathcal{R}_V$ easily, namely
\begin{equation}
	\left(\frac{\partial {\bf G}  }{\partial {\bf V}}\right) = {\bf I}  - a_{ii}~\frac{\Delta t}{\epsilon^{(i)}} \left(\frac{\partial {\bf \mathcal{R}_V}  }{\partial {\bf V}}\right).
\end{equation}

The stiff problem involves the fields ${\bf V} = (E, {F}_i, N)$, although actually it can be decoupled into a part involving ${\bf V} = (E, {F}_i)$ and a simple implicit equation for $N$ to be solved after those fields have been updated. Let us define $\kappa_{as} \equiv \kappa_a + \kappa_s$
and write $J,H_i$ as a function of $E,J_i,\chi$ for the Minerbo closure, namely
\begin{eqnarray}\label{Jminerbo}
	&& J (E,F_i) = B_0 + d_\mathrm{thin} B_\mathrm{thin} + d_\mathrm{thick} B_\mathrm{thick} \\
	\label{Hminerbo}
	&&H_i (E,F_i) = - (a_{v 0} + d_\mathrm{thin} a_{v \mathrm{thin}} + d_\mathrm{thick} a_{v \mathrm{thick}} ) v_ i  \nonumber \\
	&&- d_\mathrm{thin} a_{f \mathrm{thin}} {\hat f}_i - (a_{F 0} + d_\mathrm{thick} a_{F \mathrm{thick}} ) F_i 
\end{eqnarray}
where we have defined 
\begin{equation}
	 {\hat f}_i = \frac{F_i}{\sqrt{F_k F^k}} ~,~
	d_\mathrm{thick} = \frac{3}{2} (1- \chi) ~,~
	d_\mathrm{thin} = 1 - d_\mathrm{thick}~, \nonumber \\
\end{equation}
and the following coefficients	
\begin{eqnarray}	
	&& B_0 = W^2 \left[ E - 2 v_k F^k \right] ~~,\\
	&& B_\mathrm{thin} = W^2 E (v_k {\hat f}^k)^2 ~~,\\
	&& B_\mathrm{thick} = \frac{W^2-1}{2 W^2 + 1} \left[
	4 W^2 (v_k F^k) + (3 - 2 W^2) E  \right] ~~,\\
	&& a_{v 0} = W B_0 ~~,\\
	&& a_{v \mathrm{thin}} = W B_\mathrm{thin} ~~,\\
	&& a_{v \mathrm{thick}} = W B_\mathrm{thick} \\
	&& + \frac{W}{2 W^2 + 1}
	\left[ (2 W^2 - 1) (v_k F^k) + (3 - 2 W^2) E \right] ~~, \nonumber \\
	&& a_{f \mathrm{thin}} = W E (v_k {\hat f}^k) ~~, \\
	&& a_{F 0} = - W ~~,\\
	&& a_{F \mathrm{thick}} =  W v^2  ~~.
\end{eqnarray}
The Jacobian ${\cal J} \equiv \left(\frac{\partial {\bf \mathcal{R}_V}  }{\partial {\bf V}}\right)$ of the undensitized fields is then given by
\begin{eqnarray}
	{\cal J}_{00} &=& -\alpha W \left( \kappa_{as} - \kappa_{s} 
	\frac{\partial J}{\partial E} \right) ~,\\
	{\cal J}_{0j} &=& +\alpha W \left( \kappa_{s} \frac{\partial J}{\partial F_j} + \kappa_{as} v^j 
	\right) ~,\\
	{\cal J}_{i0} &=& -\alpha \left( \kappa_{as} \frac{\partial H_i}{\partial E}
	+ W \kappa_{a} \frac{\partial J}{\partial E} v_i
	\right) ~,\\
	{\cal J}_{ij} &=& -\alpha \left( \kappa_{as} \frac{\partial H_i}{\partial F_j}
	+ W \kappa_{a} v_i \frac{\partial J}{\partial F_j} 
	\right) ~.
\end{eqnarray}
where the necessary derivatives are
\begin{widetext}
	\begin{eqnarray}
		&&\frac{\partial J}{\partial E} = W^2 + d_\mathrm{thin} 
		(v_k {\hat f}^k)^2 W^2 
		+ d_\mathrm{thick} \frac{(3-2 W^2)(W^2 - 1)}{1 + 2 W^2}~,\\
		&&\frac{\partial J}{\partial F_j} = 2 W^2 \left( 
		-1 + d_\mathrm{thin} \frac{E (v_k {\hat f}^k)}{F} 
		+ 2 d_\mathrm{thick} \frac{W^2 - 1}{1 + 2 W^2}  \right) v^j
		- 2 d_\mathrm{thin} \frac{W^2 E (v_k {\hat f}^k)^2}{F}
		{\hat f}^j  ~,\\
		&&\frac{\partial H_i}{\partial E} = W^3 \left( 
		-1 - d_\mathrm{thin} (v_k {\hat f}^k)^2 
		+ d_\mathrm{thick} \frac{2 W^2 - 3}{1 + 2 W^2}  \right) v_i
		- d_\mathrm{thin} W (v_k {\hat f}^k) {\hat f}_i  ~,\\
		&&\frac{\partial H_i}{\partial F_j} = W \left( 
		1 - d_\mathrm{thin} \frac{E (v_k {\hat f}^k)}{F} 
		- d_\mathrm{thick} v^2  \right) \delta^j_i
		+ 2 W^3 \left[ 1 - d_\mathrm{thin} \frac{E (v_k {\hat f}^k)}{F} 
		- d_\mathrm{thick} \left( v^2 + \frac{1}{2 W^2 (1 + 2 W^2)}  \right)     \right] v_i v^j \nonumber \\
		&&+ 2 d_\mathrm{thin}  \frac{W E (v_k {\hat f}^k)}{F}  {\hat f}_i {\hat f}^j
		+ 2 d_\mathrm{thin}  \frac{W^3 E (v_k {\hat f}^k)^2}{F}  v_i {\hat f}^j
		- d_\mathrm{thin}  \frac{W E}{F}  {\hat f}_i v^j  ~~.
	\end{eqnarray}
\end{widetext}

\section{Conversion from conserved to primitive fields}
\label{reprimand}

In this appendix, we outline the procedure used to recover the primitive variables from the conserved quantities when employing a finite-temperature equation of state. 
Note that this routine is based on the recovery procedure implemented in RePrimAnd~\cite{kastaun20}, and extends it to support tabulated equations of state. Our routine, which can be compiled with CUDA, will be released as part of the open-source distribution of \mhduetX~\cite{mhduet_webpage}.

The list of primitive fields is given by $\{\rho, T, Y_e, p, v_i, B^i\}$,  with the pressure $p=p(\rho,T,Y_e)$ usually given in a tabulated form. The evolved conserved variables are defined as
\begin{eqnarray}
	D &\equiv& \rho W \nonumber \\
	S_i &\equiv& \left(h W^2 + B^2\right)v_i - \left( B^j v_j\right) B_i  \nonumber  \\
	\tau &\equiv& h W^2 + B^2 - P - \frac{1}{2}\left(\left(B^i v_i\right)^2 + \frac{B^2}{W^2}\right)
	-D \nonumber  \\
	D_Y  &\equiv& \rho W Y_e, \nonumber 
\end{eqnarray}
where the enthalpy $h=\rho (1 + \epsilon) + p$ is a function of the specific internal energy $\epsilon$. The value of $\epsilon$ is obtained from 
the EoS via a table lookup, that is, $\epsilon = \epsilon(\rho, T, Y_e)$.

The primitive solution is computed by following very closely the approach described in ~\cite{kastaun20,10.1093/mnras/stab2606}, with minor modifications to enhance the efficiency and accuracy of the solver. The procedure consists of the following steps:
\begin{enumerate}
	\item Calculate some quantities which are fixed during the iterations, namely the rescaled variables
	\begin{align}
		q \equiv \frac{\tau}{D}, \qquad r_i \equiv \frac{S_i}{D}, \qquad \mathcal{B}^i \equiv \frac{B^i}{\sqrt{D}}, 
	\end{align}
	and then the following useful relations
	\begin{align}
		{	r^2 \equiv r^i r_i~, \quad \mathcal{B}^2 \equiv \mathcal{B}^i\mathcal{B}_i~, \quad \mathcal{B}^2r^2_\perp \equiv \mathcal{B}^2 r^2 - (r^l\mathcal{B}_l)^2.}
	\end{align}
	\item In the interval $\left( 0, h_0^{-1} \right]$, solve:
	\begin{equation}\label{eq:f_mu_plus}
		f_a (\mu) = \mu \sqrt{ h_0^2 + \bar{r}^2(\mu)} - 1,
	\end{equation}
	where $h_0$ is the relativistic enthalpy lower bound over {the} entire validity region of the EoS and 
	\begin{align}
		&\bar{r}^2 (\mu) = r^2 \chi^2(\mu) + \mu \chi(\mu) \left( 1 + \chi(\mu) \right) \left( r^l \mathcal{B}_l \right)^2\label{eq:r_bar_sqr}, \\
		&\chi (\mu) = \frac{1}{1 + \mu \mathcal{B}^2}\label{eq:chi} .
	\end{align}
	Here the root of $f_a$ {in eq.\eqref{eq:f_mu_plus}} is denoted as $\mu_+$.
	We numerically solve eq.~\eqref{eq:f_mu_plus} using the Brent's method, which is usually a very efficient bracketing method.

	\item In the interval $\left( 0, \mu_+ \right]$, solve:
	\begin{equation}\label{eq:f_mu}
		f (\mu) = \mu - \frac{1}{\hat{\nu} + \mu \bar{r}^2(\mu)},
	\end{equation}
	where
	\begin{align}
		&\hat{\nu} (\mu) = \max(\nu_A (\mu),\nu_B (\mu)), \\
		&\nu_A (\mu) = \left( 1 + \hat{a} (\mu) \right) \frac{1 + \hat{\epsilon} (\mu)}{\hat{W} (\mu)}, \\
		&\nu_B (\mu) = \left( 1 + \hat{a} (\mu) \right) \left( 1 + \bar{q} (\mu) - \mu \bar{r}^2 (\mu) \right) \\
		&\hat{p} (\mu) = p\left(\hat{\rho} (\mu),\hat{\epsilon} (\mu)\right) , \qquad \hat{a} (\mu) = \frac{\hat{p} (\mu)}{\hat{\rho} (\mu) \left( 1 + \hat{\epsilon} (\mu) \right) } ,\\
		&\hat{\rho} (\mu) = \frac{D}{\hat{W} (\mu)}, \\
		&\hat{\epsilon} (\mu) = \hat{W} (\mu) \left( \bar{q} (\mu) - \mu \bar{r}^2 (\mu) \right) + \hat{v}^2 (\mu) \frac{\hat{W}^2 (\mu)}{1 + \hat{W} (\mu)} ,\\
		&\hat{v}^2 (\mu) = \min( \mu^2 \bar{r}^2 (\mu), v_0^2 ), \quad \hat{W} (\mu) = \frac{1}{\sqrt{1-\hat{v}^2 (\mu)}}, \\
		&\bar{q} (\mu) = q - \frac{1}{2} \mathcal{B}^2 - \frac{1}{2} \mu^2\chi^2 (\mu) \left( \mathcal{B}^2 r_\perp^2 \right),
	\end{align}
	$\bar{r}^2 (\mu)$ and $\chi(\mu)$ {are} defined {in} eq.~\eqref{eq:r_bar_sqr} and eq.~\eqref{eq:chi}, and the upper velocity limit square $v_0^2$ is defined as $v_0^2 \equiv r^2 / (h_0^2 + r^2) < 1$.
	We numerically solve eq.~\eqref{eq:f_mu} using again  the robust Brent's method.
	Note that during the iterations, we enforce the density $\rho$ and the specific energy $\epsilon$ fall within the validity region of the EoS, i.e., 
	we {evaluate} the updated $\rho$ and $\epsilon$ with $\hat{\rho} = \max \left( \min \left( \rho_{\max} ,\hat{\rho} \right), \rho_{\min} \right)$ and $\hat{\epsilon} = \max \left( \min \left( \epsilon_{\max}(\hat{\rho}) ,\hat{\epsilon} \right), \epsilon_{\min}(\hat{\rho}) \right)$.
	\item With the root $\mu$ of eq.~\eqref{eq:f_mu}, we can then work out the primitive variables $[\rho, \epsilon, p]$ respectively with the equations used in step 3.
	{The velocity $v^i$ can be obtained with $\mu$ by}:
	\begin{equation}
		\hat{v}^i (\mu) = \mu \chi (\mu) \left( r^i + \mu \left( r^l \mathcal{B}_l\right) \mathcal{B}^i \right).
	\end{equation}
\end{enumerate}

Note that we assume positive baryon number density, positive total energy density, and positive pressure.
Because the definitions of the specific energy $\epsilon$ and the relativistic enthalpy depend on the arbitrary choice of the mass constant $m_B$, relations such as $\epsilon > 0$ or $h \geq 1$ may not hold in general.
For example, negative specific energy $\epsilon$ is possible in nuclear physics equations of state.

Although some primitive variables are forced to fall within the validity region during the iterations, this conversion from conserved to primitive variables occasionally returns unphysical results, especially at the surfaces of neutron stars.
These errors are mostly harmless and can be corrected.
After the primitive variables are obtained, we check whether any correction is needed.
Some corrections are allowed only for low density regions or well within any black hole horizon.
Here, a point is considered within a low density region if $\rho < \rho_{\rm{low}}$ and inside a black hole if $\alpha < \alpha_{\rm{BH}}$.

The error handling process at any given point is the following:
\begin{enumerate}
	\item $\rho < \rho_{\min}$: set the point to atmosphere.
	In particular, we set the rest-mass density to be $\rho_\text{atmo}$, the velocity is set to be zero, and then update the rest of the primitive variables such as pressure $p$ and specific energy density $\epsilon$ using the equation of state.
	\item $\rho > \rho_{\max}$: a fatal error, stop the code.
	\item $\epsilon < \epsilon_{\min}$: set $\epsilon = \epsilon_{\min}$.
	\item $\epsilon > \epsilon_{\max}$: set $\epsilon = \epsilon_{\max}$ if the point is in a  low density or black hole region, and otherwise consider it a fatal error.
	\item $v > v_{\max}$: if the point is not in a low density or black hole region, consider it a fatal error. Otherwise,
	rescale the velocity such that $v = v_{\max}$ as well as the Lorentz factor $W_{\max}$.
	Here we keep the conserved density $D$ fixed, such that the rest-mass density $\rho$ increases slightly.
	\item $Y_e$ is out of range: if the point is not in a low density or black hole
    region, consider it a fatal error. Otherwise,
    set $Y_e$ to be in $\beta$-equilibrium with the atmosphere state.
\end{enumerate}

\section{Calculating the emission and absorption coefficients}
\label{app:neutrino}

We compute the source terms in the M1 equations described in Section~\ref{sec:M1} by assuming that each neutrino species $\nu_i\in\{ \nu_e,\overline{\nu}_e,\nu_x \}$ obeys a Fermi-Dirac distribution with temperature $T_{\nu}$ and chemical potential $\mu_{\nu}$. In that case, we have
\begin{equation}
	f (\epsilon_{\nu}) = \frac{1}{1+\exp[(\epsilon_{\nu} - \mu_{\nu})/(k_B T_{\nu})]}
\end{equation}
where $\epsilon_{\nu_i}$ is the neutrino energy and $k_B$ is the Boltzmann constant. Within this assumption, the gray opacities can be computed as

\begin{align}
	\kappa_{a,s} &= \frac{\int_0^\infty \kappa_{a,s} (\epsilon_{\nu})  f(\epsilon_{\nu}) \epsilon_{\nu}^3 d \epsilon_{\nu}}{\int_0^\infty  f(\epsilon_{\nu}) \epsilon_{\nu}^3 d \epsilon_{\nu}} \label{eq:gray_opacity_energy} \\
	\kappa_{a}^0 &= \frac{\int_0^\infty \kappa_{a} (\epsilon_{\nu})  f(\epsilon_{\nu}) \epsilon_{\nu}^2 d \epsilon_{\nu}}{\int_0^\infty   f(\epsilon_{\nu}) \epsilon_{\nu}^2 d \epsilon_{\nu}} \label{eq:gray_opacity_number}
\end{align}
where the spectral opacities $\kappa_{a,s}(\epsilon_{\nu})$ can be computed either by interpolation from 3D 
tables in \texttt{NuLib} format~\cite{OConnor:2014sgn},
or computed on the fly with the simplified formula of Ref.~\cite{Ruffert:1995fs}.
The blackbody functions for the neutrino energy and number density are
\begin{align}
	B_{\nu} &= g_i \frac{4 \pi}{(h c)^3} (k_B T_{\nu})^4 F_3 (\eta_{\nu}) \\
	{\cal B}_{\nu} &= g_i \frac{4 \pi}{(h c)^3} (k_B T_{\nu})^3 F_2 (\eta_{\nu})
\end{align}
where $F_n$ is the Fermi function of order $n$
and $\eta_{\nu} = \mu_{\nu}/(k_B T)$ is the degeneracy parameter of the neutrinos. Since $[hc] = \mathrm{ergs} \,cm$ and $[k_B T] = \mathrm{ergs}$, then $[B_{\nu}] = \mathrm{ergs}/cm^{3}$ and $[{\cal B_{\nu}}] = 1/cm^{3}$. Here $g_{\nu_e}=g_{\overline{\nu}_e}=1$ and $g_{\nu_x}=4$. 

The blackbody expressions could be used to compute a neutrino temperature $T_{\nu}$ and chemical potential $\mu_{\nu}$ by setting $J_{\nu}= B_{\nu}$, $n_{\nu}= {\cal B}_{\nu}$ and solving those two relations for $T_{\nu}$ and $\mu_{\nu}$, from which one can obtain gray opacities using Eqs. \eqref{eq:gray_opacity_energy} and \eqref{eq:gray_opacity_number}.

In this work, however, we adopt a simplified approach commonly used in the BNS community. We first compute the opacities $\kappa^\mathrm{eq}_{a}$ and $\kappa^\mathrm{eq}_{s}$ assuming matter and neutrinos are in local thermodynamical equilibrium (i.e., equilibrium of temperature, chemical potential,...). That means that
the neutrino temperature $T_{\nu}$ is taken to be the fluid temperature, while the neutrino chemical potentials are evaluated at equilibrium using the EoS table at the fluid density, temperature and electron fraction, separately for each neutrino flavor, as follows
\begin{equation}\label{chemical_potential}
	\mu_{\nu_e} = \mu_{e} + \mu_{p} - \mu_{n} ~~,~~
	\mu_{\bar{\nu}_e} = - \mu_{\nu_e} ~~,~~ \mu_{\nu_x} = 0~~.
\end{equation}

At this point we introduce a correction factor in the energy-integrated opacities accounting for out-of-equilibrium effects~\cite{Foucart:2016rxm}, namely
\begin{eqnarray}
	&&\kappa_{a,s} = \kappa^{\mathrm eq}_{a,s} \left[ \max \left( 1, \frac{T_{\nu}}{T} \right) \right]^2 = \kappa^{\mathrm eq}_{a,s} \left[\max \left(1,\frac{\langle\epsilon_{\nu}\rangle}{\langle\epsilon_{\mathrm eq}\rangle}\right) \right]^2  ~~,~~ \label{eq:kappa_a_correction}\\
	&& \kappa^0_{a} = \kappa^{0, \mathrm eq}_{a} \left[ \max \left( 1, \frac{T_{\nu}}{T}\right)\right]^2 = \kappa^{0, \mathrm eq}_{a}  \left[\max \left(1, \frac{\langle\epsilon_{\nu}\rangle}{\langle\epsilon_{\mathrm eq}\rangle}\right)\right]^2  ~~,~~ \label{eq:kappa_s_correction}
\end{eqnarray}
where $\langle\epsilon_{\mathrm eq}\rangle=B_{\nu}/{\cal B}_{\nu}$, that is, the average energy of the blackbody spectra of the neutrinos in equilibrium with the fluid, and $\langle \epsilon_{\nu} \rangle$ can be computed as 
\begin{equation}
	\langle \epsilon_{\nu} \rangle = \frac{J}{n}
	= \frac{\Gamma J}{N} = \frac{W (E - F^i v_i)}{N}
	\label{eq:average_epsilon}.
\end{equation}
with $\Gamma$ being the conversion factor between neutrino number density in the fluid and Eulerian frame, as defined in Eq. \eqref{eq:Gamma_nu_number}.

In the second equality of Eqs. \eqref{eq:kappa_a_correction} and \eqref{eq:kappa_s_correction}, we used the fact that the average energy is related to the neutrino temperature and chemical potential as
\begin{equation}
	\langle \epsilon_{\nu} \rangle = \frac{B_{\nu}}{{\cal B}_{\nu}} = \frac{F_3 (\eta_{\nu})}{F_2 (\eta_{\nu})} (k_B T_{\nu}) .
\end{equation}

In order to guarantee that the neutrinos equilibrate with the fluid in the optically thick regions, we compute the free-streaming emission due to charged-current reactions and the absorption due to pair processes through an energy-integrated version of Kirchhoff’s law, namely
\begin{eqnarray}
	\eta_{\nu_e} &=& c\, \kappa^\mathrm{eq}_{a,\nu_e} B_{\nu_e} ~~,\\
	\eta_{ {\bar \nu}_e} &=& c\, \kappa^\mathrm{eq}_{a,{\bar \nu}_e} B_{{\bar \nu}_e} , \label{eq:Kirchhoff_nue}\\
	c\, \kappa^\mathrm{eq}_{a,\nu_x} &=& \frac{\eta_{\nu_x}}{B_{\nu_x}}. \label{eq:Kirchhoff_nux}
\end{eqnarray}
We apply the same treatment to the neutrino number emissivities and opacities, but using ${\cal B_\nu}$ instead of $B_\nu$.

In the case of the heavy-lepton neutrinos, the main production channel is nucleon-nucleon bremsstrahlung, while there is no absorption through charge current reactions (no muons), so we need to compute an effective absorption from the emission and not the other way around.

Notice that Kirchhoff's law as formulated here is not applicable to pair processes like bremsstrahlung, as in reality the opacity for each species will be proportional to the density of neutrinos of the corresponding anti-species. 
Despite the approach of Eq. \eqref{eq:Kirchhoff_nux} guarantees a correct thermal equilibration (assuming we know $\mu_{\nu_x}$), we will need in the future to switch to some more realistic approach, see  Ref.~\cite{Betranhandy:2025amv} or~\cite{Chiesa:2024lnu}.
Although the approach presented in Eq. \eqref{eq:Kirchhoff_nux} guarantees that the neutrinos will reach thermal equilibrium (assuming we know $\mu_{\nu_x}$), we will need to switch to some more realistic approach in the future, see  Ref.~\cite{Betranhandy:2025amv} or~\cite{Chiesa:2024lnu}.

In summary, the logical procedure for $(\nu_e,\overline{\nu}_e)$ is
\begin{eqnarray}
	&& (\rho,T,Y_e) \rightarrow (\mu_e, \mu_p, \mu_n) \rightarrow \mu_{\nu}  \\
	&& (\mu_{\nu},T,Y_e) \rightarrow (\kappa_{a}^\mathrm{eq},\kappa_{s}^\mathrm{eq},B_{\nu},\mathcal{B}_{\nu}) \\
	&& (J,n,\mu_{\nu}) \rightarrow \langle\epsilon_{\nu}\rangle, T_{\nu} \\
	&& (\kappa_{a}^\mathrm{eq},\kappa_{s}^\mathrm{eq},\langle\epsilon_{\nu}\rangle,B_{\nu},\mathcal{B}_{\nu}) \rightarrow (\eta_{\nu},\kappa_{a},\kappa_{s}).
\end{eqnarray}

For simplicity, opacities are kept fixed throughout the implicit step root-finder. This can cause numerical schemes to oscillate if neutrinos are thrown out of equilibrium over a timescale,
$\tau = (c \sqrt{\kappa_a (\kappa_a + \kappa_s)})^{-1}$, small compared with $\Delta t$. However, a full inclusion of the opacity dependence on the radiation fields in the root finder presents a numerical challenge we leave for a future work.

Finally, in optically thick regions, we apply two modifications to the scheme previously described. This is done in order to damp oscillations and improve the code stability. The first one consists in suppressing the out-of-equilibrium correction to the opacities and setting directly $\kappa_{a,s}=\kappa_{a,s}^\mathrm{eq}$ (and similar for the number opacity $\kappa_{a}^0$). In such regions, deviations from equilibrium are indeed negligible and the inclusion of the opacity energy dependence would only induce oscillations worsening the thermal equilibration property. The second modification consists in computing the blackbody function using $T_\mathrm{eq}$ and $Y_{e,\mathrm{eq}}$, namely the equilibrium values of temperature and electron fraction that the fluid-radiation system should reach at the end of the time step as described in Ref.~\cite{Radice:2021jtw}. An interpolation between the two schemes is performed in gray regions according to a transition function, with the final opacities and blackbodies given by
\begin{align}
	\kappa_{a,s} &= f(\kappa_\mathrm{tot}) \kappa_{a,s}^\mathrm{eq} \left[\max \left(1, \frac{\langle\epsilon_{\nu}\rangle}{\langle\epsilon_{\mathrm eq}\rangle}\right)\right]^2  + \left[1-f(\kappa_\mathrm{tot})\right]\kappa_{a,s}^\mathrm{eq} \\
	B_\nu &= f(\kappa_\mathrm{tot}) B_\nu(T,Y_e) + \left[1-f(\kappa_\mathrm{tot})\right]B_\nu(T_\mathrm{eq},Y_{e,\mathrm{eq}})
\end{align}
with $\kappa_\mathrm{tot} = \sqrt{\kappa_a (\kappa_a + \kappa_s)}$. The same is done for $\kappa_{a}^0$ and $\mathcal{B}_\nu$. As a transition function we use
\begin{equation}
	f(\kappa_\mathrm{tot}) = e^{-\kappa_\mathrm{tot}/\kappa_\mathrm{lim}}
\end{equation}
where $\kappa_\mathrm{lim}$ is a free parameter. In our tests we set $\kappa_\mathrm{lim}=10[M_\odot^{-1}]$. Notice that our choice of $f(\kappa_\mathrm{tot})$ and $\kappa_\mathrm{lim}$ only activates such thick regime corrections when the fluid-neutrino equilibration time is much smaller than the dynamical timescale of the simulation. In this regime, the assumptions $\kappa_{a,s}=\kappa_{a,s}^\mathrm{eq}$, $T=T_\mathrm{eq}$ and $Y_e=Y_{e,\mathrm{eq}}$ are exact, leading to no bias in the simulation result. Moreover, the choice of a transition function $f(\kappa_\mathrm{tot})$ that does not depend on $\Delta t$ avoids breaking the code's convergence.

\bibliographystyle{utphys}
\bibliography{biblio}

\end{document}